\documentclass[10pt,nofootinbib]{revtex4-2}
\usepackage[utf8]{inputenc}
\usepackage{amsmath, amssymb, bm, braket,times, dsfont, amsthm,mathtools}
\usepackage{graphicx,xfrac,geometry}
\usepackage[dvipsnames]{xcolor}
\usepackage[breaklinks,colorlinks,allcolors=blue]{hyperref}
\usepackage{framed}
\usepackage{cleveref}
\usepackage{tikz-cd}
\usepackage{tikz}
\usetikzlibrary{backgrounds,intersections}
\usepackage{xcolor}
\usepackage[most]{tcolorbox}
\usepackage{setspace}
\makeatletter
\def\l@subsubsection#1#2{}
\makeatother

\usepackage[normalem]{ulem}
\usepackage{mdframed}
\usepackage{tgpagella}
\usepackage{mathpazo}
\tcbset{
    sharp corners,
    colback = white,
    before skip = 0.2cm, 
    after skip = 0.5cm  
}  
\newtcolorbox[no counter]{thmbox}[2][]{
  enhanced,
  breakable,
  sharp corners,
  boxrule=0pt,
  #1
}
\definecolor{main}{HTML}{5989cf}    
\definecolor{sub}{HTML}{cde4ff}     
\newtcolorbox{boxE}{
    enhanced, 
    boxrule = 0pt, 
    borderline = {0.75pt}{0pt}{main}, 
    borderline = {0.75pt}{2pt}{sub} 
}
\definecolor{zima}{HTML}{5BC2E7}
\definecolor{bleudefrance}{rgb}{0.19, 0.55, 0.91}
\definecolor{cyan(process)}{rgb}{0.0, 0.72, 0.92}
\definecolor{jasper}{rgb}{0.84, 0.23, 0.24}
\definecolor{frenchrose}{rgb}{0.96, 0.29, 0.54}
\definecolor{folly}{rgb}{1.0, 0.0, 0.31}
\definecolor{junglegreen}{rgb}{0.16, 0.67, 0.53}
\definecolor{mediumseagreen}{rgb}{0.24, 0.7, 0.44}
\theoremstyle{remark}
\newtheorem{theorem}{Theorem}[section]
\theoremstyle{definition}
\newtheorem{definition}{Definition}[section]
\newtheorem{lemma}{Lemma}[section]
\newtheorem{proposition}{Proposition}[section]

\newtheorem{example}{Example}[section]
\newtheorem{ccmt}{Comment}[section]

\tcbuselibrary{theorems}
\usetikzlibrary{positioning,intersections}
\usetikzlibrary{calc}
\tikzset{curve/.style={settings={#1},to path={(\tikztostart)
    .. controls ($(\tikztostart)!\pv{pos}!(\tikztotarget)!\pv{height}!270:(\tikztotarget)$)
    and ($(\tikztostart)!1-\pv{pos}!(\tikztotarget)!\pv{height}!270:(\tikztotarget)$)
    .. (\tikztotarget)\tikztonodes}},
    settings/.code={\tikzset{quiver/.cd,#1}
        \def\pv##1{\pgfkeysvalueof{/tikz/quiver/##1}}},
    quiver/.cd,pos/.initial=0.35,height/.initial=0}

\renewcommand{\>}{\rangle}
\newcommand{\<}{\langle}

\newcommand{\rep}{\text{Rep}}
\newcommand{\To}{\Rightarrow}

\newcommand{\vc}{\mathsf{Vec}}

\newcommand{\obj}{\mathsf{Obj}}

\newcommand{\symto}{SymTO~}
\newcommand{\bref}{\mathsf{B^{sym}}}
\newcommand{\bphys}{\mathsf{B^{phys}}}

\newcommand{\eA}{\e{A}}

\renewcommand{\l}{\left(}
\renewcommand{\r}{\right)}
\usepackage{enumitem}
\usepackage{euscript}
\newcommand{\e}[1]{\EuScript{#1}}
\newcommand{\bb}[1]{\mathbb{#1}}
\newcommand{\cc}[1]{\mathcal{#1}}

\newcommand{\ccat}{2-categroy~}
\newcommand{\bZ}{\mathbb{Z}}
\newcommand{\bC}{\mathbb{C}}

\renewcommand{\L}{\e{L}}
\newcommand{\cg}[2]{{\vphantom{#2}}^{#1}#2}
\newcommand{\hide}[1]{}

\newenvironment{tikzpic}[1][]{
			\begin{tikzpicture}[scale=0.5,baseline={([yshift=-.8ex]current bounding box.center)},#1]
				}{
			\end{tikzpicture} 
			}

\tikzset{sheet1/.style={draw = black!30, line width = 0.8pt}}
\tikzset{dot/.style={circle, scale=0.25, fill=black, thick, draw}}
\tikzset{label/.style={scale=0.6}}

\begin{document}
\title{String condensation and topological holography for 2+1D gapless SPT}
\author{Rui Wen}
\address{Department of Physics and Astronomy, and Stewart Blusson Quantum Matter Institute, University of British Columbia}
\email{wenrui1024@phas.ubc.ca}
\begin{abstract}
   The theory of anyon condensation is the foundation of the bulk-boundary relation and topological holography in 2+1D/1+1D. It is believed string condensation should replace anyon condensation in the 3+1D/2+1D topological holography theory. In this work we study string condensations in 3+1D topological orders and their relations to 2+1D phases. We find that a class of non-Lagrangian condensable algebras in 3+1D are exactly dual to a class of 2+1D symmetry enriched gapless phases known as gapless SPTs(gSPT). We show how topological properties of a gSPT can be fully extracted from the dual string condensation. We give an algebraic classification of this class of condensable algebras in 3+1D $G$-gauge theories that we call magnetic and simple. Through the topological holography dictionary, this maps to the classification of 2+1D $G$-symmetric phases with no topological order, including gapped and gapless SPTs. Utilizing the classification, we identify three classes of gSPTs and study their properties and gauging. Along the way, we reveal physical structures of string condensations. 
\end{abstract}
\maketitle
\tableofcontents
\newpage
\section{Introduction and summary of results}
Bulk-boundary correspondence is a foundational principle of condensed matter physics, established in the study of topological phases of matter~\cite{chen2013symmetry,chen2011two,Wen_2017,Johnson_Freyd_2022,Lan_2016,Lan_2018,Lan_2019,Kong_2020,kong2015boundarybulkrelation,Kong_2017,kong2020algebraic,ji2020categorical,ji2021unified,barkeshli2022classification,bulmash2020absolute}. For (non-chiral)2+1D topological orders, this correspondence relies on the theory of anyon condensation~\cite{kong2014anyon,kitaev2012models,burnell2018anyon,burnell2011condensation,bais2009condensate,barkeshli2013theory,barkeshli2010anyon}. Namely a gapped boundary of a 2+1D topological order is determined by how anyons of the bulk condense on the boundary. Anyon condensation on the other hand is governed by commutative algebras in braided fusion categories. With the development of this mathematical subject~\cite{davydov2013witt,etingof2016tensor}, the bulk-boundary correspondence achieved tremendous success in 2+1D/1+1D and has found application to topological quantum computation~\cite{cong2016topological,Cong_2017,kesselring2018boundaries}. 

Anyon condensation is also crucial to the topological holography paradigm, where topological properties of phases of matter in $d+1$-dimensions are claimed to be fully classified and characterized by properties of a $(d+1)+1$-dimensional topological order, called the symmetry topological order(\symto)~\cite{chatterjee2023symmetry, chatterjee2023emergent,chatterjee2023holographic} or the symmetry topological field theory(SymTFT)~\cite{freed2014relativequantumfieldtheory,kong2015boundarybulkrelation,Kong_2017,Freed:2018cec,kong2020algebraic,Gaiotto_2021,Lichtman_2021,Kaidi:2022cpf,Apruzzi_2023,Apruzzi_2022,Lin_2023,chatterjee2023emergent,chatterjee2023holographic,zhang2023anomalies,copetti2024defectchargesgappedboundary,lu2024realizing,copetti2024defectchargesgappedboundary,antinucci2024holographic}. More specifically, different phases of matter in $d+1$-D correspond to different condensation patterns of the \symto, which can be thought of as patterns of categorical symmetry breaking~\cite{kong2020algebraic,ji2020categorical,ji2021unified,chatterjee2023holographic,chatterjee2023symmetry}.The topological holography paradigm is mostly well-understood in 2+1D/1+1D, due to the well-established bulk-boundary relation of 2+1D topological orders and the theory of anyon condensation.

The topological holography paradigm has found wide success in the study of gapped phases with symmetries. It has become a standard tool in analyzing anomalies of non-invertible symmetries~\cite{Lin_2023,kaidi2023symmetry,zhang2023anomalies,apruzzi2023noninvertible,cordova2023anomalies,antinucci2023anomalies,antinucci2022holography}, various dualities were discovered and analyzed~\cite{choi2023non,bhardwaj2023non2,lu2024realizing,lu2024exploring,apruzzi2024symth,brennan2024symtft,antinucci2024anomalies}, and many novel gapped phases with generalized symmetries have been identified and analyzed through the lens of topological holography~\cite{bhardwaj2024lattice,chatterjee2024quantum,pace2023generalized,cao2024generating,Inamura_2024,bhardwaj2024illustrating,gorantla2024tensor,Delcamp_2024,fechisin2023non}. 

A much more challenging task is to systematically classify and characterize gapless phases and phase transitions. Even with ordinary invertible global symmetries, there are gapless systems with unusual bulk and boundary properties that are not described by the traditional Landau paradigm. These gapless systems are called symmetry enriched quantum criticality~\cite{Verresen_2021,Ye_2022,Mondal_2023,Yu_2022,Hidaka_2022,Tantivasadakarn_2023,wang2023stability,Ye_2024,Prembabu_2024,tantivasadakarn2023building} or gapless symmetry protected topological phases(gSPT)~\cite{Scaffidi_2017,Thorngren_2021,li2023intrinsicall,Wen_2023,wen2023classification,Li_2024,ando2024gauging,Su_2024,Yu_2024,Zhang_2024,myersonjain2024pristine,bhardwaj2023clubsandwich,bhardwaj2024hassediagrams}. It turns out that gSPTs are closely related to the concept of categorical symmetry breaking. A gapped phase may be viewed as a phase where the categorical symmetry is trivial in the IR limit, while a gSPT can be viewed as having nontrivial categorical symmetry in the IR~\cite{kong2020algebraic,ji2020categorical,ji2021unified,chatterjee2023holographic,bhardwaj2023clubsandwich,bhardwaj2024hassediagrams}. In the \symto description, gapped phases correspond to Lagrangian condensations of the \symto while gSPTs correspond to non-Lagrangian condensations. Following this principle, it was has shown that 1+1D  $G$-gSPTs are in 1-1 correspondence with magnetic non-Lagrangian anyon condensations of the \symto $\mathcal{Z}[\vc_G]$~\cite{wen2023classification,huang2023topological}. This correspondence has been generalized to 1+1D gSPTs with non-invertible symmetries~\cite{bhardwaj2023clubsandwich,bhardwaj2024hassediagrams} and fermionic symmetries~\cite{wen2024fermions,bhardwaj2024fermionic,huang2024fermionic}.

To generalize bulk-boundary correspondence or the topological holography paradigm to higher dimensions requires understanding condensation of topological excitations in higher dimensional topological orders, which requires the development and application of the theory of algebras in higher fusion categories. 3+1D topological orders support string-like topological excitations, therefore besides the conventional condensation of bosonic particles it is also possible to induce string condensation in 3+1D~\cite{Zhao_2023,kong2024highercondensationtheory}. In fact, string condensation contains particle condensation as a special case~\cite{Zhao_2023,kong2024highercondensationtheory}. Mathematically, 3+1D topological orders are described by braided fusion 2-categories~\footnote{Technically the braided fusion 2-category needs to be non-degenerated in some sense, mathematically this means it has trivial center~\cite{Johnson_Freyd_2022}.}, and it is believed that string condensations are described by condensable $E_2$-algebras in the 2-category~\cite{Zhao_2023,kong2024highercondensationtheory}. After a string condensation, the deconfined strings correspond to local modules over the condensable algebra~\cite{Zhao_2023,kong2024highercondensationtheory,D_coppet_2024}, and only recently has it been shown that local modules over a condensable algebra form a braided fusion 2-category on their own~\cite{D_coppet_2024}. This matches with the physical intuition that the deconfined strings and particles surviving the condensation form a 3+1D topological order on their own.  The condensation is called Lagrangian if it trivializes the topological order, i.e. all strings and particles are either condensed or confined, and non-Lagrangian otherwise. 

Due to the complexity of the 2-categorical structure of string condensations, very few examples have been studied~\cite{Zhao_2023,kong2024highercondensationtheory}, and it is not clear what are the physical structures of string condensations. This work provides a systematic study of string condensation in 3+1D gauge theories and apply it to the 3+1D/2+1D topological holography. 
\paragraph*{Summary of results}

We study examples of non-Lagrangian string condensations in 3+1D gauge theories and show that they are exactly dual to  2+1D gSPTs. Through the standard \symto sandwich construction, we show how topological properties of 2+1D gSPTs can be extracted from the corresponding string condensation, including the symmetry extension structure, the emergent anomaly and properties of their symmetry twist defects.

We give an explicit algebraic classification of condensable algebras in 3+1D $G$-gauge theory that are "magnetic" and "simple". The magnetic and simple conditions are dual to the no-symmetry-breaking condition and no-topological-order condition in 2+1D respectively. Via the topological holography dictionary, magnetic simple condensations are dual to gapped or gapless SPTs with symmetry $G$.  We find that a magnetic simple condensation is determined by a normal subgroup$N\lhd G$ and a quadruple data $(\alpha,\phi,\sigma,\beta)$ satisfying a set of consistency conditions. The condensation is Lagrangian if $N=G$ and non-Lagrangian otherwise. The classification reduces to $H^3[G,\bC^\times]$ in the Lagrangian case, matching with the classification of 2+1D SPTs. In the non-Lagrangian case, the quadruple $(\alpha,\phi,\sigma,\beta)$ is an element of the so-called quasi-abelian third cohomology group of the pair $(G,N)$, denoted by $H^3_{qa}[(G,N),\bC^\times]$~\cite{naidu2011crossed}, and we have $H^3_{qa}[(G,G),\bC^\times]\cong H^3[G,\bC^\times]$~\cite{naidu2011crossed}.

In general, a gSPT with symmetry $G$ has a "gapped symmetry" $N$, which is a normal subgroup of $G$. Physically, symmetry charges transforming nontrivially under $N$ are gapped in the gSPT. The normal subgroup $N$ in the classification is exactly the gapped symmetry of the gSPT. In the corresponding string condensation, $N$ is the flux sectors contained in the condensed strings. A gapped SPT corresponds to $N=G$, i.e. all symmetry charges are gapped.

We then discuss the physical interpretation of the other classifying data $\alpha,\phi,\sigma,\beta$ in the context of string condensations. 
We first analyze the structure of string condensations in the Lagrangian case($N=G$), dual to 2+1D gapped SPTs.  We find that the condensed strings in a magnetic simple condensation general take the form $mS$, where $m$ is a flux loop and $S$ is an SPT-like defect string. We give a formula that determines the condensed strings from the dual 2+1D SPT. This gives a categorical explanation of several lattice model constructions of gapped boundaries of 3d topological orders where it was found that the gapped boundary dual to the 2+1D type-III SPT of $\bZ_2^3$ condenses the strings $m_1S_{23},m_2S_{13},m_3S_{12}$~\cite{Zhu_2022,song2024magic}. 

In general specifying the set of condensed strings is not sufficient for determining the string condensation, even in abelian theories. This is clear from the 2-category theory data defining a condensable algebra. For a magnetic simple condensation defined by the quadruple data $(\alpha,\phi,\sigma,\beta)$, only the data $\phi$ is directly related to the set of strings condensed, the other data $\alpha,\sigma,\beta$ are responsible for more subtle differences between string condensations. We discuss the physical meaning of these data through examination of string condensations dual to 2+1D abelian SPTs with type I$\sim$III cocyles. We find that the extra data are associated with the fusion and braiding properties of the termination of strings on the boundary of the condensation region, which are holographically dual to braiding and fusion properties of the symmetry twist defects of the 2+1D system.

We then identify three classes of non-Lagrangian magnetic simple condensations, dual to three classes of gSPTs that we call type-I$\sim$III gSPTs. The first class has only nontrivial $\phi$ factor, corresponding to the "SPT-pump" gSPT class reported in~\cite{Wen_2023}. The second class has only nontrivial $\sigma$, corresponding to a class of gSPT where symmetry defects of the gapped symmetry carry fractional charges. The last class has only nontrivial $\alpha$, and corresponds to transitions between gapped SPTs and symmetry breaking phases. We study the properties of these gSPTs from the properties of the corresponding string condensations and via the \symto sandwich construction. 

We also study gauging of gSPTs. Gauging in the \symto framework is realized by changing the symmetry boundary. We analyze the properties of the gauged type-I$\sim$III gSPTs via this method, and find that the deconfined excitations in the gauged gSPT form a pre-modular category, i.e. a braided fusion category with degenerated braiding.

An equivalent but more abstract way of stating the classification is that a magnetic simple condensation is given by a pointed $G$-crossed braided fusion category, whose underlying $G$-graded category is $\vc_N$ for some normal subgroup $N\lhd G$. In the Lagrangian case, $N=G$, and the classification matches with the classification of 2+1D $G$-SPTs, since 2+1D $G$-SPTs are also classified by $G$-crossed braided fusion structures on $\vc_G$~\cite{barkeshli2019symmetry,lan2016classification,kong2020classification}. More precisely, in the language of~\cite{barkeshli2019symmetry}, the data $\alpha,\phi,\sigma,\beta$ are the associator, the $\eta$-symbol, the $U$-symbol and the $G$-crossed braiding of the $G$-crossed braided fusion structure on $\vc_N$.  In the non-Lagrangian case, the classification suggests 2+1D gSPTs are classified by $G$-crossed braided fusion structures on $\vc_N$ for some normal subgroup $N\lhd G$. We discuss from a 2+1D perspective how this is indeed the classification of 2+1D gSPTs, thereby completing the \symto/gSPT correspondence in 3+1D/2+1D. Nevertheless, the algebraic classification in terms of the data $(\alpha,\phi,\sigma,\beta)$ provides us a more practical and concrete method of constructing magnetic simple string condensations and therefore gSPTs. We also argue that gauging the symmetry in a gSPT amounts to equivariantization of the $G$-crossed category, similar to the gauging of SPTs. However, the equivariantization results in a pre-modular category whenever $N\neq G$. We discuss the physical origin of the non-modular-ness of gauged gSPTs. We summarize the two ways of classifying gapped and gapless SPTs, their \symto correspondence, and their gauging, in table~\ref{table:summary}.

\begin{table}[]
    \centering
    \def\arraystretch{2}
    \begin{tabular}{|p{4cm}|p{5cm}|p{5cm}|}
    \hline
       &2+1D $G$-SPT & 2+1D $G$-gSPT \\
       \hline
        Algebraic classification & $[\alpha]\in H^3[G,\bC^\times]$ & $[\alpha,\phi,\sigma,\beta]\in H^3_{qa}[(G,N),\bC^\times]$\\
        \hline
        Categorical classification & $G$-crossed braided fusion category structure on $\vc_G$ &$G$-crossed braided fusion category structure on $\vc_N$\\
        \hline 
        \symto &\multicolumn{2}{|p{10cm}|}{\hspace{4cm}$\cc{Z}[2\vc_G]$}\\
        \hline
        Dual string condensation & Lagrangian magnetic simple condensable algebra &non-Lagrangian magnetic simple condensable algebra\\
        \hline
        Gauging $G$ & \multicolumn{2}{|p{10cm}|}{\hspace{3.5cm}Equivariantization }\\
        \hline 
        Gauged theory & Modular category $\cc{Z}[\vc_G^\alpha]$ & Pre-modular category\\
        \hline
    \end{tabular}
    \caption{The classification, \symto description, and gauging of gapped and gapless SPT. Here $N\lhd G$ is the gapped symmetry of the gSPT. The SPT column can be obtained by setting $N=G$ in the gSPT column.}
    \label{table:summary}
\end{table}

For a gSPT with symmetry $G$, its gapless degrees of freedom have an effective symmetry $K$ that is a quotient of $G$: since symmetry charges transforming nontrivially under a normal subgroup $N$ are gapped, the gapless degrees freedom have an effective $K:=G/N$ symmetry. The IR symmetry of a gSPT could potentially be anomalous even if the full symmetry $G$ is not. The emergent anomaly of the IR symmetry is dual to the twist of the post-condensation phase of the \symto. In other words, condensing a non-Lagrangian algebra in $\mathcal{Z}[2\vc_G]$ may reduce it to a twisted $K$-gauge theory $\mathcal{Z}[2\vc_K^\pi]$, and the twist $\pi\in Z^4[K,U(1)]$ is dual to the emergent anomaly of the 2+1D gSPT. In this work we show with examples that the twists match with the emergent anomalies of known gSPTs. In a separate work~\cite{Wen_algebra}, we will analyze the local modules over magnetic simple condensable algebras more systematically.

Although in this work we focus on the \symto for 2+1D invertible finite symmetries $2\vc^\pi_G$, it is known that the \symto for any fusion 2-categorical symmetry is of the form $\cc{Z}[2\vc_G^\pi]$~\cite{Lan_2018,Lan_2019,bhardwaj2023generalizedcharges}, therefore we expect our results, especially the classification of simple string condensations in $\cc{Z}[2\vc_G]$, to be useful for analyzing general fusion 2-categorical symmetries as well, such as $2\rep(G)$ or 2-group symmetries $2\vc^\omega_{\bb{G}^{(2)}}$.

This work is organized as follows. In Sec.~\ref{sec:symto_review} we review basics of the topological holography, including the sandwich construction, the club sandwich construction, and string condensations. In Sec.~\ref{sec:3dTO_review}, we review basic structures of 3+1D topological orders, focusing on the braiding structures of strings. In Sec.~\ref{sec:z2z4example}, we study the \symto construction for a 2+1D $\bZ_2\times \bZ_4$-gSPT. We first review properties of this gSPT in the language of topological response theory. Then we analyze a string condensation in $\mathcal{Z}[2\vc_{\bZ_2\times \bZ_4}]$. We show that this string condensation is exactly dual to the $\bZ_2\times \bZ_4$-gSPT, via the \symto sandwich construction.  Sec.~\ref{sec:classification} is a technical section where we classify magnetic simple string condensations in $\mathcal{Z}[2\vc_G]$. Then in Sec.~\ref{sec:discussion_application} we discuss the physical interpretation and application of the classification, starting with a discussion of string condensations dual to 2+1D type-I$\sim$III SPTs. We then identify three classes of non-Lagrangian magnetic simple string condensations, and study the properties of the dual gSPTs. We then study gauging of gSPTs via the sandwich construction.  Finally, in Sec.~\ref{sec: general theory} we discuss generalities of gSPTs in 2+1D. We show that 2+1D gSPTs are classified by $G$-crossed braided fusion structures on $\vc_N$, and gauging of a gSPT is described by equivariantization of the $G$-crossed category.

\section{Generalities of the topological holography\label{sec:symto_review}}
\subsection{Categorical Landau paradigm}
In the traditional Landau paradigm, order parameters are used to characterize phases. An order parameter is a local operator that carries a definite charge(representation) of the symmetry, such that its vacuum expectation value(vev) is nonzero in a given phase, in which case we say the charge is condensed. Order parameters signal symmetry breaking, therefore phases of matter in the traditional Landau paradigm are labelled by symmetry breaking patterns.  It was later realized that local charged operators are not sufficient for distinguishing gapped phases of matter. Instead, one needs to consider a more general class of charges that are higher dimensional or attached with symmetry generators, to fully classify and characterize SPTs. In 1+1D these generalized charges are known as string order parameters, which take the form of a half-infinite symmetry operator with a local charge on the end. In spacetime dimension $d+1$, a generalized charge may have spacial dimension $p$ ranging from $0$ to $d-1$, and they carry actions of the symmetry, described by higher representations of the symmetry~\cite{bhardwaj2023generalizedcharges,Bhardwaj_2024,Bartsch_2024,bartsch2023noninvertible,bartsch2023representation}. A generalized charge may also live in the twisted sector of the symmetry, i.e. it may arise on the boundary of a symmetry generator, in which case the generalized charge is also called a twisted generalized charge~\cite{bhardwaj2023generalizedcharges,Bhardwaj_2024}. The string order parameters for 1+1D SPTs are examples of local(0d) twisted generalized charges, while decorated domain wall operators in higher dimensional SPTs are examples of twisted generalized charges of dimension $>0$. 

It is therefore natural to consider the collection of all generalized charges of various dimensions and the condensation(vev) thereof in order to fully characterize phases of matter. It turns out that a concise mathematical structure emerges from this collection of generalized charges. The symmetry defects of a $d+1$-D theory generally form a fusion $d$-category $\e{S}$, and the generalized charges of the symmetry(including the symmetry defects themselves) form a fusion $d$-category $\mathcal{Z}[\e{S}]$, called the Drinfeld center of $\e{S}$~\cite{kong2015boundarybulkrelation,Kong_2017,kong2020algebraic,Bartsch_2024,bartsch2023noninvertible,bartsch2023representation,kong2014braided,Kong_2022,kong2020classification}. $\mathcal{Z}[\e{S}]$ is naturally a \textit{braided} fusion $n$-category, therefore is naturally associated with a $(d+1)+1$-dimensional topological order. This $(d+1)+1$-D topological order $\mathcal{Z}[\e{S}]$ is called the \symto of the symmetry $\e{S}$. A condensation of generalized charges of $\e{S}$ is mapped to a condensation of objects in the center $\mathcal{Z}[\e{S}]$. The braiding structure of the center $\mathcal{Z}[\e{S}]$ contains the information needed to decide whether a set of generalized charges can be condensed or not. Mathematically, the condensation of objects in a braided fusion $d$-category is described by condensable $E_2$-algebras~\cite{Zhao_2023,kong2024highercondensationtheory}, or just condensable algebras for short. Therefore, we have turned the question of characterizing phases of matter in $d+1$-D with symmetry $\e{S}$ into the question of classification of condensable algebras in the center $\mathcal{Z}[\e{S}]$. 

It is worth mentioning that even phases of matter without symmetries, i.e. topological orders, admit a characterization in terms of condensation of objects in the center. In this case the symmetry category is trivial $\e{S}=d\vc$, and the center is also trivial $\mathcal{Z}[d\vc]=d\vc$. However the condensable $E_2$-algebras in the trivial braided fusion $d$-category  $d\vc$ are not all trivial, and they are exactly the braided fusion $d-1$-categories, i.e. (potentially anomalous)$d+1$-D topological orders~\cite{D_coppet_2024,kong2024highercondensationtheory}. Furthermore, topological orders enriched by $\e{S}$ also admit a description in terms of condensation of objects in $\mathcal{Z}[\e{S}]$~\cite{D_coppet_2024}. We will discuss more about this point from the perspective of sandwich construction later in Sec.~\ref{sec:sandwich}.

For a given condensation pattern of generalized charges, the confined generalized charges are necessarily carried by gapped excitations, which can be understood as a generalized Meissner effect~\cite{bhardwaj2023clubsandwich,bhardwaj2024hassediagrams}. On the other hand any deconfined charge is necessarily carried by gapless excitations\cite{kong2020algebraic,ji2020categorical,chatterjee2023holographic,bhardwaj2023clubsandwich,bhardwaj2024hassediagrams}. Therefore a 2+1D $\e{S}$-symmetric system is gapped only if its generalized charge condensation leaves no deconfined charges behind, in which case the corresponding condensation in the \symto trivializes the \symto and is called a Lagrangian condensation. Therefore gapped phases are described by Lagrangian condensations of their \symto, while gapless phases are described by non-Lagrangian condensations of the \symto.

From now on we focus on $d=2$, and the \symto is a 3+1D topological order, described by a non-degenerated braided fusion 2-categroy. The condensable algebras describe condensation of string-like topological excitations of the \symto, corresponding to $1d$ generalized charges of the 2+1D system. We notice that condensation of $0d$ generalized charges, corresponding to particles of the \symto, can also be phrased in terms of condensation of strings. Namely we can consider the condensation descendant of the $0d$ particle, which is a $1d$ string, then condensation of the $0d$ particle is equivalent to the condensation of the $1d$ condensation descendant~\cite{Zhao_2023,kong2024highercondensationtheory}.

After a string condensation in the \symto, the deconfined strings are described by local modules over the condensable algebra, and the confined strings are described by modules that are not local~\cite{Zhao_2023,kong2024highercondensationtheory,D_coppet_2024}. It is known that the local modules form a braided fusion 2-category on their own~\cite{D_coppet_2024}, which physically corresponds to the new 3+1D topological order after the condensation. In general the new topological order takes the form $\mathcal{Z}[\e{S}']$ for another fusion 2-category $\e{S}'$. The transition from the \symto $\mathcal{Z}[\e{S}]$ to the new topological order $\mathcal{Z}[\e{S}']$ induced by string condensation can be viewed as a categorical version of symmetry breaking. Adopting the terminology of~\cite{kong2020algebraic,ji2020categorical}, we may say that the "categorical symmetry" of $\e{S}$ is $\mathcal{Z}[\e{S}]$, and the condensation of generalized charges of $\e{S}$ breaks the categorical symmetry from $\mathcal{Z}[\e{S}]$ to $\mathcal{Z}[\e{S}']$. 

Although any generalized charge condensation can be viewed as spontaneous breaking of the categorical symmetry $\mathcal{Z}[\e{S}]$, only condensation of charges in the untwisted sector breaks the symmetry $\e{S}$ itself.  We refer to condensations of generalized charges that do not condense any charges in the untwisted sector as \textit{magnetic} condensations. Therefore 2+1D symmetric phases  correspond to magnetic string condensations of the \symto.
\subsection{Gapless SPTs from \symto}
We say a system with symmetry $\e{S}$ is in a gapless SPT phase(gSPT) if it is gapless and has no ground state degeneracy on any closed manifold. The no-GSD condition requires no topological order or symmetry breaking present. Generalized charges of a a gSPT can stay deconfined and carried by the gapless excitations of the system. On the other hand the confined charges only  appear above certain gap $\Delta_S$, called the symmetry gap of the gSPT. The spectrum gap of a gSPT vanishes in  the thermodynamic limit, but the symmetry gap $\Delta_S$ remains non-vanishing.  Different gSPTs are therefore distinguished by their patterns of condensed, confined, and deconfined generalized charges. We speak of \textit{the gSPT class} of a gSPT as the pattern of generalized charge condensation of the gSPT. Then two gapless systems with different bulk universality classes may also be in the same gSPT class if their generalized charge condensation patterns agree. Physically this means the two systems have identical behaviour as far as local/non-local order parameters of the symmetry are concerned. By the topological holographic dictionary we outlined in the previous subsection, we see that 2+1D gSPTs are described by magnetic and non-Lagrangian string condensations of the \symto.  Since the condensation is not Lagrangian, the \symto is reduced to a nontrivial 3+1D topological order $\mathcal{Z}[\e{S}']$ by the condensation. $\mathcal{Z}[\e{S}']$ is the braided fusion 2-category formed by the deconfined(and gapless) generalized charges in the gSPT, and $\e{S}'$ can be thought of as the effective symmetry for the gapless degrees of freedom of the gSPT, which we refer to as the \textit{IR symmetry} of the gSPT. We note that the IR symmetry is in some sense a "quotient" of the full symmetry $\e{S}$, in the sense that there is a canonical tensor  2-functor $\e{S}\to \e{S}'$ that is essentially surjective on morphisms at all levels.~\footnote{A subtlety is that there should be a canonical choice of $\e{S}'$ here, since different fusion 2-categories can have the same center. This canonical $\e{S}'$ can be read off from the club sandwich construction, see Sec.~\ref{sec:club_sandwich}}

We may perturb the gSPT without breaking any symmetry or closing the symmetry gap $\Delta_S$ by further condensing the deconfined charges symmetrically. Doing this will also confine some of the initially deconfined charges and reduce the IR symmetry. It may be the case that we can continue to do so until all generalized charges are either condensed or confined. If there are more gapless degrees of freedom left, they must transform trivially under the symmetry and we can perturb symmetrically to gap them out without affecting the symmetry gap, we then arrive at a trivial gapped symmetric state without ever closing the symmetry gap. If this is the case we say the gSPT is a \textit{weak gSPT}. On the other hand it could also be the case that there is an obstruction to achieving a trivial gapped state without closing the symmetry gap or introducing degeneracy. This happens when the IR symmetry of the gSPT is anomalous. Then if the symmetry gap is never closed, the anomaly of the IR symmetry requires that the degrees of freedom below the symmetry gap $\Delta_S$ must either stay gapless, break the symmetry, or have nontrivial topological order. Therefore a trivial gapped symmetric state can not be achieved unless we first close the symmetry gap, or introduce degeneracy by breaking the symmetry or inducing topological order. In this case, we say the gSPT is an \textit{intrinsically gapless SPT}(igSPT). 

As we have mentioned before, condensation of generalized charges could induce topological orders. Without any symmetry, $\e{S}=2\vc$, we have the fact that Lagrangian string condensations in the trivial \symto $\mathcal{Z}[2\vc]=2\vc$ are classified by non-degenerated braided fusion 1-categories~\cite{D_coppet_2024}. In the sandwich construction, this corresponds to a trivial bulk, a trivial symmetry boundary and a physical boundary given by a 2+1D topological order. Stacking a non-Lagrangian condensation in $\mathcal{Z}[\e{S}]$ with a nontrivial Lagrangian condensation in $\mathcal{Z}[2\vc]=2\vc$ results in a non-Lagrangian condensation in $\mathcal{Z}[\e{S}]$ that is dual to a gSPT stacked with a topological order. By our definition of gSPT, this state is not a gSPT. Therefore not all magnetic non-Lagrangian condensations correspond to gSPTs. We refer to the condensations that do not induce topological orders on the boundary as \textit{simple} condensations. Therefore gSPTs are described by magnetic, non-Lagrangian and simple condensations of the \symto.
\begin{tcolorbox}
    2+1D gapless SPTs with symmetry $\e{S}$ correspond to string condensations of the \symto $\mathcal{Z}[\e{S}]$ that are magnetic, non-Lagrangian, and simple. If the condensation reduces the \symto to $\mathcal{Z}[\e{S}']$, then $\e{S}'$ is the IR symmetry of the gSPT. 
\end{tcolorbox}

\subsection{The sandwich construction\label{sec:sandwich}}
The mathematical relation between condensations of generalized charges in 2+1D and  condensations of strings of the 3+1D \symto can be recast into a more concrete formulation known as the sandwich construction~\cite{freed2014relativequantumfieldtheory,kong2015boundarybulkrelation,Kong_2017,Freed:2018cec,kong2020algebraic,Gaiotto_2021,Lichtman_2021,Kaidi:2022cpf,Apruzzi_2023,Apruzzi_2022,Lin_2023,chatterjee2023emergent,chatterjee2023holographic,zhang2023anomalies}, to which we now turn. 
\subsubsection{Setup}
A 2+1D system $\e{T}_\e{S}$ with symmetry $\e{S}$ can always be put on the boundary of the 3+1D \symto $\mathcal{Z}[\e{S}]$. Consider putting $\mathcal{Z}[\e{S}]$ on a sandwich geometry $\mathsf{B}\times I$, with the boundary condition at $\mathsf{B}\times \{0\}$ set by the 2+1D system $\e{T}_\e{S}$, and the boundary at $\mathsf{B}\times\{1\}$ set by the canonical Dirichlet boundary condition of $\mathcal{Z}[\e{S}]$. Then by compactifying the interval $I$, we can recover the original 2+1D system $\e{T}_\e{S}$. The boundary at $\mathsf{B}\times \{1\}$ with the Dirichlet boundary condition is called the symmetry boundary, denoted by $\bref$. The symmetry boundary determines what symmetry is under consideration. Specifically, the topological operators on the reference boundary correspond to the symmetry operators of the 2+1D system. The boundary at $\mathsf{B}\times \{0\}$ is called the physical boundary, denoted by $\bphys$, which is determined by the physical system $\e{T}_\e{S}$, and can be varied to describe different $\e{S}$-phases. We take the gap of the sandwich to be infinite everywhere except for a small neighbourhood of physical boundary, which means excitations are only allowed to be created near the physical boundary. With this setup all the dynamical and topological properties of the sandwich are identical to those of the 2+1D system $\e{T}_\e{S}$. 
\begin{figure}[h]
    \centering
    \includegraphics[width=0.8\linewidth]{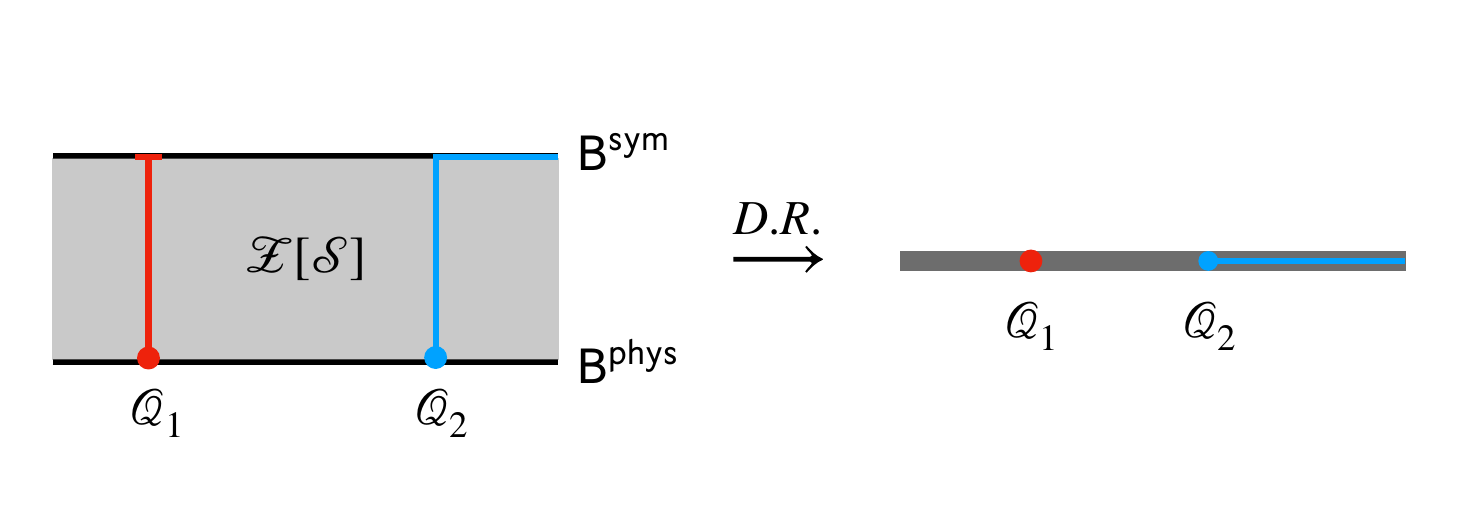}
    \caption{The sandwich construction. Generalized charges of $\e{S}$ correspond to topological excitations of the bulk $\cc{Z}[\e{S}]$. The energy gap of the sandwich is infinite everywhere except in a neighbourhood of $\bphys$. If $Q_1$ is condensed on $\bref$,  then there exists a $p+1$-d operator connecting the two boundary, such that a single $Q_1$ is created on $\bphys$(red). After compactifying the vertical direction, this operator becomes a $p$-d operator that creates a $p$-d charge. If $Q_2$ is not condensed on $\bref$, then its creation operator(blue) can not end on $\bref$, instead it continues to extend on $\bref$. After dimension reduction, this maps to a $p+1$-d half-infinite operator that creates a $p$-d charge on its boundary. By definition, nontrivial topological operators on $\bref$ are the symmetry operators. Therefore $Q_2$ is a generalized charge in the twisted sector.}
    \label{fig:sandwich}
\end{figure}

Generalized charges of $\e{T}_\e{S}$ correspond to topological excitations~\footnote{We note that we count topological defects, such as condensation descendants of particles, as excitations as well. This is natural in the 2-category description of 3+1D topological orders. We will come back to this point in Sec.~\ref{sec:3dTO_review}.} of the \symto. If a generalized charge of spacial dimension $p$ is condensed at $\bref$, the a $p+1$-d topological operator that creates it can end on $\bref$ with no excitation created, this topological operator can stretch across the sandwich and reach $\bphys$. After the interval compactification, this topological operator maps to a $p$-d operator that creates a $p$-d charge of $\e{S}$. This corresponds to a generalized charge in the untwisted sector, which can be created by operators of the same dimension. On the other hand, if a topological excitation is not condensed at $\bref$, then its creation operator can not end on $\bref$. Instead, it becomes a nontrivial topological operator on $\bref$. After compactifying the interval, this topological operator is mapped to a generalized charge living on the boundary of a half-infinite symmetry operator, i.e. a twisted generalized charge. See Fig.~\ref{fig:sandwich}. 

For instance, when $\e{S}=2\vc_{\bZ_2}$, describing an onsite $\bZ_2$ symmetry of a 2+1D system, the generalized charges are spin flips and domain walls. The spin flips are local charges carrying the nontrivial representation of $\bZ_2$, which  can be created by local operators, while the domain walls are 1d generalized charges in the twisted sector that can only by created by 2d operators. Here the \symto is the 3+1D toric code, with $\bref$ given by condensing the $e$ particle(or equivalently the Cheshire string $\bb{1}^e$). Since $e$ is condensed on $\bref$, the corresponding generalized charge, the spin-flips, can be created by 0d operators, which in the sandwich construction is realized by a Wilson line of $e$ stretching from $\bref$ to $\bphys$. On the other hand, the flux loops $m$ are not condensed on $\bref$, thus they correspond to symmetry defects, generating the $\bZ_2$ symmetry. When a $m$-membrane operator reaches $\bref$, it can not end there but must continue to extend on $\bref$. This configuration is mapped, under the interval compactification, to a domain wall creation operator that acts on half-infinite plane of the 2+1D system, and creates a 1d domain wall on its boundary. 

\begin{figure}[h]
    \centering
    \includegraphics[width=0.8\linewidth]{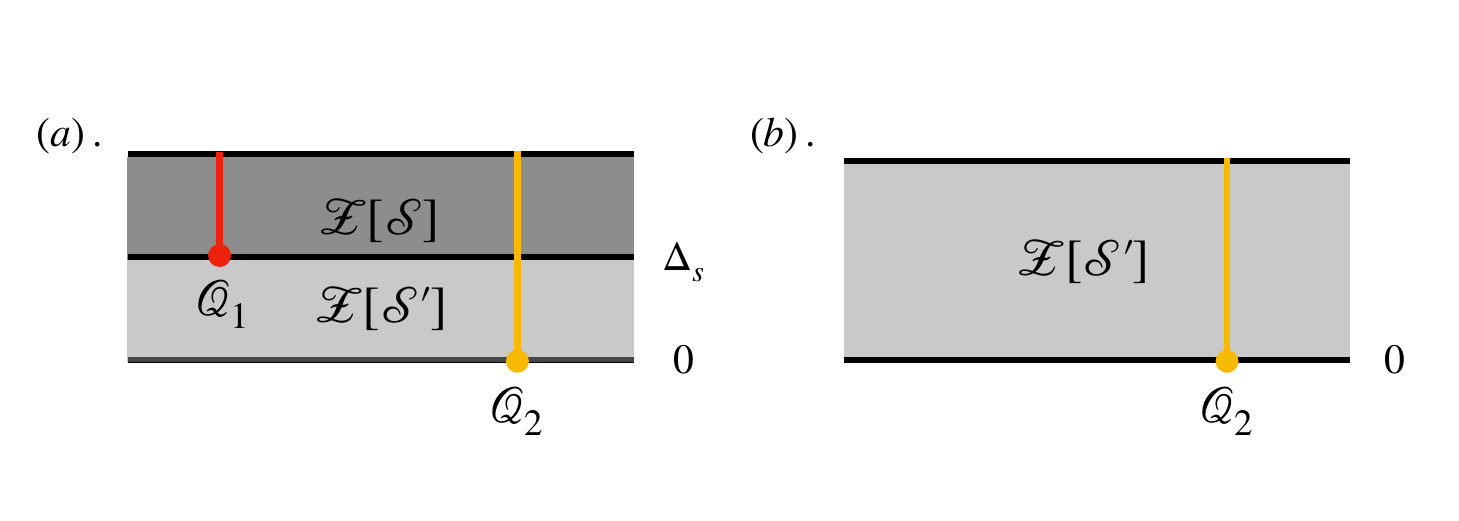}
    \caption{(a).The club sandwich construction for a gapless state. The bottom half of the sandwich is obtained by condensing a non-Lagrangian condensable algebra of $\mathcal{Z}[\e{S}]$, and takes the form $\mathcal{Z}[\e{S}']$. The bottom boundary of $\mathcal{Z}[\e{S}']$ has nothing condensed, and has energy gap 0. The domain wall in the middle has a finite energy gap $\Delta_s$, which hosts the gapped symmetry charges of the system. Everywhere else in the sandwich the gap is infinite. A confined charge(red) can only be created on the domain wall with a finite energy gap, while a deconfined charge(yellow) can be created on the bottom boundary with zero energy gap. (b). After taking $\Delta_s\to \infty$ and shrinking the top half of the sandwich, we obtain a standard sandwich construction for a gapless state with $\e{S}'$ symmetry. This shows the gapless degrees of freedom of the sandwich has an effective symmetry $\e{S}'$, and has full categorical symmetry $\mathcal{Z}[\e{S}']$ unbroken.}
    \label{fig:club_sandwich}
\end{figure}
\subsubsection{The club sandwich construction\label{sec:club_sandwich}}
If the condensation at the physical boundary is not Lagrangian, then the physical boundary is necessarily gapless, and the sandwich construction describes a 2+1D gapless state. A more transparent way of analyzing the structure of the gapless state from the structure of sandwich is to slightly extend the physical boundary into a thickened layer. If the \symto reduces to $\mathcal{Z}[\e{S}']$ after the condensation, then this thickened layer now hosts the topological order $\mathcal{Z}[\e{S}']$, with the top boundary of it being a domain wall to the original \symto $\mathcal{Z}[\e{S}]$ and the bottom boundary having "nothing condensed". This construction was named the "club sandwich" construction in~\cite{bhardwaj2023clubsandwich}. See Fig.~\ref{fig:club_sandwich} for the setup. The domain wall in the middle has a finite energy gap $\Delta_s$, which is the symmetry gap of the 2+1D system. A charge that is confined by the condensation can not enter the bottom half of the sandwich, and can only be created on the domain wall in the middle, with a finite energy gap $\Delta_s$. On the hand hand a deconfined charge can be created on the bottom boundary and is gapless. The bottom layer of the club sandwich can now be viewed as the \symto for a gapless theory with $\e{S}'$ symmetry.  If we only care about the IR properties of the sandwich, we can take $\Delta_s\to \infty$, and shrink the top layer. This gives us a single $\mathcal{Z}[\e{S}']$ bulk with a nothing-condensed gapless physical boundary, and the topological operators on the new top boundary define a symmetry $\e{S}'$. This is now a standard sandwich construction for a 2+1D system with $\e{S}'$ symmetry. Thus $\e{S}'$ corresponds to the symmetry of the gapless degrees of freedom, i.e. the IR symmetry. The advantage of the club sandwich construction is that it separates the gapped and gapless charges of the system.

\subsubsection{2+1D topological order induced by string condensation}
The condensation of strings at the physical boundary is described by condensable algebras of the \symto $\mathcal{Z}[\e{S}]$. We note that a condensable algebra contains information more than merely the set of strings condensed, it also describes the fusion and braiding structures of the endpoints of condensed strings on the boundary. For instance, for $\e{S}=2\vc_{\bZ_2}$ and \symto being the 3+1D toric code, there are two magnetic condensable algebras that condense the flux loop $m$. This is because the endpoints of $m$-strings at the boundary can be assigned a nontrivial fusion coefficient($F$-symbol) compatible with the $\bZ_2$-fusion rule of the $m$-strings, given by the nontrivial element of $H^3[\bZ_2,\bC^\times]$. These two ways of condensing $m$ are dual to the trivial and nontrivial $\bZ_2$-SPT in 2+1D. When the endpoint of the $m$-string on the boundary is assigned a nontrivial $F$-symbol, it also acquires semionic or anti-semionic self statistics. 

The endpoint of a trivial string, denoted by $\bb{1}$, can thus be thought of as a point-like excitation that can live on the boundary without attaching to any bulk string, i.e. a particle that lives freely on the boundary. If the coefficient of $\bb{1}$ in the condensable algebra is greater than 1, then it means there are different types of endpoints of the trivial string on the boundary. In other words, there are more than one type of particles living on the boundary. If nontrivial fusion and braiding structures are assigned to these particles on the boundary, the condensable algebra describes a boundary that is topologically ordered. More concretely, a magnetic condensable algebra in $\mathcal{Z}[2\vc_G]$ generally takes the form
\begin{align}
    \eA=n\bb{1}+\cdots.
\end{align}
The definition of a condensable algebra(see appendix~\ref{app:condensable algebra}) equips the trivial component $n\bb{1}$ with the structure of a braided fusion 1-category with $n$ simple objects, which physically describes the (2+1D)topological order on the boundary of the condensation region. Therefore, requiring that no topological order is formed on the boundary amounts to the condition that the coefficient of $\bb{1}$ in the condensable algebra is 1. We refer to such condensable algebras as simple, and non-simple if $n>1$~\footnote{More generally, an object in $\cc{Z}[2\vc_G]$ has a natural grading structure(graded by flux sectors), therefore $\eA=\oplus_{[g]\in \text{Cl}(G)} \eA_{[g]}$. We say the algebra is simple if its trivial degree component $\eA_1$ is a simple object.}. A non-simple condensation on the physical boundary gives a \symto construction for a 2+1D topologically ordered state. For instance, on the boundary of a 3+1D $G$-gauge theory, one can put any 2+1D $G$-enriched topological order, and couple it to the bulk gauge field~\cite{Luo_2023,Ji_2023}. Such an SET boundary is described by a string condensation that is non-simple.

\section{Generalities of 3+1D topological orders\label{sec:3dTO_review}}
Here we review aspects of 3+1D topological orders~\cite{Lan_2018,Lan_2019,Kong_2020,barkeshli2024higher,barkeshli2023codimension,Johnson_Freyd_2022,tantivasadakarn2024string}, with a focus on the string-string braiding structure which plays a crucial role in the physics of string condensations. 
\begin{itemize}
    \item (The \ccat structure.) A 3+1D topological order hosts particle-like as well as string-like excitations. There is a concise description of topological excitations in a 3+1D topological order in terms of a \ccat\cite{Zhao_2023,douglas2018fusion2,Kong_2020defect}. In this description, an object corresponds to the worldsheet of a string. A 1-morphism $f: A\to B$ is a domain wall between string $A$ and $B$. In the spacetime picture a 1-morphism is a worldline that separates two worldsheets. A 2-morphism between two 1-morphisms corresponds to a spacetime 0D instanton, such as local unitaries or measurements.  The worldsheet diagram of strings, domain walls and instantons, and their 2-category descriptions are depicted below.
\begin{align}
\setlength{\tabcolsep}{15pt}
\begin{tabular}{ccc}
\begin{tikzpic}
\draw[sheet1] (0,0) rectangle (3,3);
\node[label, above left] at (3,0){$A$};
\end{tikzpic}
&
\begin{tikzpic}
\draw[sheet1] (0,0) rectangle (3,3);
\draw[black, line width=0.8pt] (1.5,0) to (1.5,3);
\node[label, above left] at (3,0){$B$};
\node[label, above right] at (0,0){$A$};
\node[label, right] at (1.5,1.5) {$f$};
\end{tikzpic}
&
\begin{tikzpic}
\draw[sheet1] (0,0) rectangle (3,3);
\draw[black, line width=0.8pt] (1.5,0) to (1.5,3);
\node[label, above left] at (3,0){$B$};
\node[label, above right] at (0,0){$A$};
\node[label, above right] at (1.5,0) {$f$};
\node[label, below right] at (1.5,3) {$g$};
\node[dot] at (1.5,1.5) {};
\node[label, right] at (1.5, 1.5) {$\eta$};
\end{tikzpic}\label{eq:2cat1}
\\
\text{An object $A$}
&
\text{A $1$-morphism $f:A\to B$}
&
\text{A $2$-morphism $\eta:f \To g$}
\end{tabular}
\end{align}

In this description, particles correspond to certain 1-morphisms as follows. There is always a trivial string, representing the vacuum, which is denoted by $\bb{1}$. Then a particle can be viewed as a domain wall between vacuum and vacuum, i.e. a 1-morphism: $\bb{1}\to \bb{1}$. For example, we may write in the 3+1D toric code: $\bb{1}\xrightarrow{e}\bb{1}$.

We should note that the strings here not only include the deconfined excitations like flux loops in gauge theories, but also the defect strings that are condensation descendants of particles. In general these defect strings  are not eigenstates of a lattice Hamiltonian realizing the topological order. However, as we shall see soon, it is critical that one treats these defect strings on equal footing with flux loops in the theory of string condensation, in order for the theory to be complete.
\begin{example}
    In a 3+1D $G$-gauge theory, there are defect strings that are gauged versions of 1+1D symmetry breaking phases. These are called Cheshire strings, and are condensation descendants of gauge charges. They can be labelled by the unbroken subgroup, denoted as $\bb{1}^H,~H<G$, or by the gauge charges condensed on the string, denoted as $\bb{1}^{e_1,e_2\cdots}$. 

    There are also defect strings that are gauged versions of 1+1D $G$-SPTs, these are labelled by $\psi\in H^2[G,U(1)]$, and denoted by $S_\psi$. We will call them SPT-strings. There are also combinations of Cheshire and SPT strings, labelled by pairs $(F,\psi)$, where $F<G$ is the unbroken subgroup on the string, and $\psi\in H^2[F,U(1)]$ is an SPT for the remaining group $F$. 

    Unlike flux loops, the defect strings can be open. For instance, the endpoint of an SPT-string can be thought of as the edge mode of the SPT. This means there are 1-morphisms: $S_\psi \to \bb{1}$, as well as 1-morphisms $\bb{1}\to S_\psi$. Similarly there are 1-morphisms between Cheshire strings and vacuum. But there are no 1-morphisms between flux loops and vacuum, i.e. flux loops have to be closed. 

	\hfill$\blacksquare$
\end{example}
    \item (Fusion.)  The strings, domain walls and instantons can fuse. This is represented by layering worldsheets behind one another, with the convention that tensor product occurs from back to front. 
\begin{align}
    \begin{tikzpic}
        \draw[sheet1] (0,0) rectangle (4,4);
        \draw[sheet1] (1,1) rectangle (5,5);
        \node[label,above right] at (0,0){$A$};
        \node[label,above right] at (1,1){$C$};
        \node[label,above left] at (4,0){$B$};
        \node[label,above left] at (5,1){$D$};
        \draw[black, line width=0.8pt] (2,0) to (2,4);
        \draw[black, line width=0.8pt] (3,1) to (3,5);
        \node[label, above right] at (2,0) {$f$};
        \node[label, below right] at (2,4) {$g$};
        \node[label, above right] at (3,1) {$h$};
        \node[label, below right] at (3,5) {$k$};
        \node[dot] at (2,2) {};
        \node[dot] at (3,3) {};
        \node[label, right] at (3,3) {$\eta$};
        \node[label, right] at (2,2) {$\delta$};
    \end{tikzpic}
\end{align}
The monoidal product(fusion) is denoted as $\Box$. The above diagram represents a 2-morphism $\delta\Box \eta: f\Box h\to g\Box k$, and $f\Box h,g\Box k$ are both 1-morphism: $A\Box C\to B\Box D$. 

\begin{example}
    (Fusion of edge modes of SPT-strings.) The SPT-strings have domain walls to the vacuum, consider $z: S_\psi\to \bb{1}$ be a simple domain wall between the cluster chain string in a $\bZ_2^2$ gauge theory and the vacuum. Then the fusion: $z\Box z: S_\psi\Box S_\psi=\bb{1} \to \bb{1}$ must  be a gauge charge. Indeed direction computation shows $z\Box z=(1+e_1)(1+e_2): \bb{1}\to \bb{1}$.
\end{example}

Another way of representing a worldsheet diagram in a 2-category is to only draw the bottom and top horizontal
slices of the worldsheet diagram, then the 2-morphisms(instantons) in between are depicted as arrows from the bottom slice to the top slice. For instance the 2-morphism in Eq.~\eqref{eq:2cat1} can be represented as an arrow between string configurations:

\begin{align}
    \begin{tikzpic}
\draw[black, line width=0.8pt] (1.5,0) to (1.5,3);
\node[label, above right] at (1.5,0) {$A$};
\node[label, below right] at (1.5,3) {$B$};
\node[dot] at (1.5,1.5) {};
\node[label, right] at (1.5, 1.5) {$f$};
\end{tikzpic}\xRightarrow{\eta}
\begin{tikzpic}
\draw[black, line width=0.8pt] (1.5,0) to (1.5,3);
\node[label, above right] at (1.5,0) {$A$};
\node[label, below right] at (1.5,3) {$B$};
\node[dot] at (1.5,1.5) {};
\node[label, right] at (1.5, 1.5) {$g$};
\end{tikzpic}
\end{align}
We may think of the left hand side as the initial string and domain wall configuration, then a local operator is applied, resulting in the final configuration on the right. In this language, objects and 1-morphisms are represented as configurations of strings and domain walls at a time slice, and 2-morphisms are spacetime instantons that change one  configuration into another, represented as arrows between diagrams. 
\item (Braiding.) 
The braiding structures of a 3+1D topological order include particle-string braiding, string-string braiding, string self-braiding, and three-loop braiding. Particle-string braiding and three-loop braiding are relatively well-understood. Particle-string braiding can be viewed as Aharonov–Bohm phase between charges and fluxes. In abelian gauge theories, we have $\theta(e_\chi,m_g)=\chi(g),~\chi\in \rep(G),g\in G$. The three-loop braiding~\cite{Wang_2014,Wang_2015,Wan_2015,Lin_2015,Putrov_2017,Wang_2019,cheng2018loop} is tied to the dimension reduction structure of 3+1D topological orders~\cite{Wang_2014,Wang_2015,Lan_2018,Lan_2019}. Namely a 3+1D topological order with a compact direction can be viewed as direct sums of 2+1D topological orders in different flux sectors. Then the three-loop braiding reduces to particle-particle braiding of the 2d topological order resulting from dimension reduction in the flux sector determined by the base loop. For abelian twisted gauge theories, $\mathcal{Z}[2\vc^\pi_G]$, the dimension reduction in the $g$-sector is a 2+1D twisted gauge theory with the twist given by slant product:
\begin{align}
    \mathcal{Z}[2\vc^\pi_G]\xrightarrow{\text{D.R. in }g-\text{sector}} \mathcal{Z}[\vc^{i_g\pi}_G]
\end{align}
See e.g.~\cite{Wang_2014,Wang_2015,cheng2018loop} for more details. If the slant product $i_g\pi\in Z^3[G,U(1)]$ is nontrivial, then it is known that fluxes in 2+1D twisted gauge theories have nontrivial braidings, which map back to nontrivial three-loop braidings with base loop $m_g$. 

In 2+1D braiding of two particles refers to the adiabatic move of one particle around another, which is a spacetime instanton. Similarly one can define braiding of strings by adiabatic move of one string around another. However, as pointed out in~\cite{Kong_2020defect}, one can perform a wick rotation for the corresponding spacetime worldsheet diagram, and obtain an equivalent definition. In this definition, braiding of strings does not refer to a dynamical process, but a static configuration of strings, as depicted in the following diagram.
\begin{align}
    \begin{tikzpic}
        \draw[black, line width=0.8pt] (0,0) to [out=90,in=-120, looseness=0.5] (0.9,1.8);
        \draw[black, line width=0.8pt] (1.1,2.2) to [out=60, in=-90,looseness=0.5] (2,4);
        \draw[black, line width=0.8pt] (2,0) to [out=90, in=-90,looseness=0.5] (0,4);
        \node[label,below] at (0,0) {$A$};
        \node[label,below] at (2,0) {$B$};
        \node[label,above] at (2,4) {$A$};
        \node[label,above] at (0,4) {$B$};
        \draw[draw = folly,dashed, line width = 0.7pt] (-0.5,1) rectangle (2.5,3);
    \end{tikzpic}
    \hspace{0.2cm}b_{A,B}: A\Box B\to B\Box A
\end{align}
Namely, the dashed red rectangle region can be viewed as a domain wall between $A\Box B$ and $B\Box A$, corresponding to a 1-morphism.  In some cases, braiding of strings is just a particle.
\begin{example}
    Consider the untwisted $\bZ_2^2$ gauge theory in 3+1D. There is an SPT string(the cluster chain), denoted as $S_\psi$. The double braiding between $m_1$ and $S_\psi$ is depicted below. 
    \begin{align}
    \begin{tikzpic}
        \draw[black, line width=0.8pt] (0,0) to [out=90,in=-120, looseness=0.5] (0.9,1.8);
        \draw[black, line width=0.8pt] (1.1,2.2) to [out=60, in=-90,looseness=0.5] (2,4);
        \draw[black, line width=0.8pt] (2,0) to [out=90, in=-90,looseness=0.5] (0,4);
        \draw[black, line width=0.8pt] (0,4) to [out=90,in=-120, looseness=0.5] (0.9,5.8);
        \draw[black, line width=0.8pt] (1.1,6.2) to [out=60, in=-90,looseness=0.5] (2,8);
        \draw[black, line width=0.8pt] (2,4) to [out=90, in=-90,looseness=0.5] (0,8);
        \node[label,below] at (0,0) {$m_1$};
        \node[label,below] at (2,0) {$S_\psi$};
        \node[label,above] at (0,8) {$m_1$};
        \node[label,above] at (2,8) {$S_\psi$};
    \end{tikzpic}\hspace{0.1cm}=\hspace{0.1cm}
    \begin{tikzpic}
        \draw[black, line width=0.8pt] (2,0) to (2,8);
        \draw[black, line width=0.8pt] (0,0) to (0,3);
        \draw[black, line width=0.8pt] (0,3.4) to (0,8);
        \draw[black, line width=0.8pt] (-0.5,3.5) to [out=-90,in=180, looseness=0.8] (0,3.2);
        \draw[black, line width=0.8pt] (0,3.2) to [out=0,in=-90, looseness=0.8] (0.5,3.5);
        \draw[black, line width=0.8pt] (-0.5,3.5) to [out=90,in=180, looseness=0.8] (-0.1,3.8);
        \draw[black, line width=0.8pt] (0.1,3.8) to [out=0,in=90, looseness=0.8] (0.5,3.5);
        \node[label,below] at (0,0) {$m_1$};
        \node[label,below] at (2,0) {$S_\psi$};
        \node[label,above] at (0,8) {$m_1$};
        \node[label,above] at (2,8) {$S_\psi$};
        \node[label,right] at (0.5,3.8) {$S_\psi$};
    \end{tikzpic}=
    \begin{tikzpic}
        \draw[black, line width=0.8pt] (2,0) to (2,8);
        \draw[black, line width=0.8pt] (0,0) to (0,8);
        \node[label,below] at (0,0) {$m_1$};
        \node[label,below] at (2,0) {$S_\psi$};
        \node[label,above] at (0,8) {$m_1$};
        \node[label,above] at (2,8) {$S_\psi$};
        \node[dot] at (0,4) {};
        \node[label,right] at (0.5,3.8) {$e_2$};
    \end{tikzpic}
    \end{align}
\end{example}
In the last step, we used the dimension reduction relation of 1d SPTs~\cite{tantivasadakarn2017dimensional}, specifically the cluster chain has ground state charge $e_2$ when threaded an $m_1$ flux. This gives the double braiding: $b_{m_1,S_\psi}\circ b_{S_\psi,m_1}=e_2: m_1S_\psi\to m_1S_\psi$.  We shall see later that this perspective on string-string braiding is crucial for understanding the physics of string condensations. 

\hfill$\blacksquare$

We denote string-string double braiding by $\Theta(s_a,s_b):=b(s_b,s_a)\circ b(s_a,s_b)$, and string self-exchange by $\Theta(s):=b(s,s)$.
In 2+1D, we have $\theta_{ab,ab}=\theta_a\theta_{b}\theta_{a,b}$ for abelian anyons. There is a similar relation for invertible strings in 3+1D: $\Theta(s_as_b)=\Theta(s_a)\Theta(s_b)\Theta(s_a,s_b)$~\cite{johnsonfreyd2020}. As an application, we have $\Theta(m_1S_\psi)=e_2$ and $\Theta(m_2S_\psi)=e_1$. We say a string is bosonic if it has trivial self-exchange. Then the flux-SPT composites $m_iS_\psi$ are not bosonic. This is similar to how dyons in 2+1D gauge theories acquire nontrivial self-statistics.  

\begin{example}(Condensable strings)
	This new perspective on string-string braiding proves very useful for understanding string condensations. For instance, if $s_a$ is an invertible string, then it is condensable only if it is bosonic. To see this, notice that part of the definition of a condensable algebra is a braided structure(not to be confused with the braiding of the 3+1D topological order, this is an additional structure on a condensable algebra), which is a 2-isomorphism in the following diagram 
	\begin{align}
		\begin{tikzcd}[ampersand replacement=\&]
		{\eA\Box \eA} \&\& {\eA\Box \eA} \\
		\& \eA
		\arrow["{b_{\eA,\eA}}", from=1-1, to=1-3]
		\arrow[""{name=0, anchor=center, inner sep=0}, "\mu"', from=1-1, to=2-2]
		\arrow["\mu", from=1-3, to=2-2]
		\arrow["\beta"{description}, shorten <=4pt, shorten >=8pt, Rightarrow, from=1-3, to=0]
	\end{tikzcd}
	\end{align}
	where $\mu$ is the algebra product 1-morphism(See appendix~\ref{app:condensable algebra} for full definition of a condensable algebra). Now consider a string $s_a$ contained in $\eA$, and restricting to $s_a\Box s_a$, then $\beta$ is a 2-isomorphism between $\mu_{s_a,s_a}\circ \Theta(s_a)$ and $\mu_{s_a,s_a}$. If $\mu_{s_a,s_a}$, the product 1-morphism, is invertible(which is true when $s_a$ is invertible), then $\Theta(s_a)$ is 2-isomorphic to the identity.

	For instance, the strings $m_1S_\psi,m_2S_\psi$ in the 3+1D $\bZ_2^2$ gauge theory are not condensable.

	A similar derivation shows for invertible strings $s_a,s_b$ to be condensable simultaneously, we must have $\Theta(s_a,s_b)\cong 1$. Then for instance, in a 3+1D $\bZ_2^3$ gauge theory, $m_1$ and $m_2S_{13}$ can not both be condensed. 

	\hfill$\blacksquare$
\end{example}

Lastly, the braiding satisfies the hexagon equation up to two 2-isomorphisms, called the hexagonators, and the hexagonators themselves satisfy a set of coherence conditions. A nice review of the coherence conditions can be found in~\cite{Giovanni_BF2}. We shall not list them here as they are not needed for the rest of the work.

Here we summarize some key results about string-string braiding. 
\begin{example}(Flux-SPT braiding)
For an abelian gauge theory, the flux loops are labelled by group elements and denoted by $m_g,~g\in G$, while the SPT-strings are labelled by $\psi\in H^2[G,U(1)]$~\cite{Kong_2020center}. The string-string double braiding between $m_g$ and $S_\psi$ is given by the slant product
    \begin{align}
         \Theta(m_g,S_\psi)=e_{i_g\psi},
    \end{align}
    where \begin{align}
        ~i_g\psi(-):=\frac{\psi(g,-)}{\psi(-,g)}\in \rep(G)
    \end{align}
    is a $G$-charge. Notice this is nothing but the dimension reduction relation for 1+1D SPTs~\cite{tantivasadakarn2017dimensional}. \hfill $\blacksquare$
\end{example}
\begin{example}(Flux-Cheshire string braiding)
The condensation descendants of gauge charges, known as Cheshire strings, are labelled by unbroken subgroups $H<G$, and denoted by $\bb{1}^H$. They can be thought of as gauged version of 1+1D SSB phases. On such a Cheshire  string, there are point defects corresponding to domain walls between different SSB vacuua. In a 1+1D SSB phase, a domain wall may be created by acting with $g\in G$ on half of the system, but if $g\in H$, no domain walls will be created. The gauged version of these domain walls are point defects on $\bb{1}^H$, labelled by the cosets $[g]\in G/H$, and are denoted by $z_{[g]}: \bb{1}^H\to \bb{1}^H$. Braiding $\bb{1}^H$ with a flux loop $m_g$ simply creates such a point-defect:
    \begin{align}
        \Theta(m_g,\bb{1}^H)=z_{[g]}.
    \end{align} 

If the Cheshire string is defined by the set of charges condensed on it, denoted as $\bb{1}^{\{e_i\}}$, then its braiding with a flux $m_g$ is trivial if and only if the particle-flux braidings $\theta(e_i,m_g)$ are trivial. 

\hfill$\blacksquare$
\end{example}
\end{itemize}
 
 \section{2+1D gSPT from 3+1D string condensation\label{sec:z2z4example}}
In this section we present a thorough study of \symto construction for a 2+1D igSPT with symmetry $\bZ_2\times \bZ_4$. We briefly review key structures of the igSPT, then we discuss the dual string condensation and the sandwich construction. 

\subsection{Structure of the $\bZ_2\times \bZ_4$ igSPT}
The lattice model for this igSPT and various generalizations of it can be found in~\cite{Wen_2023}. We present a pure field-theoretical description here.

Consider a gapless system with $\bZ_2\times \bZ_4$ symmetry, coupled to a background $\bZ_2\times \bZ_4$ gauge field. We may decompose the $\bZ_4$ gauge field into two  gauge fields $A_2,\alpha\in C^1[M,\bZ]$, with the constraint $d\alpha=dA_2/2\mod 2,~dA_2=0\mod 2$. The gauge field $\alpha$ corresponds to the gauge field for the $\bZ_2$ subgroup of $\bZ_4$, which we denote by $\bZ_2^{gap}:=\<(0,2)\>$. And $A_2$ is the gauge field for the quotient $\bZ_4/\bZ_2^{gap}=\bZ_2$. Denote $K:=(\bZ_2\times \bZ_4)/\bZ_2^{gap}\cong \bZ_2\times \bZ_2$. We may write the partition function of the gapless system as 
\begin{align}
    Z[A_1,A_2,\alpha]
\end{align}
Now imagine that all degrees of freedom that are charged under $\bZ_2^{gap}$ are gapped, then we can integrate out these gapped degrees of freedom so that the dependence of the partition function on $\alpha$ becomes purely topological. The $\bZ_2\times \bZ_4$ igSPT is defined by the following partition function
\begin{align}
    Z_{igSPT}[A_1,A_2,\alpha]=Z_{gapless}[A_1,A_2]\exp\left(i\pi \int_{M_3} A_1A_2 \alpha\right).\label{eq: z2z4 partition function}
\end{align}
The dependence on $\alpha$ is indeed topological. $Z_{gapless}$ is the partition function of the gapless degrees of freedom, which have an effective $K=\bZ_2^2$ symmetry and are coupled to $K$ gauge fields $A_1,A_2$. Denote the topological part of the partition function as $Z_{top}[A_1,A_2,\alpha]=\exp\left(i\pi \int_{M_3}  A_1A_2 \alpha\right)$. Notice that $Z_{top}$ resembles the type-III gapped SPT with symmetry $\bZ_2^3$, where the flux of one $\bZ_2$ factor carries a projective representation of the other two. Similarly, with the partition function Eq.~\eqref{eq: z2z4 partition function} we see that the flux of $\alpha$ must carry a projective representation of $K=\bZ_2^2$. The partition function also states that the current for the $\alpha$ gauge field is 
\begin{align}
    *J_{\alpha}=\frac{1}{2} A_1A_2
\end{align}
Therefore the symmetry action by $(0,2)$ on a closed manifold $\Sigma_2$ is 
\begin{align}
    U_{(0,2)}^{\Sigma_2}=\exp\l 2\pi i \int_{\Sigma_2}*J_\alpha\r=(-1)^{\int_{\Sigma_2} A_1A_2}\label{eq:SPT_pump}
\end{align}
which we recognize is the partition function of the 1+1D $\bZ_2^2$-SPT(the cluster chain) on $\Sigma_2$. This is the field theory derivation of the SPT-pump property observed in~\cite{Wen_2023}. Namely, acting with symmetry $U_{(0,2)}$ on $\Sigma$ is equivalent to adding a cluster chain on $\Sigma$.

Notice that the topological part of the partition function is not gauge invariant. Under $A_1\to A_1+df,~A_2\to A_2+dg$, with the flatness condition $dA_1=dA_2=0 \mod 2$, we see that the change in $Z_{top}$ is 
\begin{align}
    \delta Z_{top}=\exp\l i\pi \int_{M_3} df\wedge dg \wedge \alpha \r=\exp\l \frac{i\pi}{2}\int_{M_3} f\wedge dg \wedge dA_2\r
\end{align}
where we used $d\alpha=dA_2/2$. This indicates the non-topological part of the partition function, $Z_{gapless}$, must also be not gauge invariant, in order for the full partition function $Z_{igSPT}$ to be gauge invariant. Therefore we must have
\begin{align}
    \delta Z_{gapless}[A_1,A_2]=\exp\l \frac{i\pi}{2}\int_{M_3} f\wedge dg \wedge dA_2\r.
\end{align}
We recognize that this anomaly is the boundary anomaly of a 3+1D $\bZ_2^2$-SPT with action $\frac{1}{4} \int_{C_4} A_1A_2dA_2$. Therefore the gapless degrees of freedom of the igSPT carry an anomalous symmetry action by $K=\bZ_2^2$. Candidate gapless theories with this anomaly include deconfined quantum critical point between different $\bZ_2$ symmetry breaking phases~\cite{Ji_2023}, or gapless boundary states of 3+1D twisted $\bZ_2^2$ gauge theories studied in~\cite{Chen_2016}. Unlike the usual anomaly in-flow mechanism where the anomaly is cancelled by a higher dimensional bulk, here the anomaly is cancelled by the gapped degrees of freedom, which transform under an enlarged symmetry group compared to the gapless degrees of freedom. The symmetry structure of the igSPT can be summarized by the group extension
\begin{align}
    1\to \bZ_2^{gap}\to \bZ_2\times \bZ_4\to \bZ_2\times \bZ_2\to 1
\end{align}
The anomaly of $\bZ_2\times \bZ_2$ is lifted by the group extension, and the full $\bZ_2\times \bZ_4$ symmetry is not anomalous. The effective symmetry for the gapless degrees of freedom is called the IR symmetry of the igSPT. We will usually denote the gapped symmetry of a $G$-gSPT by $N$, and the IR symmetry by $K$, which fit into the group extension
\begin{align}
    1\to N\to G\to K\to 1.
\end{align}

The key properties of the $\bZ_2\times \bZ_4$-igSPT that we wish to reconstruct from the topological holography are then 1). The symmetry extension structure. 2). The emergent anomaly of the IR symmetry. 3). The $\bZ_2^{gap}$ flux carries a projective representation of $K$. 4). The SPT-pump property Eq.\eqref{eq:SPT_pump}. We now turn to the topological holography construction.
\subsection{The \symto construction}
\subsubsection{The string condensation\label{sec:z2z4string condensation}}
Since the full symmetry of the igSPT is the non-anomalous $\bZ_2\times \bZ_4$, the \symto is the 3+1D untwisted $\bZ_2\times \bZ_4$ gauge theory, denoted by $\mathcal{Z}[2\vc_{\bZ_2\times \bZ_4}]$. We denote (generators of) gauge charges and fluxes in this theory by $e_1,e_2,m_1,m_2$. Based on our discussion in Sec.~\ref{sec:symto_review}, we look for magnetic, non-Lagrangian and simple condensations in this theory. For $\bZ_2\times \bZ_4$ there is a unique nontrivial 1+1D SPT, and the corresponding string is denoted by $S_\psi$, with the 2-cocycle $\psi$ given by 
\begin{align}
    \psi(\vec{i},\vec{j})=(-1)^{i_1j_2}.
\end{align}
We consider condensing the string $m_2^2S_\psi$. Notice this string is condensable since it is bosonic: $\Theta(m_2^2S_{\psi})=1$. On the other hand, $\Theta(m_2S_\psi)=e_1$, therefore the string $m_2S_\psi$ is not condensable unless $e_1$ is also condensed. The condensable algebra then has underlying object
\begin{align}
    \eA=\bb{1}\oplus m_2^2S_\psi. 
\end{align}
The algebra multiplication 1-morphism $\mu$ is identity on all components
\begin{align}
    \bb{1}\Box \bb{1} \xrightarrow{1_\bb{1}} \bb{1},~\bb{1}\Box m_2^2S_\psi \xrightarrow{1_{m_2^2S_\psi}}m_2^2S_\psi,~m_2^2S_\psi\Box \bb{1}\xrightarrow{1_{m_2S_\psi}} m_2^2S_\psi,~m_2^2S_\psi\Box m_2^2S_\psi \xrightarrow{1_{\bb{1}}} \bb{1}
\end{align}
The unit is $\bb{1}\xrightarrow{1\oplus 0}\bb{1}\oplus m_2^2S_\psi$ and the 2-associator and 2-unitors are identity 2-morphisms.  The braiding on $\eA$ is identity: $\beta=1_{\mu}: \mu\circ b_{\eA,\eA}\To \mu$($b_{\eA,\eA}=1_{\eA\Box \eA}$).

Next we examine the deconfined strings and particles after condensing $m_2^2S_\psi$. It is straightfoward to identify the deconfined particles. $e_2$ is confined since it has -1 braiding phase with $m_2^2S_\psi$. Therefore the deconfined particles are generated by $e_1,e_2^2$, which have a $\bZ_2\times \bZ_2$ fusion rule. From this we can already tell that the post-condensation phase must be a $\bZ_2^2$-gauge theory, based on the classification of 3+1D topological orders~\cite{Lan_2018,Lan_2019}, which states that if the particles in a 3+1D topological order form the symmetric fusion category $\rep(G)$, then the topological order must be a $G$-gauge theory. Therefore we only need to identify the twist of the this potentially twisted $\bZ_2^2$ gauge theory. To this end, we examine the deconfined strings.

Consider firstly the strings with flux sector $m_1$. $m_1$ alone is confined, due to the nontrivial string-string braiding\footnote{Mathematically, the free module $m_1\Box \eA$ is not a local module. For a free module $a\otimes \eA$ to be local, the braiding $\Theta(a,\eA)$ must be trivial~\cite{Zhao_2023}.}
\begin{align}
    \Theta(m_1,m_2^2S_\psi)=e_2^2
\end{align}
One way to trivialize the braiding is to simply condense the $e_2^2$ particle on the string $m_1$. This is equivalent to fusing with a Cheshire string of $e_2^2$. The resulting string is denoted as $m_1^{e_2^2}=m_1\otimes \bb{1}^{e_2^2}$. The string $m_1^{e_2^2}$ is then deconfined, as its braiding with the condensed string is now trivial. We can also further condense the particle $e_1$ on the string $m_1^{e_2^2}$ and obtain the string $m_1^{e_2^2,e_1}$, which is also deconfined. Therefore we find two deconfined strings with flux sector $m_1$: $m_1^{e_2^2}$ and $m_1^{e_2^2,e_1}$.

Next we consider strings with flux sector $m_2$. Similarly $m_2$ alone is confined, by the nontrivial braiding $\Theta(m_2,m_2^2S_\psi)=e_1$. To trivialize the braiding we can consider condensing $e_1$ on $m_2$ and obtain $m_2^{e_1}$. Therefore $m_2^{e_1}$ is deconfined. We may also further condense the deconfined charge $e_2^2$ and obtain $m_2^{e_1,e_2^2}$, which is also deconfined. 

There is in fact a third deconfined string with flux sector  $m_2$. To construct this string, we first condense $e_2^2$ on $m_2$, resulting in the string $m_2^{e_2^2}$. Now on this string the gauge symmetry is Higgsed down to $\bZ_2\times \bZ_2=\<(1,0),(0,2)\>$. There is a unique nontrivial 1+1D SPT for this subgroup. The corresponding SPT-string is therefore a string in the $\bZ_2\times \bZ_4$ gauge theory that can only exist when the gauge group is locally Higgsed down to the $\bZ_2^2$ subgroup on the string, which we denote by $\widetilde{S}_\phi$. Adding $\widetilde{S}_\phi$ to $m_2^{e_2^2}$, we have a string $m_2^{e_2^2}\widetilde{S}_\phi$. Notice that since $\widetilde{S}_\phi$ is the gauged version of a cluster chain of the subgroup $\<(1,0),(0,2)\>$, its braiding with an $m_2^2$-string is $e_1$. Thus the braiding of $m_2^{e_2^2}\widetilde{S}_\phi$ with $m_2^2S_\psi$ is
\begin{align}
\Theta(m_2^{e_2^2}\widetilde{S}_\phi,m_2^2S_\psi)=e_1^2=1.
\end{align}
Therefore the string $m_2^{e_2^2}\widetilde{S}_\phi$ is deconfined. In summary, there are in total three deconfined strings with flux sector $m_2$: $m_2^{e_1},m_2^{e_1,e_2^2}$ and $m_2^{e_2^2}\widetilde{S}_\psi$.

One interesting feature of these deconfined strings with nontrivial flux sectors is that they are all Cheshire, i.e. they all have some charge condensation on them. The "pure" flux loops $m_1,m_2$ are not deconfined.  This is exactly the behaviour of flux loops in a twisted gauge theory. We recall that in a twisted (abelian)gauge theory $\mathcal{Z}[2\vc_G^\pi],~\pi \in Z^4[G,U(1)]$, a flux loop $m_g$ can be thought of as (gauged version of)the boundary of a 2+1D $G$-SPT, therefore carries an anomalous symmetry action with anomaly given by slant product: $i_g\pi\in Z^3[G,U(1)]$~\cite{Else_2017}. Thus if the flux loop is gapped, it must break(Higgs) the gauge symmetry, to a subgroup $H$ such that $i_g\pi$ is trivialized when restricted to $H$. When $i_g\pi$ is nontrivial, the flux loop $m_g$ must be Cheshire, and there can be different types of fluxes with flux sector $g$ that have different charge condensation patterns~\cite{Else_2017}. 

For the deconfined flux strings we find, it is direct to verify that the Cheshire charges of these deconfined strings match with those of the fluxes in a twisted $\bZ_2^{A}\times \bZ_2^{B}$-gauge theory, with the twist given by 
\begin{align}
    \pi(\vec{i},\vec{j},\vec{k},\vec{l})=(-1)^{i_1j_2k_2l_2}.\label{eq:z2z2twist}
\end{align}
Namely, writing elements of $\bZ_2^{A}\times \bZ_2^{B}$ as $\vec{i}=(i_1,i_2),~i_1,i_2=0,1,$ we have that in this twisted $\bZ_2^2$ gauge theory strings with flux sector $(1,0)$ must condense $e_B$, and for strings with flux sector $(0,1)$, they must condense either $e_A$ or $e_B$(or both). 
Now, among the deconfined strings we find after condensing $m_2^2 S_\psi$, the ones with flux sector $m_1$ are $m_1^{e_2^2}$ and $m_1^{e_2^2,e_1}$, and the ones with flux sector $m_2$ are $m_2^{e_1},m_2^{e_2^2}\widetilde{S}_\phi,m_2^{e_1,e_2^2}$. Clearly they have the Cheshire charge structure as fluxes in the twisted $\bZ_2^2$-gauge theory, if we just relabel $e_1\to e_A, e_2^2\to e_B$. 

Since there are only four twisted $\bZ_2^2$-gauge theories, and Eq.\eqref{eq:z2z2twist} is the only twist such that the fluxes have this Cheshire charge structure, we conclude that the topological order resulting from condensing $m_2^2S_\psi$ is the twisted $\bZ_2^2$ gauge theory with twist~Eq.\eqref{eq:z2z2twist}. 

\subsubsection{The sandwich construction}
Now let us turn to the sandwich construction. The bulk of the sandwich is the untwisted $\bZ_2\times\bZ_4$ gauge theory. On the symmetry boundary, we condense the gauge charges $e_1,e_2$(or equivalently the Cheshire string $\bb{1}^{e_1,e_2}$). The symmetry twist defects are then $m_1,m_2$ strings on the symmetry boundary, generating a $\bZ_2\times \bZ_4$ symmetry.  On the physical boundary, we choose to condense $m_2^2S_\psi$. Since the condensation $1+m_2^2S_\psi$ is not Lagrangian, the sandwich describes a gapless state. Since condensing $m_2^2S_\psi$ reduces the \symto to a twisted $\bZ_2^2$-gauge theory, we see that the IR symmetry is an anomalous $\bZ_2^2$ symmetry. This matches with the symmetry structure of the $\bZ_2\times \bZ_4$-igSPT. 

Now consider an $m_2^2$ membrane operator on $\bref$, supported on a 2D manifold $\Sigma_2\subset \bref$. Since $m_2^2S_\psi$ is condensed on $\bphys$, we can pull a membrane of $m_2^2S_\phi$ out of $\bphys$, move it to $\bref$, and fuse it with the $m_2^2$-membrane. This whole process is an identity on the ground space. This shows the $m_2^2$ membrane on $\bref$ is equivalent to an $S_\phi$ membrane. Although $\phi$ is defined as a 1d $\bZ_2\times \bZ_4$ SPT, the subgroup $\bZ_2^{gap}$ acts trivially on it, and it can be viewed as a 1d SPT of the quotient $K=\bZ_2\times \bZ_4/\bZ_2^{gap}\cong \bZ_2\times \bZ_2$. This reproduces the SPT-pump property of the $\bZ_2\times \bZ_4$-igSPT. The property of the $\bZ_2^{gap}$ flux can be obtained by gauging the symmetry, which we will discuss in Sec.~\ref{sec:gauging gSPT}

\subsubsection{Dimension reduction of string condensation}
Previously we identified the phase resulting from condensing $m_2^2S_\psi$ by matching the Cheshire charges of deconfined fluxes with those of fluxes in twisted gauge theories. Another major feature of twisted gauge theories in 3+1D is nontrivial three-loop braiding of flux loops. Here we derive the post-condensation phase from the perspective of three-loop braiding and dimension reduction. 

\begin{figure}[h]
    \centering
\includegraphics[width=0.85\linewidth]{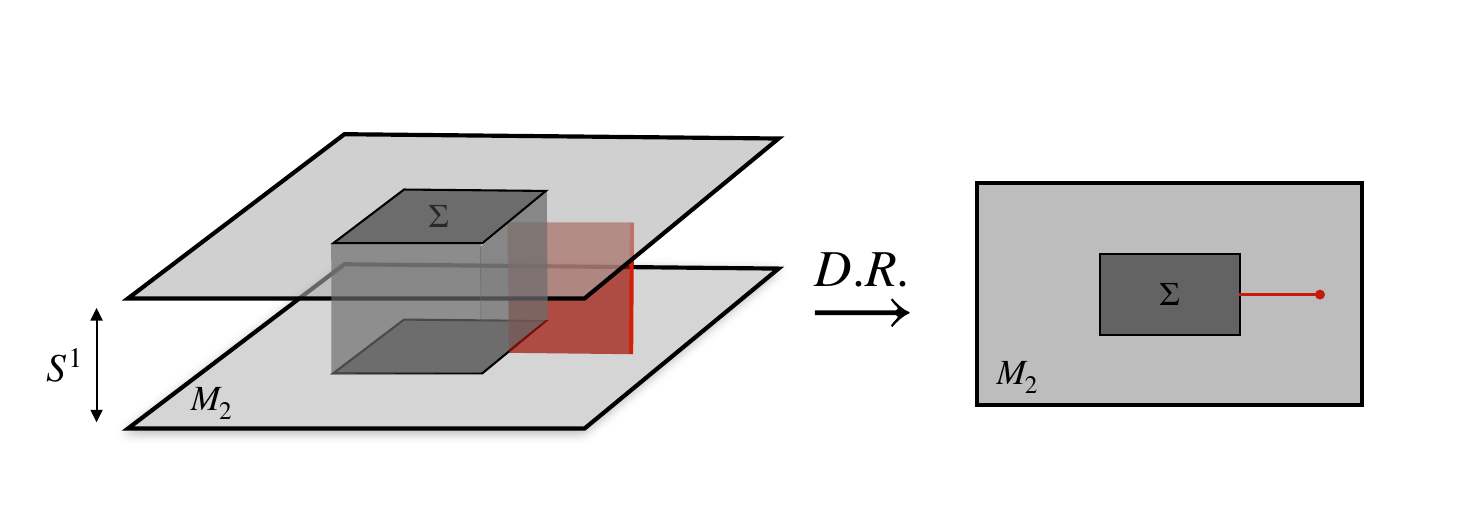}
    \caption{Dimension reduction of string condensation. On the left we show a topological order put on $M_2\times S^1$, with a volume $\Sigma\times S^1$ having a string condensation(darker region). A red membrane corresponding to a condensed string can end on the boundary of the condensation region. On the right is the picture after dimension reduction, the region $\Sigma$ now has an anyon condensation, and the red membrane reduces a red string operator that can end on the boundary of the condensation region, corresponding to a condensed anyon.}
    \label{fig:DR2}
\end{figure}

Similar to how a 3+1D topological order reduces to direct sums of 2+1D topological order upon dimension reduction, we can perform dimension reduction to a string condensation so that it becomes an anyon condensation of 2+1D topological orders. More concretely, we can consider putting a 3d topological order $\cc{C}^{3d}$ on a manifold $M_2\times S^1$. Let $\Sigma\subset M_2$ be a finite region of the slice $M_2$. Consider performing the string condensation in the volume $\Sigma\times S^1$. Upon compactifying $S^1$, the volume $\Sigma\times S^1$ becomes a region $\Sigma$ of the 2d manifold $M_2$ that has an anyon condensation.  See Fig.~\ref{fig:DR2} for the setup. Therefore a string condensation reduces to an anyon condensation after dimension reduction. 

For the current example, consider dimension reduction in the $m_1$-sector, we have that $S_\psi$ reduces to $e_{i_{(1,0)}\psi}=e_2^2$. Therefore the string condensation reduces to the anyon condensation $1+m_2^2e_2^2$ in the $m_1$-sector. This  anyon condensation reduces the 2+1D $\bZ_4$-gauge theory to the 2+1D double semion theory(DS), generated by $e_2m_2,e_2^2$. Thus the post-condensation phase reduces to $D(\bZ_2)\times \text{DS}$ in the $m_1$-sector. This is exactly the dimension reduction of the twisted 3+1D $\bZ_2^2$ gauge theory in the $m_1$-sector. In particular this means the $m_2$ flux loops have $-1$ braiding phase when linked with base loop $m_1$ in the post-condensation theory. This matches with the three-loop braiding structure of the 3+1D twisted $\bZ_2$-gauge theory. 

Similarly, the string condensation reduces to the anyon condensation $1+m_2^2e_1$ in the $m_2$-sector. This anyon condensation reduces the 2+1D $\bZ_2\times \bZ_4$ theory to the twisted  2+1D $\bZ_2\times \bZ_2$ theory, generated by $m_2,m_1e_2,e_1$. In particular, the fluxes are $m_2,m_1e_2$, which have mutual $i$ statistics. This means the flux loops $m_1,m_2$ in the post-condensation phase have $i$ braiding phase when linked with $m_2$. This also matches with the three-loop braiding structure of the 3+1D twisted $\bZ_2^2$-gauge theory. 

The dimension reduction of the string condensation can be schematically represented as a "commutative diagram", shown below.
\begin{align}
    \begin{tikzcd}[ampersand replacement=\&]
	{3d~\mathbb{Z}_2\times \mathbb{Z}_4-\text{theory}} \&\& {3d~(\mathbb{Z}_2\times \mathbb{Z}_2)^\pi-\text{theory}} \\
	{2d~\mathbb{Z}_2\times \mathbb{Z}_4-\text{theory}} \&\& {2d~(\mathbb{Z}_2\times \mathbb{Z}_2)^{i_g\pi}-\text{theory}}
	\arrow["{\text{condense}~m_2^2S_\psi}", from=1-1, to=1-3]
	\arrow["{\text{D.R.~in}~ g-\text{sector}}"', from=1-1, to=2-1]
	\arrow["{\text{D.R.~in}~ g-\text{sector}}", from=1-3, to=2-3]
	\arrow["{\text{condense}~ m_2^2e_{i_g\psi}}"', from=2-1, to=2-3]
\end{tikzcd}\label{fig: dr of condensation}
\end{align}
\section{Magnetic simple condensable algebras in $\mathcal{Z}[2\vc_G]$ \label{sec:classification}}
Starting in this section we turn to general study of condensable algebras in 3+1D gauge theories. We first give a classification of magnetic simple condensation in $\mathcal{Z}[2\vc_G]$. We will see that a magnetic simple condensation is determined by a quadruple data $(\alpha,\phi,\sigma,\beta)$ satisfying a set of consistency conditions. We then discuss the physical meaning and applications of the classification in the next section. Definition of a condensable algebra is reviewed in appendix~\ref{app:condensable algebra}. Details of the braided fusion 2-category $\mathcal{Z}[2\vc_G]$ is reviewed in appendix~\ref{app:Z[2vcG]}. 

A condensable algebra in $\mathcal{Z}[2\vc_G]$ is a $G$-crossed braided multi-fusion 1-category such that the  $G$-action on the monoidal unit permutes all the simple summands transitively~\cite{D_coppet_2024}.  Denote the algebra by $\eA$, with decomposition $\eA=\oplus_g \eA_g$ as a $G$-graded category. The unit component $\eA_1$ is a braided multi-fusion 1-category with $G$-action. If $\eA_1$ is not fusion, then the monoidal unit of $\eA$ is necessarily a Cheshire string with multiple simple objects(vaccua) permuted by $G$-action transitively. Since charge condensation corresponds to $G$-SSB phases, condensable algebras dual to $G$-symmetric phases must correspond to a $G$-crossed braided fusion 1-category. In this case the unit component $\eA_1$ is equipped with a braided fusion 1-category structure, that describes the topological order of the dual 2+1D system. For condensable algebras dual to phases with no topological order, $\eA_1$ must be the trivial braided fusion category, i.e. $\eA_1=\vc$ and $\bC_1\in \vc$ is the unit of the algebra. In this case, it is a standard result that if the trivial degree component of a $G$-crossed fusion category is trivial, then the $G$-crossed fusion category is pointed. Thus the underlying category of $\eA$ must be $\vc_N$ for some group $N<G$. Since the $G$-action maps $g*(\eA_h)\subset \eA_{\cg{g}{h}}$, the support of $\eA$ must be a normal subgroup of $G$. Therefore a magnetic simple condensable algebra of $\mathcal{Z}[2\vc_G]$ is a $G$-crossed braided fusion category whose underlying category is $\vc_N$, with $N\lhd G$ a normal subgroup. 

The $G$-crossed braided fusion category structure on $\vc_N$ is defined by algebraic data $(\alpha, \phi, \sigma, \beta)$ as follows. Denote the simple objects of $\vc_N$ as $\bC_n,~n\in N$. The monoidal product on $\vc_N$ is denoted as $\diamond$, and we have 
\begin{align}
    \bC_n\diamond \bC_m=\bC_{nm}.
\end{align}
This corresponds to the product 1-morphism of the algebra: $\mu: \eA\Box \eA\to \eA$.
The monoidal unit is $\bC_{1}$. $\alpha$ is the associator 
\begin{align}
    \alpha(n_1,n_2,n_3):(\bC_{n_1}\diamond \bC_{n_2})\diamond \bC_{n_3}\to \bC_{n_1}\diamond(\bC_{n_2}\diamond \bC_{n_3})
\end{align}
which corresponds to the 2-associator of the algebra product, depicted in the diagram below.
\begin{align}
    \begin{tikzcd}[ampersand replacement=\&]
	{\eA\Box \eA\Box \eA} \& {\eA\Box \eA} \\
	{\eA\Box \eA} \& \eA
	\arrow["{1_{\eA}\Box \mu}", from=1-1, to=1-2]
	\arrow["{\mu\Box 1_{\eA}}"', from=1-1, to=2-1]
	\arrow["\mu", from=1-2, to=2-2]
	\arrow["\alpha"{description}, Rightarrow, from=2-1, to=1-2]
	\arrow["\mu"', from=2-1, to=2-2]
\end{tikzcd}
\end{align}
The $G$-action is denoted by $g*$, which must act on simple objects as
\begin{align}
    g*\bC_{n}=\bC_{\cg{g}{n}}\label{eq: G-action 1}
\end{align}
The $G$-action is linear in $G$ up to an isomorphism, denoted as $\phi$, defined as 
\begin{align}
    \phi(g_1,g_2|n)\in \bC^\times: g_1*g_2*\bC_n\to g_1g_2*\bC_n\label{eq: G-action 2}
\end{align}
The $G$-action and the isomorphism $\phi$ are part of the definition of $\eA$ as an object of $\mathcal{Z}[2\vc_G]$. The $G$-action should be compatible with the monoidal product up to an isomorphism, denoted by $\sigma$, defined as 
\begin{align}
    \sigma(g|n_1,n_2)\in\bC^\times: g*\bC_{n_1}\diamond g*\bC_{n_2}\to g*(\bC_{n_1}\diamond \bC_{n_2})
\end{align}
This is part of the definition of $\mu:\eA\Box\eA\to \eA$ as a 1-morphism in $\mathcal{Z}[2\vc_G]$.
The $G$-crossed braiding is defined as 
\begin{align}
    \beta(n_1,n_2): n_1*\bC_{n_2}\diamond \bC_{n_1}\to \bC_{n_1}\diamond \bC_{n_2}
\end{align}
which comes from the braided algebra structure on $\eA$, i.e. a 2-isomorphism in the following diagram
\begin{align}
    \begin{tikzcd}[ampersand replacement=\&]
	{\eA\Box \eA} \&\& {\eA\Box \eA} \\
	\& \eA
	\arrow["{b_{\eA,\eA}}", from=1-1, to=1-3]
	\arrow[""{name=0, anchor=center, inner sep=0}, "\mu"', from=1-1, to=2-2]
	\arrow["\mu", from=1-3, to=2-2]
	\arrow["\beta"{description}, shorten <=4pt, shorten >=8pt, Rightarrow, from=1-3, to=0]
\end{tikzcd}
\end{align}
The conditions for $\eA$ to have a braided algebra structure(see Appendix~\ref{app:condensable algebra}), require that the data $(\alpha,\phi,\sigma,\beta)$ satisfy the following equations
\begin{align}
    d_N\alpha=1, ~ d_G \phi=1,~ d_N \phi& =d_G\sigma,   ~d_N\sigma=d_G\alpha\nonumber\\
     \beta(\cg{g}{n_1},\cg{g}{n_2})\sigma(g|n_1,n_2)\phi(g,n_1|n_2)&=\phi(\cg{g}{n_1},g|n_2)\sigma(g|\cg{n_1}{n_2},n_1)\beta(n_1,n_2)\nonumber\\
     \alpha(n_1,n_2,n_3)\alpha(\cg{n_1}{n_2},\cg{n_1}{n_3},n_1)\beta(n_1,n_2)\beta(n_1,n_3)&=\alpha(\cg{n_1}{n_2},n_1,n_3)\beta(n_1,n_2n_3)\sigma(n_1|n_2,n_3)\nonumber\\
    \alpha(n_1,n_2,n_3)\alpha(\cg{n_1n_2}{n_3},n_1,n_3)\beta(n_1n_2,n_3)\phi(n_1,n_2|n_3)&=\alpha(n_1,\cg{n_2}{n_3},n_2)\beta(n_1,\cg{n_2}{n_3})\beta(n_2,n_3)\label{eq: cocycle condition}
\end{align}
where $d_N$ and $d_G$ are the standard horizontal and vertical differentials of the double complex $C^p[G,C^q[N,\bC^\times]]$, with $G$ acting on $N$ by conjugation.  Following~\cite{naidu2011crossed}, we make the following definition.
\begin{definition}
    A quadruple $\kappa=(\alpha, \phi,\sigma,\beta)$ that satisfies the above conditions is called a quasi-abelian 3-cocyle of the pair $(G, N)$. Quasi-abelian 3-cocyles of $(G,N)$ form an abelian group under poin-twise multiplication, which is denoted by $Z^3_{qa}[(G,N),\bC^\times]$. 
\end{definition}
Therefore we see that a magnetic simple condensation in $\mathcal{Z}[2\vc_G]$  is determined by a quasi-abelian 3-cocyle $\kappa\in Z^3_{qa}[(G,N),\bC^\times]$ for some normal subgroup $N\lhd G$. Denoted the corresponding condensable algebra as $\eA[\kappa]=\eA[N,\alpha,\phi,\sigma,\beta]$.
As in ordinary group-cohomology, there is a concept of quasi-abelian coboundary~\cite{naidu2011crossed}.
\begin{definition}
    A quasi-abelian 3-coboundary of the pair $(G,N)$ is a quadruple $(\alpha, \phi,\sigma,\beta)$ such that there exists $\eta\in C^2[N,\bC^\times]$, ~$\epsilon\in C^1[N,C^1[G,\bC^\times]]$, and
\begin{align}
    &\alpha=d_N \eta,~ \phi=d_G\epsilon,~ \sigma=d_N \epsilon d_G \eta,~\nonumber\\
    &\beta(n_1,n_2)=\epsilon(n_1,n_2)\frac{\eta(\cg{n_1}{n_2},n_1)}{\eta(n_1,n_2)}\label{eq: coboundary}
\end{align}
The set of quasi-abelian 3-coboundaries of $(G,N)$ is denoted by $B_{qa}^3[(G,N),\bC^\times]$, which is an abelian group under point-wise multiplication.
\end{definition}

By direction computation, a 3-coboundary of $(G,N)$ is automatically a 3-cocycle of $(G,N)$, therefore we can form the quotient $Z^3_{qa}[(G,N),\bC^\times]/B^3_{qa}[(G,N),\bC^\times]$, which is denoted by $H^3_{qa}[(G,N),\bC^\times]$, called the third quasi-abelian cohomology group of the pair $(G,N)$. As shown in~\cite{naidu2011crossed}, two quasi-abelian cocyles $\kappa, \kappa'\in Z^3_{qa}[(G,N),U(1)]$ define isomorphic $G$-crossed braided fusion category structures on $\vc_N$ if and only if they differ by a coboundary. Therefore we have
\begin{tcolorbox}
    Isomorphism classes of magnetic simple condensable algebras in $\mathcal{Z}[2\vc_G]$ are classified by the pairs $(N,\kappa)$, where $N\lhd G$ is a normal subgroup, and $\kappa\in H^3_{qa}[(G,N),\bC^\times]$.
\end{tcolorbox}

\section{Discussion and application of the classification\label{sec:discussion_application}}
We now discuss the physical meaning of the above classification. According to our discussion in Sec.~\ref{sec:symto_review}, the above classification is holographically dual to the classification of 2+1D gapped and gapless SPT. 

Firstly, we can read off the set of condensed strings from the data $\phi$ as follows. Recall that an object in $\mathcal{Z}[2\vc_G]$ is a $G$-crossed finite semi-simple category. For the algebra $\eA[N,\alpha,\phi,\sigma,\beta]$, the underlying $G$-graded category is $\vc_N$, and the $G$-action is defined by Eq.\eqref{eq: G-action 1} and Eq.\eqref{eq: G-action 2}, which is determined by $\phi$. By decomposing the algebra into sum of simple objects, we see that the underlying object of $\eA[N,\alpha,\phi,\sigma,\beta]$ is
\begin{align}
    \eA=\bigoplus_{[n]\in \text{Cl}(G), n\in N}m_{[n]} S_{\phi[n]},~\phi[n]:=\phi(-,-|n)|_{C_G(n)}\in Z^2[C_G(n),\bC^\times],\label{eq:strings}
\end{align}
where $\text{Cl}(G)$ stands for the set of conjugacy classes of $G$ and $C_G(n)$ is the centralizer of $n$ in $G$. We see that the meaning of $\phi$ is clear: it determines the SPT-strings bound to flux loops in the condensable algebra. 

Then it is clear that when $N\neq G$ the condensation is not Lagrangian. This is obtained by analyzing the deconfined charges. Since the charge-flux braiding is unaffected by the SPT-string bound to the flux, and $N$ is the flux sectors contained in the condensed strings, we see that a gauge charge is deconfined if and only if $N$ acts trivially on it. The collection of such charges form a fusion category $\rep(G/N)$, which is a sub-category of $\rep(G)$. Therefore from the classification of 3+1D topological orders~\cite{Lan_2018,Lan_2019}, we can tell that the post-condensation phase must be a $G/N$-gauge theory, possibly twisted. Holographically, this means the condensable algebra $\eA[N,\alpha,\phi,\sigma,\beta]$ describes a gSPT with an IR symmetry $G/N$, possibly anomalous. When $N=G$, all gauge charges are confined, and the condensing is Lagrangian.

We first consider the case $N=G$, corresponding to Lagrangian condensations and dual to gapped SPTs. 
\subsection{Lagrangian condensation and 2+1D SPTs}
 When $N=G$,  we have $H^3_{qa}[(G,G),\bC^\times]=H^3[G,\bC^\times]$, recovering the classification of 2+1D bosonic SPTs with finite unitary symmetry. Indeed, when $N=G$, $\beta$ can be trivialized by modifying the cocycle by a coboundary with $\epsilon=\beta$, then the last two conditions of ~Eq.\eqref{eq: cocycle condition} state
\begin{align}
        &\alpha(g_1,g_2,g_3)\alpha(\cg{g_1}{g_2},\cg{g_1}{g_3},g_1)=\alpha(\cg{g_1}{g_2},g_1,g_3)\sigma(g_1|g_2,g_3)\label{eq:N=G1}\\
    &\alpha(g_1,g_2,g_3)\alpha(\cg{g_1g_2}{g_3},g_1,g_2)\phi(g_1,g_2|g_3)=\alpha(g_1,\cg{g_2}{g_3},g_2)\label{eq:N=G2}
\end{align}
which show that $\phi,\sigma$ are completely determined by $\alpha\in Z^3[G,\bC^\times]$. 

We notice that the symmetry twist defects in a 2+1D $G$-SPT form a pointed $G$-crossed braided fusion category, which matches with the structure of Lagrangian magnetic simple condensation algebras. Namely when $N=G$ the data $(\alpha,\phi,\sigma,\beta)$ are exactly what are needed to define a $G$-crossed braided fusion category structure on the underlying pointed category $\vc_G$. In the convention of~\cite{barkeshli2019symmetry}, $\alpha$ is the associator of the $G$-crossed braided fusion category, $\phi$ is the $\eta$-symbol, $\sigma$ is the $U$-symbol, and $\beta$ is the $G$-crossed braiding. Therefore it is natural to conjecture that the string condensation associated with $(\alpha,\phi,\sigma,\beta)$ must describe the braiding and fusion properties of the symmetry twist defects of the dual 2+1D SPT. We analyze the string condensations dual to type I$\sim$III SPTs to illustrate.

 First of all, we can now determine the condensed strings from the dual 2+1D SPT: when $N=G$, $\phi$ is determined by $\alpha$ via Eq.\eqref{eq:N=G2}, thus by Eq.\eqref{eq:strings} we have
\begin{tcolorbox}
    A 2+1D SPT with cocycle $[\alpha]\in H^3[G,U(1)]$ is dual to a string condensation in $\mathcal{Z}[2\vc_G]$ that condenses the strings
    \begin{align}
         \eA^\alpha=\bigoplus_{[g]\in \text{Cl}(G)}m_{[g]} S_{i_g\alpha},~i_g\alpha(h_1,h_2):=\frac{\alpha(h_1,g,h_2)}{\alpha(h_1,h_2,g)\alpha(g,h_1,h_2)}~\forall h_1,h_2\in C_G(g).
    \end{align}
\end{tcolorbox}

\subsubsection{Type-III SPT\label{sec:type-III SPT}}
The slant product $i_g\alpha$ is nontrivial for type-III SPTs. A 2+1D type-III SPT has symmetry $\bZ_{N_1}\times \bZ_{N_2}\times \bZ_{N_3}$ and cocycle
\begin{align}
    \alpha_p(\vec{i},\vec{j},\vec{k}):=\exp\left(\frac{2\pi i p}{N_{123}}i_1j_2k_3\right),~p=1,\cdots, N_{123}-1,
\end{align}
where $N_{123}=\text{GCD}(N_1,N_2,N_3)$. Computing the slant product, we have
\begin{align}
i_{(1,0,0)}\alpha(\vec{i},\vec{j})=\exp\left(\frac{2\pi i p}{N_{123}}i_2j_3\right),~i_{(0,1,0)}\alpha(\vec{i},\vec{j})=\exp\left(\frac{2\pi i p}{N_{123}}i_1j_3\right),~i_{(0,0,1)}\alpha(\vec{i},\vec{j})=\exp\left(\frac{2\pi i p}{N_{123}}i_2j_3\right).
\end{align}
Labelling the generating 1+1D SPT of $\bZ_{N_i}\times \bZ_{N_j}$ as $S_{ij}$, which has order $N_{ij}:=\text{GCD}(N_i,N_j)$. We see that the corresponding string condensation condenses
\begin{align}
    m_1S_{23}^{\frac{pN_{23}}{N_{123}}},~m_2S_{13}^{\frac{pN_{13}}{N_{123}}},~m_3S_{12}^{\frac{pN_{12}}{N_{123}}}
\end{align}
The simplest case is when $G=\bZ_2^{(1)}\times \bZ_2^{(2)}\times \bZ_2^{(3)}$, then the type-III SPT of $\bZ_2^3$ is dual to the string condensation that condenses $m_1S_{23},m_2S_{13},m_3S_{12}$. The gapped boundary that has this string condensation has been constructed in various lattice models of $\mathcal{Z}[2\vc_{\bZ_2^3}]$, including the color code model~\cite{song2024magic} and the Vasmer-Browne model~\cite{Zhu_2022}.

Now we consider a \symto sandwich construction with the standard Dirichlet symmetry boundary(all-charge-condensed), and a physical boundary where $m_1S_{23},m_2S_{13},m_3S_{12}$ are condensed. It is clear that with this choice of physical boundary, the corresponding 2+1D SPT has the decorated domain wall structure of a type-III SPT. Consider for instance a vertical membrane of $m_1S_{23}$ than ends on the physical boundary. This membrane can not end on the symmetry boundary, as $m_1S_{23}$ is not condensed there. However, on the symmetry boundary we have $m_1S_{23}\sim m_1$. This is because when gauge charges are condensed, so are all the condensation descendants of charges. Therefore the membrane of $m_1S_{23}$ becomes the symmetry generator of $\bZ_2^{(1)}$ on the symmetry boundary. Following the general discussion of sandwich construction in Sec.~\ref{sec:sandwich}, we see that this means the $\bZ_2^{(1)}$-domain wall creation operator decorated with $S_{23}$ on the boundary is condensed in the ground state of the 2+1D phase, which matches with the decorated domain wall structure of the type-III SPT.

We can also probe the corresponding 2+1D phase by gauging the symmetry. In the \symto sandwich construction, gauging a symmetry amounts to changing the symmetry boundary from the all-charge-condensed one to the all-flux-condensed one. Consider now a symmetry boundary with all fluxes $m_i$ condensed, and a physical boundary with $m_1S_{23},m_2S_{13},m_3S_{12}$ condensed. With this setup, the gauge charges in the bulk of the sandwich can no longer end on the symmetry boundary. Consequently, when viewed as excitations in a 2+1D system, the gauge charges in the sandwich no longer correspond to symmetry charges of a $\bZ_2^3$ symmetry that can be created by local operators. Instead they are now anyonic excitations that can only be created by string operators in pairs.  Now a vertical string of $m_1S_{23}$ can end on the physical boundary, notice that it can also end on the symmetry boundary: although $S_{23}$ is not condensed on $\bref$, it can still end on $\bref$ because SPT-strings can be open. This vertical $m_1S_{23}$ maps, under the interval compactification, to an anyon of a 2+1D system, denoted by $\widetilde{m}_1$. Similarly we denote the dimension reduction of vertical strings of $m_2S_{13},m_3S_{12}$ by $\widetilde{m}_2,\widetilde{m}_3$. They have nontrivial braiding with gauge charges: $\theta(\widetilde{m}_i,e_i)=-1$, which are inherited from the string-particle braidings. Now the upper end of a vertical $m_1S_{23}$ at $\bref$ carries an edge mode of the $S_{23}$ string, i.e. 1-morphisms $z:S_{23}\to \bb{1}$. As a result, the anyon $\widetilde{m}_1$ has fusion rule: $\widetilde{m}_1\otimes \widetilde{m}_1=(1+e_2)(1+e_3)$. Similarly we have $\widetilde{m_2}\otimes \widetilde{m_2}=(1+e_1)(1+e_3),~\widetilde{m_3}\otimes \widetilde{m_3}=(1+e_1)(1+e_2)$. These are precisely the fusion rules of gauge fluxes in the gauged type-III SPT. We conclude that the sandwich now describes the gauged type-III SPT.

However, specifying the set of condensed strings is in general not sufficient for determining the string condensation. Famously for the 3d toric code/$\mathcal{Z}[2\vc_{\bZ_2}]$ there are two condensable algebras that both condense the flux loop $m$, which are not isomorphic and dual to the trivial and nontrivial 2+1D SPT of $\bZ_2$~\cite{Zhao_2023}. The other data $\alpha,\sigma,\beta$ in the classification distinguish these different condensations sharing the same set of condensed strings. We discuss two more classes of string condensations to illustrate. 

\subsubsection{Type-II SPT\label{sec:type-II SPT}}
\begin{figure}[h]
    \centering
    \includegraphics[width=0.8\linewidth]{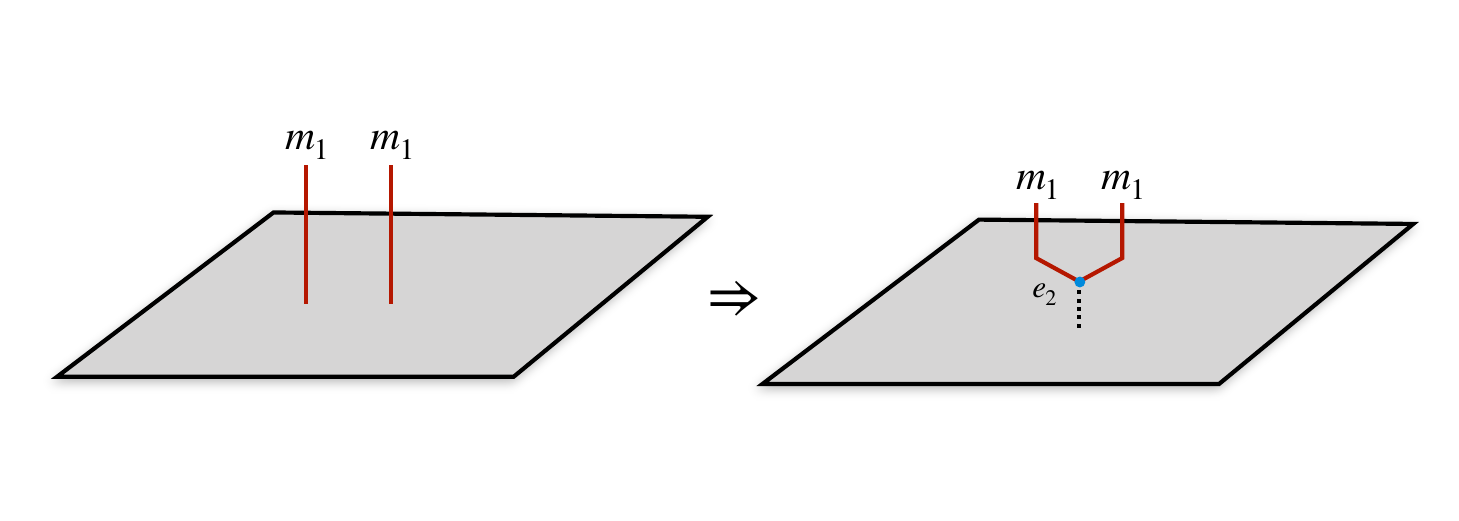}
    \vspace{-1cm}
    \caption{Physical interpretation of the product 1-morphism $\mu: m_1\Box m_1\xrightarrow{e_2} \bb{1}$. On the left two $m_1$ strings terminate on the boundary where $m_1,m_2$ condense with the  nontrivial $\sigma$ Eq.\eqref{eq:sigma}. On the right the configuration is deformed into one where the two $m_1$ fuse into identify. A charge $e_2$ is created in the process. A similar relation holds for $m_2$ and $e_1$. These relations match with the fusion rules of the gauged SPT: $m_2^2=e_1,~m_1^2=e_2$.}
    \label{fig:sigma}
\end{figure}

A  2+1D type-II SPT has symmetry $\bZ_{N_1}\times \bZ_{N_2}$, and a type-II cocycle represented by
\begin{align}
    \alpha_p(\vec{i},\vec{j},\vec{k})=\exp\left(\frac{2\pi i p}{N_{1}N_2}i_1(j_2+k_2-[j_2+k_2])\right),~p=1,\cdots, N_{12}-1.\label{eq:typeII cocycle}
\end{align}
For clarity of presentation we take $N_1=N_2=2, p=1$, then computing the factors $\phi, \sigma$ according to Eq.\eqref{eq:N=G1},Eq.\eqref{eq:N=G2}, we find 
\begin{align}
    &\phi(\vec{i},\vec{j}|\vec{k})=(-1)^{i_2j_2k_1},~\sigma(\vec{i}|\vec{j},\vec{k})=(-1)^{i_1j_2k_2}.
\end{align}
Although $\phi[\vec{k}]$ is not identity, it is cohomologically trivial. The flux loops in the theory $\mathcal{Z}[2\vc_G]$ have $\phi=1$, therefore we set $\phi$ to 1 by modifying with a coboundary
\begin{align}
    \epsilon(\vec{i}|\vec{k})=i^{i_2k_1}.
\end{align}
Modifying the quadruple $(\alpha,\phi,\sigma,\beta)$ by the coboundary $\epsilon$(recall the definition in~Eq.\eqref{eq: coboundary}), $\phi$ is set to $1$, and $\sigma$ becomes
\begin{align}
    \sigma(\vec{i}|\vec{j},\vec{k})=(-1)^{i_1j_2k_2+i_2j_1k_1}.\label{eq:sigma}
\end{align}
We now provide an interpretation of the data $\sigma$. Recall that part of the definition of a condensable algebra is a product 1-morphism: $\mu: \eA\Box \eA\to \eA$. For a magnetic simple condensable algebra, the product 1-morphism is determined by $\sigma$. For the $\sigma$ above, we have the following algebra product 1-morphism,
\begin{align}
 \mu:~  &\bb{1}\Box \bb{1}\xrightarrow{1}\bb{1},~\bb{1}\Box m_i \xrightarrow{1} m_i,~ m_i\Box  \bb{1}\xrightarrow{1} m_i,\nonumber\\
   &m_1\Box m_1\xrightarrow{e_2} \bb{1},~ m_2\Box m_2\xrightarrow{e_1} \bb{1}.
\end{align}
We interpret the nontrivial product 1-morphism as the physical process depicted in Fig.~\ref{fig:sigma}. Namely fusing the endpoints of two $m_1$ strings on the boundary results in an $e_2$ particle. Similarly, fusing endpoints~\footnote{In the rest of the work when we say "the endpoint/termination of a condensed string", we always mean the endpoint of a condensed string on a boundary where the string condenses.} of two $m_2$ strings produces an $e_1$. 

A subtlety regarding this interpretation is that we must choose a "gauge" for the termination of every type of condensed string. For instance, in Fig.\ref{fig:sigma}, if we redefine the termination of one of the two $m_1$ string by attaching a charge $e_2$ to it, then the fusion of endpoints would be trivial. After fixing a choice for the termination of the $m_1$-string, we can demand that the left $m_1$-string in Fig.~\ref{fig:sigma} is the "parallel transport" of the right one. In this case, even if we redefine the terminations of both $m_1$-strings in Fig~\ref{fig:sigma} by an $e_2$, the fusion of string endpoints still results in an $e_2$. Mathematically, different choices of  endpoints of condensed strings correspond to isomorphic but not identical condensable algebras, here we are working with a fixed condensable algebra, therefore a fixed choice of endpoints of condensed strings. 

\begin{figure}[h]
    \centering
    \includegraphics[width=0.8\linewidth]{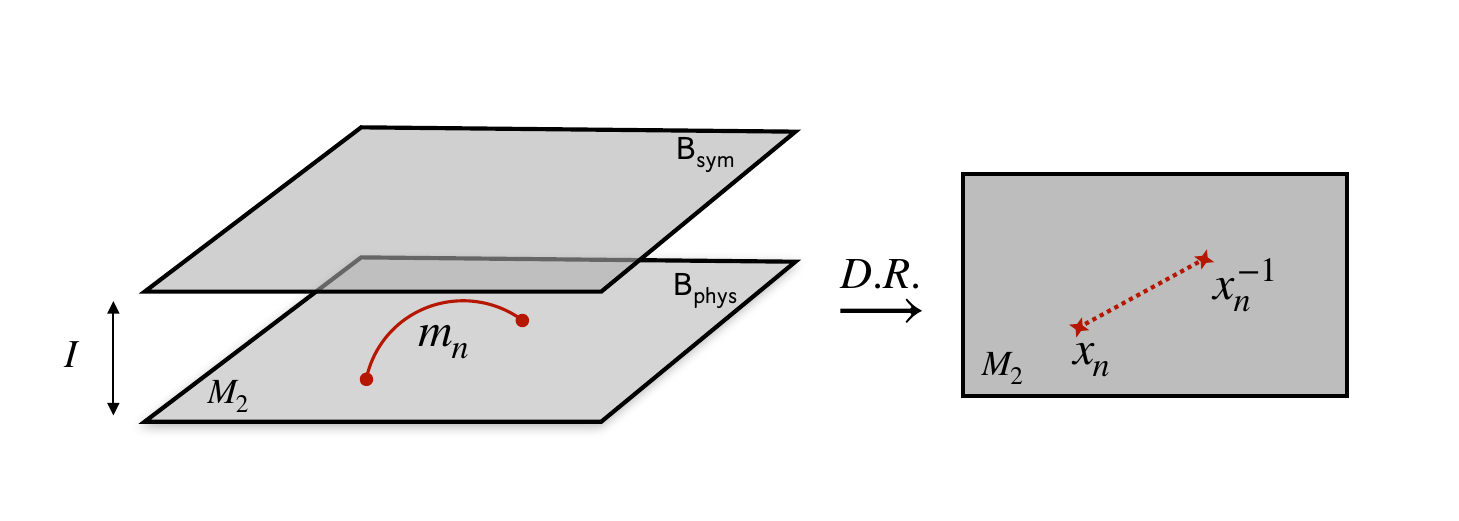}
    \caption{Sandwich construction for a pair of symmetry twist defect. On the left the physical boundary has $m_n$ condensed, therefore an arc of $m_n$ can end on it. Compactifying the vertical direction, this arc is mapped to a pair of symmetry twist defects of $n$ connected by a defect line(dashed red). }
    \label{fig:defect_line}
\end{figure}

The nontrivial fusion rule of the endpoints of condensed fluxes can be understood as the nontrivial fusion rule of symmetry twist defects of the corresponding 2+1D SPT. Namely an arc of string ending on the physical boundary of the sandwich may be viewed as the sandwich construction for a pair of symmetry twist defects, see Fig.~\ref{fig:defect_line}. It is known that in the type-II $\bZ_2^2$ SPT the symmetry twist defects fuse into symmetry charges. If we gauge the type-II SPT, the gauge fluxes have fusion rules $m_1\otimes m_1=e_2,~m_2\otimes m_2=e_1$~\cite{barkeshli2019symmetry}. The fusion rules of the string endpoints precisely capture this nontrivial fusion rule. This will be become even clearer if we perform gauging in the sandwich construction, which means we change the symmetry boundary to the all-flux-condensed one, with a trivial data $\alpha=1$. Then a vertical $m_1$ string can end on both boundaries and it corresponds to an $m_1$ gauge flux of the gauged 2+1D SPT. Then the string endpoints fusion rule associated with the bottom physical boundary implies the fusion rule $m_1^2=e_2$. In appendix~\ref{sec:coupled layer}, we also give a coupled layer construction for this gapped boundary and confirm the nontrivial fusion rules of string endpoints.

\begin{figure}[h]
    \centering
    \includegraphics[width=0.9\linewidth]{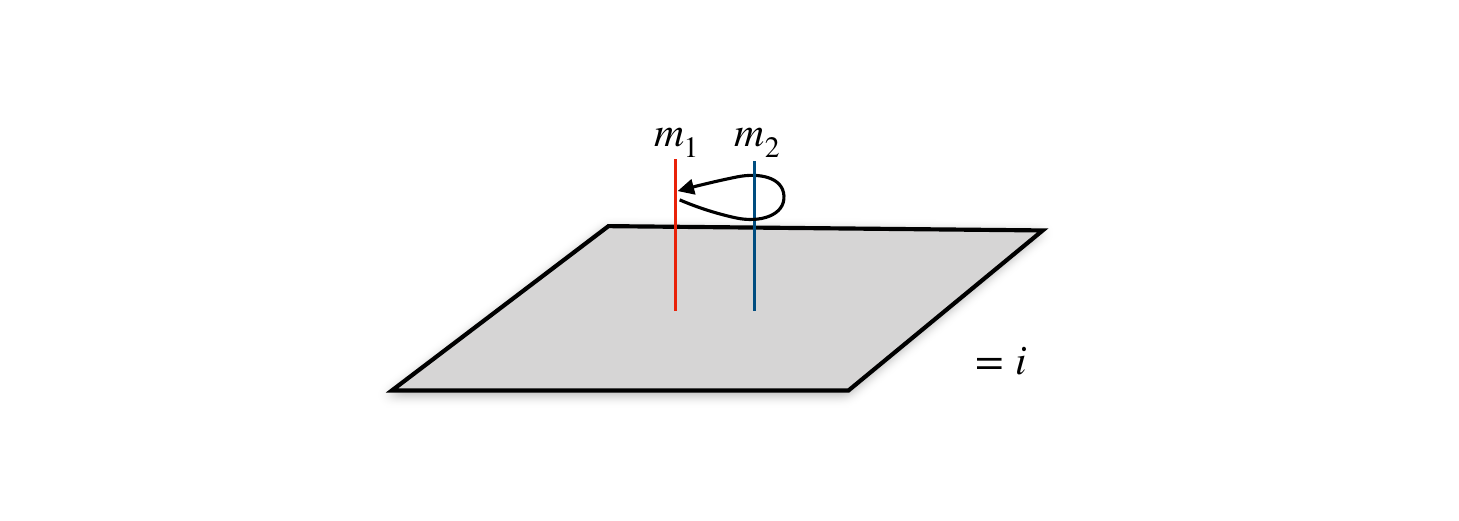}
    \vspace{-1cm}
    \caption{When the boundary condenses $m_1,m_2$ with a nontrivial $\sigma$ Eq.~\eqref{eq:sigma}, braiding the endpoints of $m_1$ and $m_2$ results in a phase $i$.}
    \label{fig:typeII3}
\end{figure}

Notice that when modifying the quasi-abelian cocycle by the coboundary $\epsilon$, we also obtain a nontrivial $\beta$ factor:
\begin{align}
    \beta(\vec{i},\vec{j})=\epsilon(\vec{i}|\vec{j})=i^{i_2j_1}.
\end{align}
This factor can be interpreted as the phase of braiding the endpoints of two strings, an observation ready made in~\cite{Zhao_2023}. In general $\beta(n_1,n_2)$ is the exchange phase of exchanging the endpoints of two strings with flux sector $n_1$ and $n_2$. In the current case we see that the double braiding of endpoints of an $m_1$ string and an $m_2$ string is $\beta(m_2,m_1)\beta(m_1,m_2)=i$. See Fig.~\ref{fig:typeII3}.
This is exactly the braiding of the corresponding gauge fluxes $m_1,m_2$ in the gauged SPT. ~\footnote{We note that $\epsilon=(-i)^{i_2k_1}$ is equally valid for trivializing $\phi$, which would result in $\beta(m_2,m_1)\beta(m_1,m_2)=-i$. These two choices differ by attaching an $e_1$ particle to the end point of $m_1$(or attaching an $e_2$ particle to the end point of $m_2$). In the gauged 2+1D SPT, this corresponds to relabelling $m_1$ by $m_1e_2$ or $m_2$ by $m_2e_1$.}
 
Finally, it is possible that the data $\phi,\sigma$ are both trivial yet there are more than one ways of condensing the same set of strings. To illustrate, we consider the string condensations dual to 2+1D type-I SPTs.
\subsubsection{Type I SPT}
A 2+1D type-I SPT has symmetry $\bZ_N$ and 3-cocycle below
\begin{align}
    \alpha(i,j,k)=\exp\left(\frac{2\pi i p}{N^2}i(j+k-[j+k]) \right)\label{eq:type-I cocycle}
\end{align}
Computing $\phi,\sigma$ according to Eq.\eqref{eq:N=G1},~Eq.\eqref{eq:N=G2}, we find $\phi=\sigma=1$. Therefore the condensed strings are pure fluxes $m_{i},~i=0,\cdots, N-1$, and there is no modification of fusion rule of termination of flux tubes.
However, the non-trivial $\alpha$ alone distinguishes the condensable algebra from the trivial one with $\alpha=1$. $\alpha$ determines the fusion $F$-symbol of the termination of fluxes on the boundary: if $x_i,y_j,z_k$ are symmetry twist defects created by open flux strings $m_i,m_j,m_k$, then  $\alpha(i,j,k): (x_i\otimes y_j)\otimes z_k\to x_i\otimes (y_j\otimes z_k)$.  

\begin{figure}[h]
    \centering
    \includegraphics[width=0.8\linewidth]{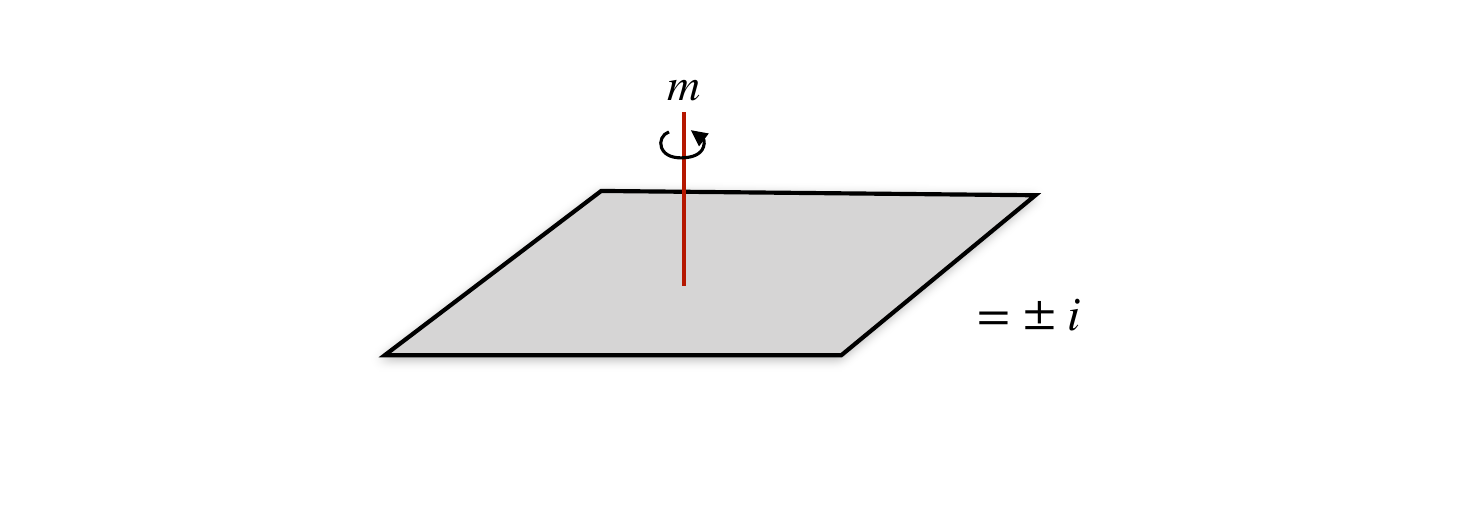}
    \vspace{-1cm}
    \caption{The twisted $m$-condensed boundary of 3d toric code. Twisting an $m$-string that ends the boundary by $2\pi$ produces a phase $\pm i$.}
    \label{fig:alpha}
\end{figure}

The simplest example is when the symmetry is $\bZ_2$, and the \symto is the 3d toric code. There are two ways of condensing the flux loop $m$, corresponding to the trivial and nontrivial choice of $\alpha$~\cite{Zhao_2023}. These two string condensations are dual to the trivial $\bZ_2$ paramagnet and the Levin-Gu phase in 2+1D respectively. The nontrivial $\alpha$ case is also called the twisted $m$ condensation. Since the $F$-symbols are subject to gauge equivalences, a perhaps more gauge-invariant way of distinguishing different condensations with different $\alpha$s is to look at the spin of the termination of the condensed strings. For instance, with a nontrivial $\alpha$ for the $m$ in the 3d toric code, the termination of $m$ on the boundary must also acquire semionic or anti-semionic statistics. This can be understood from the fact that the last two conditions of~Eq.\eqref{eq: cocycle condition} dictate that for fixed $\phi=\sigma=1$ and $\alpha$ a type-I cocyle, the phase $\beta(m,m)$ must be $\pm i$. Therefore twisting an $m$-string ending on the boundary by $2\pi$ results in a phase $\pm i$, as depicted in Fig.~\ref{fig:alpha}. 
\subsection{Non-Lagrangian condensation and 2+1D gSPTs}
For $N\neq G$, the condensable algebra $\eA[N,\alpha,\phi,\sigma, \beta]$ is non-Lagrangian, and is holographically dual to a 2+1D gSPT with symmetry extension structure
\begin{align}
    1\to N\to G\to K\to 1\label{eq:extension}
\end{align}
where $N$ is the gapped symmetry and $K$ is the IR symmetry. In this case the data $\phi,\sigma$ are no longer determined by $\alpha$, and we have a much richer structure compared to the Lagrangian case. It is worth mentioning that the classification of 1+1D and 0+1D gSPTs are in fact $H^2_{qa}[(G,N),\bC^\times]$ and $H^1_{qa}[(G,N),\bC^\times]$, therefore the classification of 2+1D gSPTs in terms of $H^3_{qa}[(G,N),\bC^\times]$ is a natural generalization. We discuss this point in more details in appendix~\ref{app:low_d_gSPT}.

Next we identify three classes of gSPTs by finding three classes of solutions to Eq.\eqref{eq: cocycle condition}. The first one has $\phi$ being nontrivial and is therefore fully determined by the set of strings condensed, this class is dual to the SPT-pump gSPTs studied in~\cite{Wen_2023}. The second one has $\sigma$ being nontrivial, the corresponding gSPT has not been systematically studied, but see~\cite{Wen_2023} appendix.B for an example. 

\subsubsection{Type-III gSPT}
A simple class of nontrivial solution to the cocycle condition~Eq.\eqref{eq: cocycle condition} is provided by $\alpha=\sigma=\beta=1$, and $\phi(-,-|-)\in Z^2[G,Z^1[N,U(1)]]$. The last three equations in~Eq.\eqref{eq: cocycle condition} demand 
\begin{align}
    \phi(g,n_1|n_2)=\phi(\cg{g}{n_1},g|n_2),~\phi(n_1,n_2|n_3)=1.
\end{align}
which can be satisfied by demanding that $\phi(g,h|n)=1$ whenever $g$ or $h$ is in $N$. This allows us to view $\phi\in Z^2[K,Z^1[N,U(1)]]$. As discussed in~\ref{sec:type-III SPT}, $S_{\phi[n]}$ is the SPT-string bound to the flux $m_n$ in the condensable algebra. The above condition says we can essentially view $\phi[n]$ as a 1+1D SPT of $K$. This class of condensable algebra condenses the strings $m_nS_{\phi[n]}$ and is fully determined by the set of condensed strings. Similar to the sandwich construction in~\ref{sec:z2z4example}, we see that the dual gSPT has the property that acting with a gapped symmetry $n\in N$ is equivalent to inserting a 1+1D SPT of $K$, $\phi[n]$. Such "SPT-pump" property has been observed in lattice model constructions in~\cite{Wen_2023}.  Due to the structural similarity with type-III gapped SPTs(they both have nontrivial $\phi$ factor), we call this class of gSPTs with nontrivial $\phi$ "type-III gSPTs". 

In \cite{Wen_2023}, it was also shown with lattice model construction that for SPT-pump gSPTs, the emergent anomaly of the IR symmetry has a simple expression as a cup product. Namely, the extension Eq.\eqref{eq:extension} is associated with an extension class $\mathsf{e}_2\in Z^2[K,N]$, therefore we can form the cup product $\phi\cup \mathsf{e}_2$:
\begin{align}
    \phi\cup \mathsf{e}_2(k_1,k_2,k_3,k_4):=\phi(k_1,k_2|\mathsf{e}_2(k_3,k_4)).~\label{eq: cup product}
\end{align}
which is a 4-cocycle of $K$. In Part II of this series of papers, we will prove that the dual string condensation indeed reduces the untwisted $G$-gauge theory to a twisted $K$-gauge theory with the twist given by E.q.\eqref{eq: cup product}. 

The $\bZ_2\times \bZ_4$ example we studied in~\ref{sec:z2z4example} belongs to this class, and we have indeed seen that the corresponding string condensation reduces the \symto to a twisted $\bZ_2\times \bZ_2$ theory. This example may be generalized. Consider $K=\bZ_{N_1}\times\bZ_{N_2}, N=\bZ_{N_{12}}$, with extension class $\mathsf{e}_2(\vec{i},\vec{j})=\frac{i_2+j_2-[i_2+j_2]}{N_2}$. $\mathsf{e}_2$ determines $G$ to be $\bZ_{N_1}\times \bZ_{N_2 N_{12}}$. Consider condensing $m_2^{N_{12}}S_{12}^{\text{GCD}(N_1,N_2N_{12})/N_{12}}$, where $S_{12}$ is the generating 1+1D SPT of $\bZ_{N_1}\times \bZ_{N_2N_{12}}$, which has order $\text{GCD}(N_1,N_2N_{12})$. The corresponding $\phi$ is 
\begin{align}
    \phi(\vec{i},\vec{j}|n)=\exp\left(\frac{2\pi i }{N_{12}}i_1j_2 n\right),~\vec{i},\vec{j}\in G, n\in N
\end{align}
The method of matching Cheshire charges of fluxes in Sec.~\ref{sec:z2z4example} can be straightforwardly applied to identify the twist of the post-condensation $K$ gauge theory, one finds
\begin{align}
    \pi^K_4=\exp\left(\frac{2\pi i}{N_{12}N_2}i_1j_2 \left(i_2+j_2-[i_2+j_2]\right)\right)
\end{align}
which is a nontrivial element of $H^4[K,U(1)]$. 
\subsubsection{Type-II gSPT}
We now consider another class of non-Lagrangian magnetic simple condensations that have nontrivial $\sigma$. We call this class of gSPTs type-II due to its structural similarity to type-II gapped SPTs.  Setting $\alpha=\phi=\beta=1$ in the cocycle conditions~Eq.\eqref{eq: cocycle condition}, we see that $\sigma(-|-,-)\in Z^1[G,Z^2[N,U(1)]]$ and $\sigma(g|n_1,n_2)=1$ if $g\in N$. This means the first variable of $\sigma$ can be viewed as $K=G/N$-valued and we can view $\sigma(-|n_1,n_2)$ as essentially an element of $Z^1[K,U(1)]$, i.e. a charge of $K$. As discussed in~\ref{sec:type-II SPT}, $\sigma$ determines the algebra product 1-morphism, which physically describes the fusion rules of the endpoints of condensed strings. To be concrete, we study a $G=D_4$ example. 

Denote $D_4=\<r,s|r^4=s^2=1,srs=r^3\>$, a central subgroup is $N=\<r^2\>\cong \bZ_2$. We denote the gauge charges by $e_{\pm,\pm}$ and $e_\chi$, where $e_{\pm,\pm}$ are the four 1d-irreps of $D_4$ where $r\to \pm$ and $s\to \pm$, and $e_\chi$ is the 2d irrep of $D_4$.  The quotient $K=D_4/N=\<[r],[s]|[r]^2=[s]^2=1,[s][r]=[r][s]\>\cong \bZ_2^2$. The identification of $D_4/N$ with $\bZ_2^2$ is $[rs]\sim (1,0),~ [s]\sim (0,1)$, and the extension class is $\mathsf{e}_2(\vec{i},\vec{j})=i_1j_2$. Consider $\sigma$ to be
\begin{align}
    \sigma(\vec{i}|n,m)=(-1)^{i_1nm},~\vec{i}\in K,~m,n\in N.\label{eq:D4_sigma}
\end{align}
This defines a non-trivial algebra product: $\mu: m_{r^2}\Box m_{r^2}\xrightarrow{e_{-,+}} \bb{1}$, which means the endpoints of two $m_{r^2}$ now fuse into a charge $e_{-,+}$. This means a $r^2$-symmetry twist defect in the gSPT must carry half of $e_{-,+}$. 

Notice that $\sigma$ satisfies the consistency conditions ~Eq.\eqref{eq: cocycle condition} for $G$ being any $\bZ_2$ extension of $\bZ_2^2$. The reason for considering the extension class $i_1j_2$ is that, with this choice, there is a nontrivial emergent anomaly of $K=\bZ_2^2$. This was shown via lattice model construction in~\cite{Wen_2023}, and the emergent anomaly is
\begin{align}
    \pi^K(\vec{i},\vec{j},\vec{k},\vec{l})=(-1)^{i_1j_1k_1l_2}\label{eq:anomaly_of_D4}
\end{align}

This example can be generalized to $K=\bZ_{N_1}\times \bZ_{N_2},~N=\bZ_{N_{12}}$, with $\mathsf{e}_2(\vec{i},\vec{j})=[i_1j_2]_{N_{12}}$. The extension class determines $G$ to be $(\bZ_{N_1}\times \bZ_{N_{12}})\rtimes \bZ_{N_2}$. $G$ can be written as  $\<r,s,t|r^{N_1}=s^{N_2}=t^{N_{12}}=1, rt=tr, st=ts,rs=tsr\>$, the central element $t$ generates the central subgroup $\<t\>\cong Z_{N_{12}}$. The string condensation data is $N=\<t\>$(i.e. $m_t$ is condensed) and 
\begin{align}
    \sigma(\vec{i}|n,m)=\exp\left(\frac{2\pi i}{N_1N_{12}} i_1\left(n+m-[n+m]_{N_{12}}\right)\right),~\vec{i}\in K,~n,m\in N.
\end{align}

The form of the $\sigma$ factor states that if $x_t$ is the symmetry defect of $t$, corresponding to the termination of $m_t$ string on the physical boundary, then $x_t^{N_{12}}=e_1$, where $e_1$ is a charge of $G$, corresponding to the 1d representation $r\sim e^{\frac{2\pi i}{N_1}},s\sim 1, t\sim 1$. This means each $x_t$ must carry a $e_1/N_{12}$ charge of $G$. 

Similar to the $D_4$ example, the above choice of $\sigma$ is valid for $G$ being any $\bZ_{N_{12}}$ extension of $\bZ_{N_1}\times \bZ_{N_2}$, the reason for considering the extension class $[i_1j_2]_{N_{12}}$ is that this choice leads to a nontrivial emergent anomaly of $K$. In other words, the condensation results in a twisted $\bZ_{N_1}\times \bZ_{N_2}$ gauge theory.  We leave the analysis of the post-condensation phase to a separate work~\cite{Wen_algebra}.
\subsubsection{Type-I gSPT}
Finally we introduce a class of gSPTs with only nontrivial $\alpha$. Similar to the case of type-I gapped SPTs, $\alpha\in Z^3[N,U(1)]$ stands for the fusion $F$-symbols for the $N$-symmetry twist defects of the gSPT. In other words, it describes the SPT-class of the gapped symmetry $N$. In order to have only nontrivial $\alpha$ while other data remain trivial, we take $N=\bZ_n$ and $\alpha$ to be the type-I cocycles Eq.~\eqref{eq:type-I cocycle}. Then the cocycle conditions Eq.~\eqref{eq: cocycle condition} can be satisfied by $\phi=\sigma=1$. We can take $G=\bZ_{k\cdot n}$, and $K=G/N=\bZ_k$. 

In this case the there is no emergent anomaly of $K=\bZ_k$, as $H^4[\bZ_k,U(1)]=1$. Therefore the gSPT is a weak one. We note that this gSPT can be viewed as a gapped SPT to symmetry breaking transition. Notice the $N$-cocycle $\alpha$ can be viewed as the restriction of a $G$-cocycle to $N$, i.e. there exists $\widetilde{\alpha}\in Z^3[G,U(1)]$, such that $\widetilde{\alpha}|_N=\alpha$. Therefore if we start with a gapped $G$-SPT with cocycle $\widetilde{\alpha}$, and tune to the critical point where $G$ is about to break to $N$, then at the critical point only $N$-charges are still gapped, and the $N$-symmetry twist defects would have the fusion $F$-symbol given by $\alpha$.  we see that type-I gSPTs are characterized by the nontrivial-SPT class of the gapped degrees of freedom, and they are SPT-to-SSB transitions.
\subsection{Gauging of 2+1D gSPT and pre-modular category\label{sec:gauging gSPT}}
Gauging symmetry is a powerful tool for analyzing properties of SPTs. It is therefore natural to expect that gauging gSPT would help us understand the properties of gSPTs, which is the focus of this subsection. For 2+1D gapped SPTs the gauging process results in a topological order, described by a non-degenerated braided fusion category, also known as a modular category.  We find that gauging 2+1D gSPTs leads to phases where the deconfined excitations form a \textit{degenerated} braided fusion category, also known as pre-modular category.

The topological holography framework provides us an extremely convenient tool for analyzing the effect of gauging, where gauging a symmetry is simply realized by changing the symmetry boundary. For the \symto $\mathcal{Z}[2\vc_G]$, corresponding to symmetry $G$, the standard symmetry boundary is the all-charge-condensed one. Gauging $G$ in the sandwich construction amounts to changing the symmetry boundary to the all-flux-condensed one. See~\cite{ji2024topological} for examples with $G=\bZ_2$. We study the gauging of the three classes of gSPTs discussed previously.
\subsubsection{Gauging of type-III gSPT}
To illustrate let us consider the $G=\bZ_2\times \bZ_4$ igSPT. The \symto is $\mathcal{Z}[2\vc_{\bZ_2\times \bZ_4}]$ and the corresponding string condensation is $1+m_2^2S_\phi$. Consider the sandwich construction for this igSPT, and change the top symmetry boundary to be the $\<m_1,m_2\>$-condensed boundary(with all data $\alpha,\sigma,\beta$ trivial). Notice that with this choice of symmetry boundary, the charges $e_1,e_2$ can no longer end on the symmetry boundary. Consequently, they no longer correspond to symmetry charges of $\bZ_2\times \bZ_4$ that can be created by local operators. They have become anyonic excitations that can only be created by string operators in pairs. Now a vertical string of $m_2^2S_{\phi}$ can end on the physical boundary. Notice that it can also end on the symmetry boundary: unlike flux loops, the defect string $S_{\phi}$ can be open. 

Now if we dimension reduce the sandwich, this vertical $m_2^2S_\phi$ string becomes a deconfined particle of a 2+1D system as well, which we denote as $\widetilde{m}$. It braids with the gauge charges: $\theta(\widetilde{m},e_2)=-1$, which is inherited from the string-particle braiding between $m_2^2S_\phi$ and $e_2$. However the fusion rules for $\widetilde{m}$ is non-abelian, due to the edge modes of the $S_\phi$ string at the symmetry boundary. Since fusing two projective representations of $\bZ_2\times \bZ_4$ gives a linear representation corresponding to $(1+e_1)(1+e_2^2)$, we have the fusion rule:
\begin{align}
    \widetilde{m}\otimes \widetilde{m}=(1+e_1)(1+e_2^2)
\end{align}
Therefore, if we consider the sandwich to be a 2+1D system, the deconfined excitations are generated by $e_1,e_2,\widetilde{m}$, which has a degenerated braiding since $e_1$ and $e_2^2$ are transparent. And the fusion rule for $\widetilde{m}$ is non-abelian. Other flux strings in the sandwich can not end on the physical boundary, therefore do not correspond to deconfined excitations of the 2+1D system. Thus we find that the deconfined excitations of the gauged 2+1D $\bZ_2\times \bZ_4$ igSPT form a pre-modular category. 

\subsubsection{Gauging of type-II gSPT}
Next we consider gauging of type-II gSPTs, where the symmetry twist defects carry fractional symmetry charges. As an example, we study gauging of the $D_4$-igSPT. The \symto sandwich for the gauged $D_4$-igSPT is obtained by condensing all fluxes of $D_4$ on the symmetry boundary, with all trivial data $\alpha=\phi=\sigma=\beta=1$, and condensing $m_{r^2}$ on the physical boundary, with nontrivial $\sigma$ given in~Eq.\eqref{eq:D4_sigma}. Now all the gauge charges in the bulk of the sandwich become nontrivial anyons of the dimension reduced 2+1D system, and they form the fusion category $\rep(D_4)$.  We denote the gauge charges by $e_{\pm,\pm}$ and $e_\chi$, where $e_{\pm,\pm}$ are the four 1d-irreps of $D_4$ where $r\to \pm$ and $s\to \pm$, and $e_\chi$ is the 2d irrep of $D_4$. A vertical string of $m_{r^2}$ can now end on both boundaries, which corresponds to an anyon of the dimension reduced 2+1D theory, denoted as $\widetilde{m}$. It braids nontrivially with the gauge charge $e_\chi$, $\theta(\widetilde{m},e_\chi)=-1$, which is inherited from the string-particle braiding between $m_{r^2}$ and $e_\chi$. Given the nontrivial fusion rules of symmetry twist defects of the $D_4$-igSPT, we have nontrivial fusion rules for the anyon $\widetilde{m}$: $\widetilde{m}\otimes \widetilde{m}=e_{-,+}$. 

Therefore the deconfined excitations of the gauged $D_4$-igSPT form a fusion category that contains $\rep(D_4)$ as a sub-category, and contains another object $\widetilde{m}$ with fusion rule $\widetilde{m}\otimes \widetilde{m}=e_{-,+}$. The only nontrivial braiding is $\theta(\widetilde{m},e_\chi)=-1$. It is a pre-modular category, as $e_{\pm,\pm}$ are transparent. 

\subsubsection{Gauging of type-I gSPT}
Finally we consider gauging of type-I gSPTs. Let us consider the example where $G=\bZ_4,~N=\bZ_2,~K=\bZ_2$ and $\alpha$ is the nontrivial element of $H^3[N,U(1)]=\bZ_2$. In this case the strings $m^2$ are condensed on the physical boundary, and the termination of $m^2$ on the physical boundary has nontrivial fusion $F$-symbol. It is known that for an order-2 anyon with nontrivial fusion $F$-symbol, it must has semionic or anti-semionic self statistics. Now we gauge the $\bZ_4$, corresponding to a symmetry boundary having $m$-condensed with a trivial $\alpha$. Then a vertical string of $m^2$ can end on both boundaries, and when viewed as an anyon of a 2+1D system, it has semionic statistics. Therefore, we see that the sandwich, viewed as a 2+1D system, hosts anyons $e,m^2$, with a group-like $\bZ_4\times \bZ_2$ fusion rule, mutual statistics $\theta(e,m^2)=-1$, and self statistics $\theta(m^2)=i$. This is again a pre-modular theory, as the charge $e^2$ is transparent. 

We notice that this theory can be embedded into a modular theory. Namely consider a twisted $\bZ_4$ gauge theory in 2+1D, which has charges $e',m'$, with mutual statistics $\theta(e',m')=i$, and self-statistics $\theta(m')=\exp\left(\frac{i\pi}{8}\right)$. Then we may identify $e$ with $e'$ and $m^2$ with $m'^2$. In general, if a gSPT is the transition between an SPT and a (partial)SSB order, then the gauged gSPT can be embedded into the gauged SPT. We believe that a characteristic of 2+1D igSPT is that the gauged igSPT(as a pre-modular theory) can not be embedded into a twisted $G$-gauge theory.  
\section{A general theory of 2+1D gSPTs and their gauging\label{sec: general theory}}
In this subsection we discuss some generalities of  2+1D gSPTs and their gauging. The goal is to argue that a 2+1D gSPT is characterized by a $G$-crossed braided fusion category structure on $\vc_N$-- the category of the gapped symmetry.

Recall that in a 2+1D symmetry enriched gapped phase, we can fully characterize the topological properties of the phase by studying the fusion and braiding of anyons as well as the symmetry defects. Although the symmetry defects are extrinsic defects that are not quantum fluctuating, one can still define their fusion and braiding with each other as well as with the anyons in the system. This structure is captured by a $G$-crossed braided fusion category~\cite{barkeshli2019symmetry}, $\cc{C}=\oplus_g\cc{C}_g$, with the trivial degree component $\cc{C}_e$ describing the underlying topological order of the 2+1D system.  The grading of the category $\cc{C}$ is faithful, since the braiding and fusion of symmetry defects of any value $g\in G$ are well-defined in a fully gapped system.

In the SPT case, the trivial degree component $\cc{C}_e=\vc$ is trivial.  In this case the category $\cc{C}$ reduces to $\vc_G$, describing only the symmetry twist defects. The fusion and braiding structure on $\vc_G$ turn out to be determined by the associator $\alpha\in Z^3[G,U(1)]$ alone, and equivalent classes of braided fusion structures on $\vc_G$ turn out to be classified by $H^3[G,U(1)]$. This reproduces the cohomology classification of 2+1D SPTs~\cite{barkeshli2019symmetry}.

Now consider a gapless SPT with gapped symmetry $N\lhd G$. Naturally we want to characterize its property by the braiding and fusion properties of the symmetry twist defects. However, in this case only the braiding and fusion of symmetry defects valued in the gapped symmetry $N$ are well-defined. For instance, even slowly braiding a symmetry defect with value $g\notin N$ could excite gapless symmetry charges with nontrivial $g$-action and render the braiding outcome non-topological. Therefore the only topologically robust properties of the symmetry twist defects are the braiding and fusion of $N$-symmetry twist defects, which form a fusion category $\vc_N^\alpha$. The $G$ symmetry action on the twist defects equips $\vc_N^\alpha$ with the structure of a $G$-crossed category, and the braiding and fusion must be compatible with the $G$-action in obvious ways. We come to the conclusion that the topological properties of the symmetry defects in a gSPT are captured by a $G$-crossed braided fusion category structure on $\vc_N$. We believe that a 2+1D gSPT is completely determined by this $G$-crossed braided fusion category, that is,
\begin{tcolorbox}
    A 2+1D gSPT with full symmetry $G$ and gapped symmetry $N\lhd G$ is fully characterized by a $G$-crossed braided fusion category whose underlying $G$-graded category is $\vc_N$.
\end{tcolorbox}
This then matches with the classification of magnetic simple condensable algebras in 3+1D $G$-gauge theory straightforwardly: as we mentioned in Sec.~\ref{sec:classification}, a magnetic simple condensable algebra in $\cc{Z}[2\vc_G]$ is just a $G$-crossed braided fusion category with the underlying $G$-graded category being $\vc_N$. The classifying data $\alpha,\phi,\sigma,\beta$ are exactly what define the $G$-crossed fusion and braiding structures on $\vc_N$.

Recall that gauging $G$ in a 2+1D $G$-enriched gapped phase is mathematically described by equivariantization of the corresponding $G$-crossed braided fusion category. In the SPT case, the equivariantization of the $G$-crossed braided fusion category $\vc_G^\alpha$ results in $\cc{Z}[\vc_G^\alpha]$, i.e. the 2+1D twisted $G$-gauge theory, which is exactly the gauged SPT. Physically, gauging amounts to promoting the symmetry defects to 
quantum fluctuating gauge fluxes, and taking into account the braiding and fusion with gauge charges.  The physics of gauging is the same for gSPTs, except that only the gauge charges with non-trivial $N$-action are gapped. Therefore, the braiding and fusion involving gauge fluxes not valued in $N$ are not well-defined(similar to how the braiding and fusion of $N$-symmetry defects are not well-defined prior to gauging). Therefore, the topologically robust properties of the gauged SPT only involve the braiding and fusion of all $G$-gauge charges and $N$-gauge fluxes, which is clearly a degenerated braided fusion category, since some gauge fluxes are "missing". Indeed, the equivariantization of the $G$-crossed braided fusion category $\eA[N,\alpha,\phi,\sigma,\beta]$ results in a degenerated braided fusion category whenever $N\neq G$~\cite{naidu2011crossed}. It also seems that a characteristic of igSPT is that the corresponding gauged igSPT can not be embedded into any gauged $G$-SPT. We leave the exploration of this to future works.

Just like for gapped SPTs, gauging is powerful for analyzing properties of gSPTs. Here we present an argument for unusual boundary properties of gSPTs based on gauging. Recall that the boundary anomaly of gapped SPTs can be argued from unusual properties of gauge fluxes in the gauged SPT. For instance, in the gauged type-I SPT, the flux is a semion, therefore can not condense at a boundary. In fact, in any 2+1D twisted $G$ gauge theory, there is not a boundary where all gauge fluxes condense. This is the reflection of the absence of a symmetric gapped boundary of the SPT. Had we had a symmetric gapped boundary for the SPT, gauging the SPT would result in a gapped boundary where all fluxes condense.  

Similarly, the unusual symmetry and fusion properties of the gauge fluxes in the gauged gSPT imply absence of an $N$-gapped and symmetric boundary. As an example, consider the type-II $D_4$-gSPT. We know that everywhere in the bulk of the gSPT, $N$-charges are gapped. Assume that $N$-charges are also gapped at a boundary, then after gauging $G$, the boundary should condense all $N$-fluxes. Since $\widetilde{m}^2=e_{-,+}$, this means $e_{-,+}$ also condenses, which corresponds to breaking $D_4$ to $\bZ_2\times \bZ_2$ on the boundary. Therefore there is no boundary of the $D_4$-gSPT that is $D_4$-symmetric and $N=\bZ_2^{gap}$-gapped. In other words, although $N$-charges are absent in the bulk in the IR, they appear on the boundary as gapless modes.
\section{Discussion and outlook\label{sec:outlook}}
In this work we studied string condensations in 3+1D topological orders and their relations to 2+1D phases with symmetry. We find that a class of string condensations of the 3+1D $G$-gauge theory, which we call magnetic and simple, are dual to 2+1D gapped and gapless SPTs, via the topological holography framework. We systematically classified this class of string condensations and studied their physical properties. Through the study of string condensations dual to 2+1D gapped SPTs, we were able to identify the physical meaning of the classifying data. We found that the structure of a condensable algebra determines the properties of the termination of open strings on the physical boundary, which are dual to the properties of symmetry twist defects of the 2+1D system.  We identified three families of non-Lagrangian magnetic simple string condensations, dual to three families of gSPTs. We studied the physical structure of these three families of gSPT and discussed their gauging. We found that gauging gSPTs results in pre-modular categories, which describe the braiding and fusion of deconfined excitations of the gSPT. We argued that general 2+1D $G$-gSPTs are described $G$-crossed braided fusion category structures on $\vc_N$, the category of gapped symmetry defects, which completes the \symto/gSPT correspondence in 3+1D/2+1D. Gauging gSPT is then described by equivariantization of the $G$-crossed category, which results in a pre-modular categort whenever $N\neq G$.


There are many future directions worth exploring and we mention a few here. 

 \paragraph{String condensation in 3+1D topological orders with fermion charges} We have been focusing on string condensations in $\mathcal{Z}[2\vc_G]$, a theory where the point charges are all bosons. It is known that all 3+1D topological orders are gauged SPTs~\cite{Lan_2018,Lan_2019,Johnson_Freyd_2022}, and $\mathcal{Z}[2\vc_G^\pi]$ are gauged bosonic SPTs. Another class of 3+1D topological order is gauged fermionic SPTs, where some gauge charges are fermionic. It is of great importance to the theory of string condensation and topological holography to have a systematically understanding of string condensations in these gauged fermionic SPTs. For instance, the \symto for an anomalous fermionic symmetry $G^f$ is exactly a gauged fermionic SPT, with a fermionic symmetry boundary in the sandwich construction. Unlike the \symto for 1+1D phases, the 3+1D \symto for a 2+1D fermionic symmetry is  different from the \symto for any 2+1D bosonic symmetry. 
Recently it was pointed out defect strings in these gauged fermionic SPTs, including the Kitaev chain defect, should play a role in the theory of string condensation and topological holography for 2+1D fermionic phases~\cite{ji2024topological}. In~\cite{ji2024topological} the authors argue that the string condensation dual to the 2+1D $p+ip$ superconductor condenses $mS$, where $m$ is the gauge flux, and $S$ is the Kitaev chain defect in the 3+1D toric code with a fermionic charge. However, there should be in total 16 string condensations in the fermionic toric code, dual to the 16 topological superconductors in 2+1D.

 We suspect that there are extra structures needed to define the condensation of flux in the fermionic toric code, similar to the data $\sigma,\alpha,\beta$ studied in this paper. Physically, the fusion of the flux endpoints  may be nontrivial, and the flux endpoints may have nontrivial spin. This gives us 16 ways of condensing fluxes by inspection. We intent to investigate string condensations in 3+1D topological orders with fermionic charges systematically in the future.

\paragraph{non-simple string condensations} In this work we focused on simple string condensations, which are dual to 2+1D phases with no topological order. It would be interesting to study string condensations corresponding to topologically ordered states. For Lagrangian condensations it has already been established that a Lagrangian magnetic condensation in $\mathcal{Z}[2\vc_G]$ is given by non-degenerated $G$-crossed braided fusion category~\cite{D_coppet_2024}, which corresponds to 2+1D symmetry enriched topological orders(SET). A general Lagrangian condensations in $\mathcal{Z}[2\vc_G]$ is determined by a subgroup $H<G$, and an 2+1D $H$-SET, which matches with the classification of gapped $G$-phases in 2+1D. Thus the remaining case to be studied is where the string condensation is non-Lagrangian and non-simple, which should correspond to a 2+1D gapless state whose gapped excitations are anyonic. For instance we can stack a 2+1D gSPT with a 2+1D SET to obtain such a state. Such a state will have different ground state degeneracies on different manifolds. The corresponding string condensation is given by a $G$-crossed braided fusion category that is degenerated, and whose trivial degree component is a nontrivial braided fusion category. It would be interesting to identify examples that are not simple stacking of gSPTs and SETs, and study the properties of them from either a 2+1D point of view or a \symto point of view. It appears the deconfined excitations in such a gapless state do not need to form a modular category, instead a pre-modular category would suffice. This indicates there is a very rich structure for gapless \textit{and} topologically ordered states. 

\paragraph{\symto/gSPT correspondence for continuous symmetries.}  Topological holography for continuous symmetries have attracted much attention recently~\cite{antinucci2024holographic,apruzzi2024symth,brennan2024symtft,antinucci2024anomalies}, a picture where any symmetry has a holographic description starts to emerge from this series of works. A new ingredient in the topological holography for continuous symmetries is a new type of topological field theory with gauge
fields valued in $U(1)$ or  $\bb{R}$ that are beyond the standard TQFTs framework.  A theory of condensation in such TQFTs is needed in order to have a complete topological holography dictionary for continuous symmetries, which is currently nonexistent. 

 \paragraph{Lattice and field theory model for 2+1D gSPTs} In this work we studied the topological properties of 2+1D gSPTs through the lens of topological holography. However we have not given a lattice or field theory account. For instance, the $\bZ_2\times \bZ_4$-igSPT has an anomalous $\bZ_2\times \bZ_2$ IR symmetry, but there are many gapless theories that can match this anomaly. For instance the deconfined quantum critical point between different $\bZ_2$ symmetry breaking phases~\cite{Ji_2023}, or gapless boundary states of 3+1D twisted $\bZ_2^2$ gauge theory studied in~\cite{Chen_2016}. It would be interesting to construct lattice or field theory models of 2+1D gSPTs and study the topological as well as dynamical properties, and potential interplay between the two.

\paragraph*{Acknowledgement}
We thank Drew Potter, Weicheng Ye, Joseph Sullivan, Fiona Burnell, Wilbur Shirley, Yu-An Chen, Nathanan Tantivasadakarn, Ryohei Kobayashi  for insightful discussions. We thank Thibault D. Décoppet for valuable feedback on a draft of the paper. We thank the Perimeter Institute for hosting the workshop “Higher Categorical Tools for Quantum Phases of Matter,” where part of this work was completed. We thank the authors of~\cite{bhardwaj2024gapped} for coordinating submission of their related independent work. This work was supported by NSERC and the European Commission under the Grant Foundations of Quantum Computational Advantage.
\bibliography{v5}

\begin{thebibliography}{129}%
\makeatletter
\providecommand \@ifxundefined [1]{%
 \@ifx{#1\undefined}
}%
\providecommand \@ifnum [1]{%
 \ifnum #1\expandafter \@firstoftwo
 \else \expandafter \@secondoftwo
 \fi
}%
\providecommand \@ifx [1]{%
 \ifx #1\expandafter \@firstoftwo
 \else \expandafter \@secondoftwo
 \fi
}%
\providecommand \natexlab [1]{#1}%
\providecommand \enquote  [1]{``#1''}%
\providecommand \bibnamefont  [1]{#1}%
\providecommand \bibfnamefont [1]{#1}%
\providecommand \citenamefont [1]{#1}%
\providecommand \href@noop [0]{\@secondoftwo}%
\providecommand \href [0]{\begingroup \@sanitize@url \@href}%
\providecommand \@href[1]{\@@startlink{#1}\@@href}%
\providecommand \@@href[1]{\endgroup#1\@@endlink}%
\providecommand \@sanitize@url [0]{\catcode `\\12\catcode `\$12\catcode
  `\&12\catcode `\#12\catcode `\^12\catcode `\_12\catcode `\%12\relax}%
\providecommand \@@startlink[1]{}%
\providecommand \@@endlink[0]{}%
\providecommand \url  [0]{\begingroup\@sanitize@url \@url }%
\providecommand \@url [1]{\endgroup\@href {#1}{\urlprefix }}%
\providecommand \urlprefix  [0]{URL }%
\providecommand \Eprint [0]{\href }%
\providecommand \doibase [0]{https://doi.org/}%
\providecommand \selectlanguage [0]{\@gobble}%
\providecommand \bibinfo  [0]{\@secondoftwo}%
\providecommand \bibfield  [0]{\@secondoftwo}%
\providecommand \translation [1]{[#1]}%
\providecommand \BibitemOpen [0]{}%
\providecommand \bibitemStop [0]{}%
\providecommand \bibitemNoStop [0]{.\EOS\space}%
\providecommand \EOS [0]{\spacefactor3000\relax}%
\providecommand \BibitemShut  [1]{\csname bibitem#1\endcsname}%
\let\auto@bib@innerbib\@empty
\bibitem [{\citenamefont {Chen}\ \emph {et~al.}(2013)\citenamefont {Chen},
  \citenamefont {Gu}, \citenamefont {Liu},\ and\ \citenamefont
  {Wen}}]{chen2013symmetry}%
  \BibitemOpen
  \bibfield  {author} {\bibinfo {author} {\bibfnamefont {X.}~\bibnamefont
  {Chen}}, \bibinfo {author} {\bibfnamefont {Z.-C.}\ \bibnamefont {Gu}},
  \bibinfo {author} {\bibfnamefont {Z.-X.}\ \bibnamefont {Liu}},\ and\ \bibinfo
  {author} {\bibfnamefont {X.-G.}\ \bibnamefont {Wen}},\ }\bibfield  {title}
  {\bibinfo {title} {Symmetry protected topological orders and the group
  cohomology of their symmetry group},\ }\href@noop {} {\bibfield  {journal}
  {\bibinfo  {journal} {Physical Review B—Condensed Matter and Materials
  Physics}\ }\textbf {\bibinfo {volume} {87}},\ \bibinfo {pages} {155114}
  (\bibinfo {year} {2013})}\BibitemShut {NoStop}%
\bibitem [{\citenamefont {Chen}\ \emph {et~al.}(2011)\citenamefont {Chen},
  \citenamefont {Liu},\ and\ \citenamefont {Wen}}]{chen2011two}%
  \BibitemOpen
  \bibfield  {author} {\bibinfo {author} {\bibfnamefont {X.}~\bibnamefont
  {Chen}}, \bibinfo {author} {\bibfnamefont {Z.-X.}\ \bibnamefont {Liu}},\ and\
  \bibinfo {author} {\bibfnamefont {X.-G.}\ \bibnamefont {Wen}},\ }\bibfield
  {title} {\bibinfo {title} {Two-dimensional symmetry-protected topological
  orders and their protected gapless edge excitations},\ }\href@noop {}
  {\bibfield  {journal} {\bibinfo  {journal} {Physical Review B—Condensed
  Matter and Materials Physics}\ }\textbf {\bibinfo {volume} {84}},\ \bibinfo
  {pages} {235141} (\bibinfo {year} {2011})}\BibitemShut {NoStop}%
\bibitem [{\citenamefont {Wen}(2017)}]{Wen_2017}%
  \BibitemOpen
  \bibfield  {author} {\bibinfo {author} {\bibfnamefont {X.-G.}\ \bibnamefont
  {Wen}},\ }\bibfield  {title} {\bibinfo {title} {Colloquium : Zoo of
  quantum-topological phases of matter},\ }\bibfield  {journal} {\bibinfo
  {journal} {Reviews of Modern Physics}\ }\textbf {\bibinfo {volume} {89}},\
  \href {https://doi.org/10.1103/revmodphys.89.041004}
  {10.1103/revmodphys.89.041004} (\bibinfo {year} {2017})\BibitemShut {NoStop}%
\bibitem [{\citenamefont {Johnson-Freyd}(2022)}]{Johnson_Freyd_2022}%
  \BibitemOpen
  \bibfield  {author} {\bibinfo {author} {\bibfnamefont {T.}~\bibnamefont
  {Johnson-Freyd}},\ }\bibfield  {title} {\bibinfo {title} {On the
  classification of topological orders},\ }\href
  {https://doi.org/10.1007/s00220-022-04380-3} {\bibfield  {journal} {\bibinfo
  {journal} {Communications in Mathematical Physics}\ }\textbf {\bibinfo
  {volume} {393}},\ \bibinfo {pages} {989–1033} (\bibinfo {year}
  {2022})}\BibitemShut {NoStop}%
\bibitem [{\citenamefont {Lan}\ \emph {et~al.}(2016{\natexlab{a}})\citenamefont
  {Lan}, \citenamefont {Kong},\ and\ \citenamefont {Wen}}]{Lan_2016}%
  \BibitemOpen
  \bibfield  {author} {\bibinfo {author} {\bibfnamefont {T.}~\bibnamefont
  {Lan}}, \bibinfo {author} {\bibfnamefont {L.}~\bibnamefont {Kong}},\ and\
  \bibinfo {author} {\bibfnamefont {X.-G.}\ \bibnamefont {Wen}},\ }\bibfield
  {title} {\bibinfo {title} {Theory of (2+1)-dimensional fermionic topological
  orders and fermionic/bosonic topological orders with symmetries},\ }\bibfield
   {journal} {\bibinfo  {journal} {Physical Review B}\ }\textbf {\bibinfo
  {volume} {94}},\ \href {https://doi.org/10.1103/physrevb.94.155113}
  {10.1103/physrevb.94.155113} (\bibinfo {year}
  {2016}{\natexlab{a}})\BibitemShut {NoStop}%
\bibitem [{\citenamefont {Lan}\ \emph {et~al.}(2018)\citenamefont {Lan},
  \citenamefont {Kong},\ and\ \citenamefont {Wen}}]{Lan_2018}%
  \BibitemOpen
  \bibfield  {author} {\bibinfo {author} {\bibfnamefont {T.}~\bibnamefont
  {Lan}}, \bibinfo {author} {\bibfnamefont {L.}~\bibnamefont {Kong}},\ and\
  \bibinfo {author} {\bibfnamefont {X.-G.}\ \bibnamefont {Wen}},\ }\bibfield
  {title} {\bibinfo {title} {Classification of 3+1d bosonic topological orders:
  The case when pointlike excitations are all bosons},\ }\bibfield  {journal}
  {\bibinfo  {journal} {Physical Review X}\ }\textbf {\bibinfo {volume} {8}},\
  \href {https://doi.org/10.1103/physrevx.8.021074} {10.1103/physrevx.8.021074}
  (\bibinfo {year} {2018})\BibitemShut {NoStop}%
\bibitem [{\citenamefont {Lan}\ and\ \citenamefont {Wen}(2019)}]{Lan_2019}%
  \BibitemOpen
  \bibfield  {author} {\bibinfo {author} {\bibfnamefont {T.}~\bibnamefont
  {Lan}}\ and\ \bibinfo {author} {\bibfnamefont {X.-G.}\ \bibnamefont {Wen}},\
  }\bibfield  {title} {\bibinfo {title} {Classification of 3+1d bosonic
  topological orders (ii): The case when some pointlike excitations are
  fermions},\ }\bibfield  {journal} {\bibinfo  {journal} {Physical Review X}\
  }\textbf {\bibinfo {volume} {9}},\ \href
  {https://doi.org/10.1103/physrevx.9.021005} {10.1103/physrevx.9.021005}
  (\bibinfo {year} {2019})\BibitemShut {NoStop}%
\bibitem [{\citenamefont {Kong}\ \emph
  {et~al.}(2020{\natexlab{a}})\citenamefont {Kong}, \citenamefont {Lan},
  \citenamefont {Wen}, \citenamefont {Zhang},\ and\ \citenamefont
  {Zheng}}]{Kong_2020}%
  \BibitemOpen
  \bibfield  {author} {\bibinfo {author} {\bibfnamefont {L.}~\bibnamefont
  {Kong}}, \bibinfo {author} {\bibfnamefont {T.}~\bibnamefont {Lan}}, \bibinfo
  {author} {\bibfnamefont {X.-G.}\ \bibnamefont {Wen}}, \bibinfo {author}
  {\bibfnamefont {Z.-H.}\ \bibnamefont {Zhang}},\ and\ \bibinfo {author}
  {\bibfnamefont {H.}~\bibnamefont {Zheng}},\ }\bibfield  {title} {\bibinfo
  {title} {Classification of topological phases with finite internal symmetries
  in all dimensions},\ }\bibfield  {journal} {\bibinfo  {journal} {Journal of
  High Energy Physics}\ }\textbf {\bibinfo {volume} {2020}},\ \href
  {https://doi.org/10.1007/jhep09(2020)093} {10.1007/jhep09(2020)093} (\bibinfo
  {year} {2020}{\natexlab{a}})\BibitemShut {NoStop}%
\bibitem [{\citenamefont {Kong}\ \emph {et~al.}(2015)\citenamefont {Kong},
  \citenamefont {Wen},\ and\ \citenamefont
  {Zheng}}]{kong2015boundarybulkrelation}%
  \BibitemOpen
  \bibfield  {author} {\bibinfo {author} {\bibfnamefont {L.}~\bibnamefont
  {Kong}}, \bibinfo {author} {\bibfnamefont {X.-G.}\ \bibnamefont {Wen}},\ and\
  \bibinfo {author} {\bibfnamefont {H.}~\bibnamefont {Zheng}},\ }\href
  {https://arxiv.org/abs/1502.01690} {\bibinfo {title} {Boundary-bulk relation
  for topological orders as the functor mapping higher categories to their
  centers}} (\bibinfo {year} {2015}),\ \Eprint
  {https://arxiv.org/abs/1502.01690} {arXiv:1502.01690 [cond-mat.str-el]}
  \BibitemShut {NoStop}%
\bibitem [{\citenamefont {Kong}\ \emph {et~al.}(2017)\citenamefont {Kong},
  \citenamefont {Wen},\ and\ \citenamefont {Zheng}}]{Kong_2017}%
  \BibitemOpen
  \bibfield  {author} {\bibinfo {author} {\bibfnamefont {L.}~\bibnamefont
  {Kong}}, \bibinfo {author} {\bibfnamefont {X.-G.}\ \bibnamefont {Wen}},\ and\
  \bibinfo {author} {\bibfnamefont {H.}~\bibnamefont {Zheng}},\ }\bibfield
  {title} {\bibinfo {title} {Boundary-bulk relation in topological orders},\
  }\href {https://doi.org/10.1016/j.nuclphysb.2017.06.023} {\bibfield
  {journal} {\bibinfo  {journal} {Nuclear Physics B}\ }\textbf {\bibinfo
  {volume} {922}},\ \bibinfo {pages} {62–76} (\bibinfo {year}
  {2017})}\BibitemShut {NoStop}%
\bibitem [{\citenamefont {Kong}\ \emph
  {et~al.}(2020{\natexlab{b}})\citenamefont {Kong}, \citenamefont {Lan},
  \citenamefont {Wen}, \citenamefont {Zhang},\ and\ \citenamefont
  {Zheng}}]{kong2020algebraic}%
  \BibitemOpen
  \bibfield  {author} {\bibinfo {author} {\bibfnamefont {L.}~\bibnamefont
  {Kong}}, \bibinfo {author} {\bibfnamefont {T.}~\bibnamefont {Lan}}, \bibinfo
  {author} {\bibfnamefont {X.-G.}\ \bibnamefont {Wen}}, \bibinfo {author}
  {\bibfnamefont {Z.-H.}\ \bibnamefont {Zhang}},\ and\ \bibinfo {author}
  {\bibfnamefont {H.}~\bibnamefont {Zheng}},\ }\bibfield  {title} {\bibinfo
  {title} {Algebraic higher symmetry and categorical symmetry: A holographic
  and entanglement view of symmetry},\ }\href@noop {} {\bibfield  {journal}
  {\bibinfo  {journal} {Physical Review Research}\ }\textbf {\bibinfo {volume}
  {2}},\ \bibinfo {pages} {043086} (\bibinfo {year}
  {2020}{\natexlab{b}})}\BibitemShut {NoStop}%
\bibitem [{\citenamefont {Ji}\ and\ \citenamefont
  {Wen}(2020)}]{ji2020categorical}%
  \BibitemOpen
  \bibfield  {author} {\bibinfo {author} {\bibfnamefont {W.}~\bibnamefont
  {Ji}}\ and\ \bibinfo {author} {\bibfnamefont {X.-G.}\ \bibnamefont {Wen}},\
  }\bibfield  {title} {\bibinfo {title} {Categorical symmetry and noninvertible
  anomaly in symmetry-breaking and topological phase transitions},\ }\href@noop
  {} {\bibfield  {journal} {\bibinfo  {journal} {Physical Review Research}\
  }\textbf {\bibinfo {volume} {2}},\ \bibinfo {pages} {033417} (\bibinfo {year}
  {2020})}\BibitemShut {NoStop}%
\bibitem [{\citenamefont {Ji}\ and\ \citenamefont {Wen}(2021)}]{ji2021unified}%
  \BibitemOpen
  \bibfield  {author} {\bibinfo {author} {\bibfnamefont {W.}~\bibnamefont
  {Ji}}\ and\ \bibinfo {author} {\bibfnamefont {X.-G.}\ \bibnamefont {Wen}},\
  }\bibfield  {title} {\bibinfo {title} {A unified view on symmetry, anomalous
  symmetry and non-invertible gravitational anomaly},\ }\href@noop {}
  {\bibfield  {journal} {\bibinfo  {journal} {arXiv preprint arXiv:2106.02069}\
  } (\bibinfo {year} {2021})}\BibitemShut {NoStop}%
\bibitem [{\citenamefont {Barkeshli}\ \emph {et~al.}(2022)\citenamefont
  {Barkeshli}, \citenamefont {Chen}, \citenamefont {Hsin},\ and\ \citenamefont
  {Manjunath}}]{barkeshli2022classification}%
  \BibitemOpen
  \bibfield  {author} {\bibinfo {author} {\bibfnamefont {M.}~\bibnamefont
  {Barkeshli}}, \bibinfo {author} {\bibfnamefont {Y.-A.}\ \bibnamefont {Chen}},
  \bibinfo {author} {\bibfnamefont {P.-S.}\ \bibnamefont {Hsin}},\ and\
  \bibinfo {author} {\bibfnamefont {N.}~\bibnamefont {Manjunath}},\ }\bibfield
  {title} {\bibinfo {title} {Classification of (2+ 1) d invertible fermionic
  topological phases with symmetry},\ }\href@noop {} {\bibfield  {journal}
  {\bibinfo  {journal} {Physical Review B}\ }\textbf {\bibinfo {volume}
  {105}},\ \bibinfo {pages} {235143} (\bibinfo {year} {2022})}\BibitemShut
  {NoStop}%
\bibitem [{\citenamefont {Bulmash}\ and\ \citenamefont
  {Barkeshli}(2020)}]{bulmash2020absolute}%
  \BibitemOpen
  \bibfield  {author} {\bibinfo {author} {\bibfnamefont {D.}~\bibnamefont
  {Bulmash}}\ and\ \bibinfo {author} {\bibfnamefont {M.}~\bibnamefont
  {Barkeshli}},\ }\bibfield  {title} {\bibinfo {title} {Absolute anomalies in
  (2+ 1) d symmetry-enriched topological states and exact (3+ 1) d
  constructions},\ }\href@noop {} {\bibfield  {journal} {\bibinfo  {journal}
  {Physical Review Research}\ }\textbf {\bibinfo {volume} {2}},\ \bibinfo
  {pages} {043033} (\bibinfo {year} {2020})}\BibitemShut {NoStop}%
\bibitem [{\citenamefont {Kong}(2014)}]{kong2014anyon}%
  \BibitemOpen
  \bibfield  {author} {\bibinfo {author} {\bibfnamefont {L.}~\bibnamefont
  {Kong}},\ }\bibfield  {title} {\bibinfo {title} {Anyon condensation and
  tensor categories},\ }\href@noop {} {\bibfield  {journal} {\bibinfo
  {journal} {Nuclear Physics B}\ }\textbf {\bibinfo {volume} {886}},\ \bibinfo
  {pages} {436} (\bibinfo {year} {2014})}\BibitemShut {NoStop}%
\bibitem [{\citenamefont {Kitaev}\ and\ \citenamefont
  {Kong}(2012)}]{kitaev2012models}%
  \BibitemOpen
  \bibfield  {author} {\bibinfo {author} {\bibfnamefont {A.}~\bibnamefont
  {Kitaev}}\ and\ \bibinfo {author} {\bibfnamefont {L.}~\bibnamefont {Kong}},\
  }\bibfield  {title} {\bibinfo {title} {Models for gapped boundaries and
  domain walls},\ }\href@noop {} {\bibfield  {journal} {\bibinfo  {journal}
  {Communications in Mathematical Physics}\ }\textbf {\bibinfo {volume}
  {313}},\ \bibinfo {pages} {351} (\bibinfo {year} {2012})}\BibitemShut
  {NoStop}%
\bibitem [{\citenamefont {Burnell}(2018)}]{burnell2018anyon}%
  \BibitemOpen
  \bibfield  {author} {\bibinfo {author} {\bibfnamefont {F.~J.}\ \bibnamefont
  {Burnell}},\ }\bibfield  {title} {\bibinfo {title} {Anyon condensation and
  its applications},\ }\href@noop {} {\bibfield  {journal} {\bibinfo  {journal}
  {Annual Review of Condensed Matter Physics}\ }\textbf {\bibinfo {volume}
  {9}},\ \bibinfo {pages} {307} (\bibinfo {year} {2018})}\BibitemShut {NoStop}%
\bibitem [{\citenamefont {Burnell}\ \emph {et~al.}(2011)\citenamefont
  {Burnell}, \citenamefont {Simon},\ and\ \citenamefont
  {Slingerland}}]{burnell2011condensation}%
  \BibitemOpen
  \bibfield  {author} {\bibinfo {author} {\bibfnamefont {F.}~\bibnamefont
  {Burnell}}, \bibinfo {author} {\bibfnamefont {S.~H.}\ \bibnamefont {Simon}},\
  and\ \bibinfo {author} {\bibfnamefont {J.}~\bibnamefont {Slingerland}},\
  }\bibfield  {title} {\bibinfo {title} {Condensation of achiral simple
  currents in topological lattice models: Hamiltonian study of topological
  symmetry breaking},\ }\href@noop {} {\bibfield  {journal} {\bibinfo
  {journal} {Physical Review B—Condensed Matter and Materials Physics}\
  }\textbf {\bibinfo {volume} {84}},\ \bibinfo {pages} {125434} (\bibinfo
  {year} {2011})}\BibitemShut {NoStop}%
\bibitem [{\citenamefont {Bais}\ and\ \citenamefont
  {Slingerland}(2009)}]{bais2009condensate}%
  \BibitemOpen
  \bibfield  {author} {\bibinfo {author} {\bibfnamefont {F.~A.}\ \bibnamefont
  {Bais}}\ and\ \bibinfo {author} {\bibfnamefont {J.}~\bibnamefont
  {Slingerland}},\ }\bibfield  {title} {\bibinfo {title} {Condensate-induced
  transitions between topologically ordered phases},\ }\href@noop {} {\bibfield
   {journal} {\bibinfo  {journal} {Physical Review B—Condensed Matter and
  Materials Physics}\ }\textbf {\bibinfo {volume} {79}},\ \bibinfo {pages}
  {045316} (\bibinfo {year} {2009})}\BibitemShut {NoStop}%
\bibitem [{\citenamefont {Barkeshli}\ \emph {et~al.}(2013)\citenamefont
  {Barkeshli}, \citenamefont {Jian},\ and\ \citenamefont
  {Qi}}]{barkeshli2013theory}%
  \BibitemOpen
  \bibfield  {author} {\bibinfo {author} {\bibfnamefont {M.}~\bibnamefont
  {Barkeshli}}, \bibinfo {author} {\bibfnamefont {C.-M.}\ \bibnamefont
  {Jian}},\ and\ \bibinfo {author} {\bibfnamefont {X.-L.}\ \bibnamefont {Qi}},\
  }\bibfield  {title} {\bibinfo {title} {Theory of defects in abelian
  topological states},\ }\href@noop {} {\bibfield  {journal} {\bibinfo
  {journal} {Physical Review B—Condensed Matter and Materials Physics}\
  }\textbf {\bibinfo {volume} {88}},\ \bibinfo {pages} {235103} (\bibinfo
  {year} {2013})}\BibitemShut {NoStop}%
\bibitem [{\citenamefont {Barkeshli}\ and\ \citenamefont
  {Wen}(2010)}]{barkeshli2010anyon}%
  \BibitemOpen
  \bibfield  {author} {\bibinfo {author} {\bibfnamefont {M.}~\bibnamefont
  {Barkeshli}}\ and\ \bibinfo {author} {\bibfnamefont {X.-G.}\ \bibnamefont
  {Wen}},\ }\bibfield  {title} {\bibinfo {title} {Anyon condensation and
  continuous topological phase transitions<? format?> in non-abelian fractional
  quantum hall states},\ }\href@noop {} {\bibfield  {journal} {\bibinfo
  {journal} {Physical review letters}\ }\textbf {\bibinfo {volume} {105}},\
  \bibinfo {pages} {216804} (\bibinfo {year} {2010})}\BibitemShut {NoStop}%
\bibitem [{\citenamefont {Davydov}\ \emph {et~al.}(2013)\citenamefont
  {Davydov}, \citenamefont {M{\"u}ger}, \citenamefont {Nikshych},\ and\
  \citenamefont {Ostrik}}]{davydov2013witt}%
  \BibitemOpen
  \bibfield  {author} {\bibinfo {author} {\bibfnamefont {A.}~\bibnamefont
  {Davydov}}, \bibinfo {author} {\bibfnamefont {M.}~\bibnamefont {M{\"u}ger}},
  \bibinfo {author} {\bibfnamefont {D.}~\bibnamefont {Nikshych}},\ and\
  \bibinfo {author} {\bibfnamefont {V.}~\bibnamefont {Ostrik}},\ }\bibfield
  {title} {\bibinfo {title} {The witt group of non-degenerate braided fusion
  categories},\ }\href@noop {} {\bibfield  {journal} {\bibinfo  {journal}
  {Journal f{\"u}r die reine und angewandte Mathematik (Crelles Journal)}\
  }\textbf {\bibinfo {volume} {2013}},\ \bibinfo {pages} {135} (\bibinfo {year}
  {2013})}\BibitemShut {NoStop}%
\bibitem [{\citenamefont {Etingof}\ \emph {et~al.}(2016)\citenamefont
  {Etingof}, \citenamefont {Gelaki}, \citenamefont {Nikshych},\ and\
  \citenamefont {Ostrik}}]{etingof2016tensor}%
  \BibitemOpen
  \bibfield  {author} {\bibinfo {author} {\bibfnamefont {P.}~\bibnamefont
  {Etingof}}, \bibinfo {author} {\bibfnamefont {S.}~\bibnamefont {Gelaki}},
  \bibinfo {author} {\bibfnamefont {D.}~\bibnamefont {Nikshych}},\ and\
  \bibinfo {author} {\bibfnamefont {V.}~\bibnamefont {Ostrik}},\ }\href@noop {}
  {\emph {\bibinfo {title} {Tensor categories}}},\ Vol.\ \bibinfo {volume}
  {205}\ (\bibinfo  {publisher} {American Mathematical Soc.},\ \bibinfo {year}
  {2016})\BibitemShut {NoStop}%
\bibitem [{\citenamefont {Cong}\ \emph {et~al.}(2016)\citenamefont {Cong},
  \citenamefont {Cheng},\ and\ \citenamefont {Wang}}]{cong2016topological}%
  \BibitemOpen
  \bibfield  {author} {\bibinfo {author} {\bibfnamefont {I.}~\bibnamefont
  {Cong}}, \bibinfo {author} {\bibfnamefont {M.}~\bibnamefont {Cheng}},\ and\
  \bibinfo {author} {\bibfnamefont {Z.}~\bibnamefont {Wang}},\ }\href
  {https://arxiv.org/abs/1609.02037} {\bibinfo {title} {Topological quantum
  computation with gapped boundaries}} (\bibinfo {year} {2016}),\ \Eprint
  {https://arxiv.org/abs/1609.02037} {arXiv:1609.02037 [quant-ph]} \BibitemShut
  {NoStop}%
\bibitem [{\citenamefont {Cong}\ \emph {et~al.}(2017)\citenamefont {Cong},
  \citenamefont {Cheng},\ and\ \citenamefont {Wang}}]{Cong_2017}%
  \BibitemOpen
  \bibfield  {author} {\bibinfo {author} {\bibfnamefont {I.}~\bibnamefont
  {Cong}}, \bibinfo {author} {\bibfnamefont {M.}~\bibnamefont {Cheng}},\ and\
  \bibinfo {author} {\bibfnamefont {Z.}~\bibnamefont {Wang}},\ }\bibfield
  {title} {\bibinfo {title} {Universal quantum computation with gapped
  boundaries},\ }\bibfield  {journal} {\bibinfo  {journal} {Physical Review
  Letters}\ }\textbf {\bibinfo {volume} {119}},\ \href
  {https://doi.org/10.1103/physrevlett.119.170504}
  {10.1103/physrevlett.119.170504} (\bibinfo {year} {2017})\BibitemShut
  {NoStop}%
\bibitem [{\citenamefont {Kesselring}\ \emph {et~al.}(2018)\citenamefont
  {Kesselring}, \citenamefont {Pastawski}, \citenamefont {Eisert},\ and\
  \citenamefont {Brown}}]{kesselring2018boundaries}%
  \BibitemOpen
  \bibfield  {author} {\bibinfo {author} {\bibfnamefont {M.~S.}\ \bibnamefont
  {Kesselring}}, \bibinfo {author} {\bibfnamefont {F.}~\bibnamefont
  {Pastawski}}, \bibinfo {author} {\bibfnamefont {J.}~\bibnamefont {Eisert}},\
  and\ \bibinfo {author} {\bibfnamefont {B.~J.}\ \bibnamefont {Brown}},\
  }\bibfield  {title} {\bibinfo {title} {The boundaries and twist defects of
  the color code and their applications to topological quantum computation},\
  }\href@noop {} {\bibfield  {journal} {\bibinfo  {journal} {Quantum}\ }\textbf
  {\bibinfo {volume} {2}},\ \bibinfo {pages} {101} (\bibinfo {year}
  {2018})}\BibitemShut {NoStop}%
\bibitem [{\citenamefont {Chatterjee}\ and\ \citenamefont
  {Wen}(2023{\natexlab{a}})}]{chatterjee2023symmetry}%
  \BibitemOpen
  \bibfield  {author} {\bibinfo {author} {\bibfnamefont {A.}~\bibnamefont
  {Chatterjee}}\ and\ \bibinfo {author} {\bibfnamefont {X.-G.}\ \bibnamefont
  {Wen}},\ }\bibfield  {title} {\bibinfo {title} {Symmetry as a shadow of
  topological order and a derivation of topological holographic principle},\
  }\href@noop {} {\bibfield  {journal} {\bibinfo  {journal} {Physical Review
  B}\ }\textbf {\bibinfo {volume} {107}},\ \bibinfo {pages} {155136} (\bibinfo
  {year} {2023}{\natexlab{a}})}\BibitemShut {NoStop}%
\bibitem [{\citenamefont {Chatterjee}\ \emph {et~al.}(2023)\citenamefont
  {Chatterjee}, \citenamefont {Ji},\ and\ \citenamefont
  {Wen}}]{chatterjee2023emergent}%
  \BibitemOpen
  \bibfield  {author} {\bibinfo {author} {\bibfnamefont {A.}~\bibnamefont
  {Chatterjee}}, \bibinfo {author} {\bibfnamefont {W.}~\bibnamefont {Ji}},\
  and\ \bibinfo {author} {\bibfnamefont {X.-G.}\ \bibnamefont {Wen}},\ }\href
  {https://arxiv.org/abs/2212.14432} {\bibinfo {title} {Emergent generalized
  symmetry and maximal symmetry-topological-order}} (\bibinfo {year} {2023}),\
  \Eprint {https://arxiv.org/abs/2212.14432} {arXiv:2212.14432
  [cond-mat.str-el]} \BibitemShut {NoStop}%
\bibitem [{\citenamefont {Chatterjee}\ and\ \citenamefont
  {Wen}(2023{\natexlab{b}})}]{chatterjee2023holographic}%
  \BibitemOpen
  \bibfield  {author} {\bibinfo {author} {\bibfnamefont {A.}~\bibnamefont
  {Chatterjee}}\ and\ \bibinfo {author} {\bibfnamefont {X.-G.}\ \bibnamefont
  {Wen}},\ }\bibfield  {title} {\bibinfo {title} {Holographic theory for
  continuous phase transitions: Emergence and symmetry protection of
  gaplessness},\ }\href@noop {} {\bibfield  {journal} {\bibinfo  {journal}
  {Physical Review B}\ }\textbf {\bibinfo {volume} {108}},\ \bibinfo {pages}
  {075105} (\bibinfo {year} {2023}{\natexlab{b}})}\BibitemShut {NoStop}%
\bibitem [{\citenamefont {Freed}\ and\ \citenamefont
  {Teleman}(2014)}]{freed2014relativequantumfieldtheory}%
  \BibitemOpen
  \bibfield  {author} {\bibinfo {author} {\bibfnamefont {D.~S.}\ \bibnamefont
  {Freed}}\ and\ \bibinfo {author} {\bibfnamefont {C.}~\bibnamefont
  {Teleman}},\ }\href {https://arxiv.org/abs/1212.1692} {\bibinfo {title}
  {Relative quantum field theory}} (\bibinfo {year} {2014}),\ \Eprint
  {https://arxiv.org/abs/1212.1692} {arXiv:1212.1692 [hep-th]} \BibitemShut
  {NoStop}%
\bibitem [{\citenamefont {Freed}\ and\ \citenamefont
  {Teleman}(2022)}]{Freed:2018cec}%
  \BibitemOpen
  \bibfield  {author} {\bibinfo {author} {\bibfnamefont {D.~S.}\ \bibnamefont
  {Freed}}\ and\ \bibinfo {author} {\bibfnamefont {C.}~\bibnamefont
  {Teleman}},\ }\bibfield  {title} {\bibinfo {title} {{Topological dualities in
  the Ising model}},\ }\href {https://doi.org/10.2140/gt.2022.26.1907}
  {\bibfield  {journal} {\bibinfo  {journal} {Geom. Topol.}\ }\textbf {\bibinfo
  {volume} {26}},\ \bibinfo {pages} {1907} (\bibinfo {year} {2022})},\ \Eprint
  {https://arxiv.org/abs/1806.00008} {arXiv:1806.00008 [math.AT]} \BibitemShut
  {NoStop}%
\bibitem [{\citenamefont {Gaiotto}\ and\ \citenamefont
  {Kulp}(2021)}]{Gaiotto_2021}%
  \BibitemOpen
  \bibfield  {author} {\bibinfo {author} {\bibfnamefont {D.}~\bibnamefont
  {Gaiotto}}\ and\ \bibinfo {author} {\bibfnamefont {J.}~\bibnamefont {Kulp}},\
  }\bibfield  {title} {\bibinfo {title} {Orbifold groupoids},\ }\bibfield
  {journal} {\bibinfo  {journal} {Journal of High Energy Physics}\ }\textbf
  {\bibinfo {volume} {2021}},\ \href {https://doi.org/10.1007/jhep02(2021)132}
  {10.1007/jhep02(2021)132} (\bibinfo {year} {2021})\BibitemShut {NoStop}%
\bibitem [{\citenamefont {Lichtman}\ \emph {et~al.}(2021)\citenamefont
  {Lichtman}, \citenamefont {Thorngren}, \citenamefont {Lindner}, \citenamefont
  {Stern},\ and\ \citenamefont {Berg}}]{Lichtman_2021}%
  \BibitemOpen
  \bibfield  {author} {\bibinfo {author} {\bibfnamefont {T.}~\bibnamefont
  {Lichtman}}, \bibinfo {author} {\bibfnamefont {R.}~\bibnamefont {Thorngren}},
  \bibinfo {author} {\bibfnamefont {N.~H.}\ \bibnamefont {Lindner}}, \bibinfo
  {author} {\bibfnamefont {A.}~\bibnamefont {Stern}},\ and\ \bibinfo {author}
  {\bibfnamefont {E.}~\bibnamefont {Berg}},\ }\bibfield  {title} {\bibinfo
  {title} {Bulk anyons as edge symmetries: Boundary phase diagrams of
  topologically ordered states},\ }\bibfield  {journal} {\bibinfo  {journal}
  {Physical Review B}\ }\textbf {\bibinfo {volume} {104}},\ \href
  {https://doi.org/10.1103/physrevb.104.075141} {10.1103/physrevb.104.075141}
  (\bibinfo {year} {2021})\BibitemShut {NoStop}%
\bibitem [{\citenamefont {Kaidi}\ \emph
  {et~al.}(2023{\natexlab{a}})\citenamefont {Kaidi}, \citenamefont {Ohmori},\
  and\ \citenamefont {Zheng}}]{Kaidi:2022cpf}%
  \BibitemOpen
  \bibfield  {author} {\bibinfo {author} {\bibfnamefont {J.}~\bibnamefont
  {Kaidi}}, \bibinfo {author} {\bibfnamefont {K.}~\bibnamefont {Ohmori}},\ and\
  \bibinfo {author} {\bibfnamefont {Y.}~\bibnamefont {Zheng}},\ }\bibfield
  {title} {\bibinfo {title} {{Symmetry TFTs for Non-invertible Defects}},\
  }\href {https://doi.org/10.1007/s00220-023-04859-7} {\bibfield  {journal}
  {\bibinfo  {journal} {Commun. Math. Phys.}\ }\textbf {\bibinfo {volume}
  {404}},\ \bibinfo {pages} {1021} (\bibinfo {year} {2023}{\natexlab{a}})},\
  \Eprint {https://arxiv.org/abs/2209.11062} {arXiv:2209.11062 [hep-th]}
  \BibitemShut {NoStop}%
\bibitem [{\citenamefont {Apruzzi}\ \emph
  {et~al.}(2023{\natexlab{a}})\citenamefont {Apruzzi}, \citenamefont {Bonetti},
  \citenamefont {García~Etxebarria}, \citenamefont {Hosseini},\ and\
  \citenamefont {Schäfer-Nameki}}]{Apruzzi_2023}%
  \BibitemOpen
  \bibfield  {author} {\bibinfo {author} {\bibfnamefont {F.}~\bibnamefont
  {Apruzzi}}, \bibinfo {author} {\bibfnamefont {F.}~\bibnamefont {Bonetti}},
  \bibinfo {author} {\bibfnamefont {I.}~\bibnamefont {García~Etxebarria}},
  \bibinfo {author} {\bibfnamefont {S.~S.}\ \bibnamefont {Hosseini}},\ and\
  \bibinfo {author} {\bibfnamefont {S.}~\bibnamefont {Schäfer-Nameki}},\
  }\bibfield  {title} {\bibinfo {title} {Symmetry tfts from string theory},\
  }\href {https://doi.org/10.1007/s00220-023-04737-2} {\bibfield  {journal}
  {\bibinfo  {journal} {Communications in Mathematical Physics}\ }\textbf
  {\bibinfo {volume} {402}},\ \bibinfo {pages} {895–949} (\bibinfo {year}
  {2023}{\natexlab{a}})}\BibitemShut {NoStop}%
\bibitem [{\citenamefont {Apruzzi}(2022)}]{Apruzzi_2022}%
  \BibitemOpen
  \bibfield  {author} {\bibinfo {author} {\bibfnamefont {F.}~\bibnamefont
  {Apruzzi}},\ }\bibfield  {title} {\bibinfo {title} {Higher form symmetries
  tft in 6d},\ }\bibfield  {journal} {\bibinfo  {journal} {Journal of High
  Energy Physics}\ }\textbf {\bibinfo {volume} {2022}},\ \href
  {https://doi.org/10.1007/jhep11(2022)050} {10.1007/jhep11(2022)050} (\bibinfo
  {year} {2022})\BibitemShut {NoStop}%
\bibitem [{\citenamefont {Lin}\ \emph {et~al.}(2023)\citenamefont {Lin},
  \citenamefont {Okada}, \citenamefont {Seifnashri},\ and\ \citenamefont
  {Tachikawa}}]{Lin_2023}%
  \BibitemOpen
  \bibfield  {author} {\bibinfo {author} {\bibfnamefont {Y.-H.}\ \bibnamefont
  {Lin}}, \bibinfo {author} {\bibfnamefont {M.}~\bibnamefont {Okada}}, \bibinfo
  {author} {\bibfnamefont {S.}~\bibnamefont {Seifnashri}},\ and\ \bibinfo
  {author} {\bibfnamefont {Y.}~\bibnamefont {Tachikawa}},\ }\bibfield  {title}
  {\bibinfo {title} {Asymptotic density of states in 2d cfts with
  non-invertible symmetries},\ }\bibfield  {journal} {\bibinfo  {journal}
  {Journal of High Energy Physics}\ }\textbf {\bibinfo {volume} {2023}},\ \href
  {https://doi.org/10.1007/jhep03(2023)094} {10.1007/jhep03(2023)094} (\bibinfo
  {year} {2023})\BibitemShut {NoStop}%
\bibitem [{\citenamefont {Zhang}\ and\ \citenamefont
  {C{\'o}rdova}(2023)}]{zhang2023anomalies}%
  \BibitemOpen
  \bibfield  {author} {\bibinfo {author} {\bibfnamefont {C.}~\bibnamefont
  {Zhang}}\ and\ \bibinfo {author} {\bibfnamefont {C.}~\bibnamefont
  {C{\'o}rdova}},\ }\bibfield  {title} {\bibinfo {title} {Anomalies of $(1+ 1)
  d $ categorical symmetries},\ }\href@noop {} {\bibfield  {journal} {\bibinfo
  {journal} {arXiv preprint arXiv:2304.01262}\ } (\bibinfo {year}
  {2023})}\BibitemShut {NoStop}%
\bibitem [{\citenamefont
  {Copetti}(2024)}]{copetti2024defectchargesgappedboundary}%
  \BibitemOpen
  \bibfield  {author} {\bibinfo {author} {\bibfnamefont {C.}~\bibnamefont
  {Copetti}},\ }\href {https://arxiv.org/abs/2408.01490} {\bibinfo {title}
  {Defect charges, gapped boundary conditions, and the symmetry tft}} (\bibinfo
  {year} {2024}),\ \Eprint {https://arxiv.org/abs/2408.01490} {arXiv:2408.01490
  [hep-th]} \BibitemShut {NoStop}%
\bibitem [{\citenamefont {Lu}\ \emph {et~al.}(2024{\natexlab{a}})\citenamefont
  {Lu}, \citenamefont {Sun},\ and\ \citenamefont {You}}]{lu2024realizing}%
  \BibitemOpen
  \bibfield  {author} {\bibinfo {author} {\bibfnamefont {D.-C.}\ \bibnamefont
  {Lu}}, \bibinfo {author} {\bibfnamefont {Z.}~\bibnamefont {Sun}},\ and\
  \bibinfo {author} {\bibfnamefont {Y.-Z.}\ \bibnamefont {You}},\ }\href
  {https://arxiv.org/abs/2405.14939} {\bibinfo {title} {Realizing triality and
  $p$-ality by lattice twisted gauging in (1+1)d quantum spin systems}}
  (\bibinfo {year} {2024}{\natexlab{a}}),\ \Eprint
  {https://arxiv.org/abs/2405.14939} {arXiv:2405.14939 [cond-mat.str-el]}
  \BibitemShut {NoStop}%
\bibitem [{\citenamefont {Antinucci}\ \emph {et~al.}(2024)\citenamefont
  {Antinucci}, \citenamefont {Benini},\ and\ \citenamefont
  {Rizi}}]{antinucci2024holographic}%
  \BibitemOpen
  \bibfield  {author} {\bibinfo {author} {\bibfnamefont {A.}~\bibnamefont
  {Antinucci}}, \bibinfo {author} {\bibfnamefont {F.}~\bibnamefont {Benini}},\
  and\ \bibinfo {author} {\bibfnamefont {G.}~\bibnamefont {Rizi}},\ }\bibfield
  {title} {\bibinfo {title} {Holographic duals of symmetry broken phases},\
  }\href@noop {} {\bibfield  {journal} {\bibinfo  {journal} {arXiv preprint
  arXiv:2408.01418}\ } (\bibinfo {year} {2024})}\BibitemShut {NoStop}%
\bibitem [{\citenamefont {Kaidi}\ \emph
  {et~al.}(2023{\natexlab{b}})\citenamefont {Kaidi}, \citenamefont {Nardoni},
  \citenamefont {Zafrir},\ and\ \citenamefont {Zheng}}]{kaidi2023symmetry}%
  \BibitemOpen
  \bibfield  {author} {\bibinfo {author} {\bibfnamefont {J.}~\bibnamefont
  {Kaidi}}, \bibinfo {author} {\bibfnamefont {E.}~\bibnamefont {Nardoni}},
  \bibinfo {author} {\bibfnamefont {G.}~\bibnamefont {Zafrir}},\ and\ \bibinfo
  {author} {\bibfnamefont {Y.}~\bibnamefont {Zheng}},\ }\bibfield  {title}
  {\bibinfo {title} {Symmetry tfts and anomalies of non-invertible
  symmetries},\ }\href@noop {} {\bibfield  {journal} {\bibinfo  {journal}
  {Journal of High Energy Physics}\ }\textbf {\bibinfo {volume} {2023}},\
  \bibinfo {pages} {1} (\bibinfo {year} {2023}{\natexlab{b}})}\BibitemShut
  {NoStop}%
\bibitem [{\citenamefont {Apruzzi}\ \emph
  {et~al.}(2023{\natexlab{b}})\citenamefont {Apruzzi}, \citenamefont {Bah},
  \citenamefont {Bonetti},\ and\ \citenamefont
  {Sch{\"a}fer-Nameki}}]{apruzzi2023noninvertible}%
  \BibitemOpen
  \bibfield  {author} {\bibinfo {author} {\bibfnamefont {F.}~\bibnamefont
  {Apruzzi}}, \bibinfo {author} {\bibfnamefont {I.}~\bibnamefont {Bah}},
  \bibinfo {author} {\bibfnamefont {F.}~\bibnamefont {Bonetti}},\ and\ \bibinfo
  {author} {\bibfnamefont {S.}~\bibnamefont {Sch{\"a}fer-Nameki}},\ }\bibfield
  {title} {\bibinfo {title} {Noninvertible symmetries from holography and
  branes},\ }\href@noop {} {\bibfield  {journal} {\bibinfo  {journal} {Physical
  review letters}\ }\textbf {\bibinfo {volume} {130}},\ \bibinfo {pages}
  {121601} (\bibinfo {year} {2023}{\natexlab{b}})}\BibitemShut {NoStop}%
\bibitem [{\citenamefont {Cordova}\ \emph {et~al.}(2023)\citenamefont
  {Cordova}, \citenamefont {Hsin},\ and\ \citenamefont
  {Zhang}}]{cordova2023anomalies}%
  \BibitemOpen
  \bibfield  {author} {\bibinfo {author} {\bibfnamefont {C.}~\bibnamefont
  {Cordova}}, \bibinfo {author} {\bibfnamefont {P.-S.}\ \bibnamefont {Hsin}},\
  and\ \bibinfo {author} {\bibfnamefont {C.}~\bibnamefont {Zhang}},\ }\bibfield
   {title} {\bibinfo {title} {Anomalies of non-invertible symmetries in (3+ 1)
  d},\ }\href@noop {} {\bibfield  {journal} {\bibinfo  {journal} {arXiv
  preprint arXiv:2308.11706}\ } (\bibinfo {year} {2023})}\BibitemShut {NoStop}%
\bibitem [{\citenamefont {Antinucci}\ \emph {et~al.}(2023)\citenamefont
  {Antinucci}, \citenamefont {Benini}, \citenamefont {Copetti}, \citenamefont
  {Galati},\ and\ \citenamefont {Rizi}}]{antinucci2023anomalies}%
  \BibitemOpen
  \bibfield  {author} {\bibinfo {author} {\bibfnamefont {A.}~\bibnamefont
  {Antinucci}}, \bibinfo {author} {\bibfnamefont {F.}~\bibnamefont {Benini}},
  \bibinfo {author} {\bibfnamefont {C.}~\bibnamefont {Copetti}}, \bibinfo
  {author} {\bibfnamefont {G.}~\bibnamefont {Galati}},\ and\ \bibinfo {author}
  {\bibfnamefont {G.}~\bibnamefont {Rizi}},\ }\bibfield  {title} {\bibinfo
  {title} {Anomalies of non-invertible self-duality symmetries:
  fractionalization and gauging},\ }\href@noop {} {\bibfield  {journal}
  {\bibinfo  {journal} {arXiv preprint arXiv:2308.11707}\ } (\bibinfo {year}
  {2023})}\BibitemShut {NoStop}%
\bibitem [{\citenamefont {Antinucci}\ \emph {et~al.}(2022)\citenamefont
  {Antinucci}, \citenamefont {Benini}, \citenamefont {Copetti}, \citenamefont
  {Galati},\ and\ \citenamefont {Rizi}}]{antinucci2022holography}%
  \BibitemOpen
  \bibfield  {author} {\bibinfo {author} {\bibfnamefont {A.}~\bibnamefont
  {Antinucci}}, \bibinfo {author} {\bibfnamefont {F.}~\bibnamefont {Benini}},
  \bibinfo {author} {\bibfnamefont {C.}~\bibnamefont {Copetti}}, \bibinfo
  {author} {\bibfnamefont {G.}~\bibnamefont {Galati}},\ and\ \bibinfo {author}
  {\bibfnamefont {G.}~\bibnamefont {Rizi}},\ }\bibfield  {title} {\bibinfo
  {title} {The holography of non-invertible self-duality symmetries},\
  }\href@noop {} {\bibfield  {journal} {\bibinfo  {journal} {arXiv preprint
  arXiv:2210.09146}\ } (\bibinfo {year} {2022})}\BibitemShut {NoStop}%
\bibitem [{\citenamefont {Choi}\ \emph {et~al.}(2023)\citenamefont {Choi},
  \citenamefont {C{\'o}rdova}, \citenamefont {Hsin}, \citenamefont {Lam},\ and\
  \citenamefont {Shao}}]{choi2023non}%
  \BibitemOpen
  \bibfield  {author} {\bibinfo {author} {\bibfnamefont {Y.}~\bibnamefont
  {Choi}}, \bibinfo {author} {\bibfnamefont {C.}~\bibnamefont {C{\'o}rdova}},
  \bibinfo {author} {\bibfnamefont {P.-S.}\ \bibnamefont {Hsin}}, \bibinfo
  {author} {\bibfnamefont {H.~T.}\ \bibnamefont {Lam}},\ and\ \bibinfo {author}
  {\bibfnamefont {S.-H.}\ \bibnamefont {Shao}},\ }\bibfield  {title} {\bibinfo
  {title} {Non-invertible condensation, duality, and triality defects in 3+ 1
  dimensions},\ }\href@noop {} {\bibfield  {journal} {\bibinfo  {journal}
  {Communications in Mathematical Physics}\ }\textbf {\bibinfo {volume}
  {402}},\ \bibinfo {pages} {489} (\bibinfo {year} {2023})}\BibitemShut
  {NoStop}%
\bibitem [{\citenamefont {Bhardwaj}\ \emph
  {et~al.}(2023{\natexlab{a}})\citenamefont {Bhardwaj}, \citenamefont
  {Bottini}, \citenamefont {Sch{\"a}fer-Nameki},\ and\ \citenamefont
  {Tiwari}}]{bhardwaj2023non2}%
  \BibitemOpen
  \bibfield  {author} {\bibinfo {author} {\bibfnamefont {L.}~\bibnamefont
  {Bhardwaj}}, \bibinfo {author} {\bibfnamefont {L.~E.}\ \bibnamefont
  {Bottini}}, \bibinfo {author} {\bibfnamefont {S.}~\bibnamefont
  {Sch{\"a}fer-Nameki}},\ and\ \bibinfo {author} {\bibfnamefont
  {A.}~\bibnamefont {Tiwari}},\ }\bibfield  {title} {\bibinfo {title}
  {Non-invertible symmetry webs},\ }\href@noop {} {\bibfield  {journal}
  {\bibinfo  {journal} {SciPost Physics}\ }\textbf {\bibinfo {volume} {15}},\
  \bibinfo {pages} {160} (\bibinfo {year} {2023}{\natexlab{a}})}\BibitemShut
  {NoStop}%
\bibitem [{\citenamefont {Lu}\ \emph {et~al.}(2024{\natexlab{b}})\citenamefont
  {Lu}, \citenamefont {Sun},\ and\ \citenamefont {Zhang}}]{lu2024exploring}%
  \BibitemOpen
  \bibfield  {author} {\bibinfo {author} {\bibfnamefont {D.-C.}\ \bibnamefont
  {Lu}}, \bibinfo {author} {\bibfnamefont {Z.}~\bibnamefont {Sun}},\ and\
  \bibinfo {author} {\bibfnamefont {Z.}~\bibnamefont {Zhang}},\ }\href
  {https://arxiv.org/abs/2406.12151} {\bibinfo {title} {Exploring $g$-ality
  defects in 2-dim qfts}} (\bibinfo {year} {2024}{\natexlab{b}}),\ \Eprint
  {https://arxiv.org/abs/2406.12151} {arXiv:2406.12151 [hep-th]} \BibitemShut
  {NoStop}%
\bibitem [{\citenamefont {Apruzzi}\ \emph {et~al.}(2024)\citenamefont
  {Apruzzi}, \citenamefont {Bedogna},\ and\ \citenamefont
  {Dondi}}]{apruzzi2024symth}%
  \BibitemOpen
  \bibfield  {author} {\bibinfo {author} {\bibfnamefont {F.}~\bibnamefont
  {Apruzzi}}, \bibinfo {author} {\bibfnamefont {F.}~\bibnamefont {Bedogna}},\
  and\ \bibinfo {author} {\bibfnamefont {N.}~\bibnamefont {Dondi}},\ }\bibfield
   {title} {\bibinfo {title} {Symth for non-finite symmetries},\ }\href@noop {}
  {\bibfield  {journal} {\bibinfo  {journal} {arXiv preprint arXiv:2402.14813}\
  } (\bibinfo {year} {2024})}\BibitemShut {NoStop}%
\bibitem [{\citenamefont {Brennan}\ and\ \citenamefont
  {Sun}(2024)}]{brennan2024symtft}%
  \BibitemOpen
  \bibfield  {author} {\bibinfo {author} {\bibfnamefont {T.~D.}\ \bibnamefont
  {Brennan}}\ and\ \bibinfo {author} {\bibfnamefont {Z.}~\bibnamefont {Sun}},\
  }\bibfield  {title} {\bibinfo {title} {A symtft for continuous symmetries},\
  }\href@noop {} {\bibfield  {journal} {\bibinfo  {journal} {arXiv preprint
  arXiv:2401.06128}\ } (\bibinfo {year} {2024})}\BibitemShut {NoStop}%
\bibitem [{\citenamefont {Antinucci}\ and\ \citenamefont
  {Benini}(2024)}]{antinucci2024anomalies}%
  \BibitemOpen
  \bibfield  {author} {\bibinfo {author} {\bibfnamefont {A.}~\bibnamefont
  {Antinucci}}\ and\ \bibinfo {author} {\bibfnamefont {F.}~\bibnamefont
  {Benini}},\ }\bibfield  {title} {\bibinfo {title} {Anomalies and gauging of u
  (1) symmetries},\ }\href@noop {} {\bibfield  {journal} {\bibinfo  {journal}
  {arXiv preprint arXiv:2401.10165}\ } (\bibinfo {year} {2024})}\BibitemShut
  {NoStop}%
\bibitem [{\citenamefont {Bhardwaj}\ \emph
  {et~al.}(2024{\natexlab{a}})\citenamefont {Bhardwaj}, \citenamefont
  {Bottini}, \citenamefont {Schafer-Nameki},\ and\ \citenamefont
  {Tiwari}}]{bhardwaj2024lattice}%
  \BibitemOpen
  \bibfield  {author} {\bibinfo {author} {\bibfnamefont {L.}~\bibnamefont
  {Bhardwaj}}, \bibinfo {author} {\bibfnamefont {L.~E.}\ \bibnamefont
  {Bottini}}, \bibinfo {author} {\bibfnamefont {S.}~\bibnamefont
  {Schafer-Nameki}},\ and\ \bibinfo {author} {\bibfnamefont {A.}~\bibnamefont
  {Tiwari}},\ }\bibfield  {title} {\bibinfo {title} {Lattice models for phases
  and transitions with non-invertible symmetries},\ }\href@noop {} {\bibfield
  {journal} {\bibinfo  {journal} {arXiv preprint arXiv:2405.05964}\ } (\bibinfo
  {year} {2024}{\natexlab{a}})}\BibitemShut {NoStop}%
\bibitem [{\citenamefont {Chatterjee}\ \emph {et~al.}(2024)\citenamefont
  {Chatterjee}, \citenamefont {Aksoy},\ and\ \citenamefont
  {Wen}}]{chatterjee2024quantum}%
  \BibitemOpen
  \bibfield  {author} {\bibinfo {author} {\bibfnamefont {A.}~\bibnamefont
  {Chatterjee}}, \bibinfo {author} {\bibfnamefont {{\"O}.~M.}\ \bibnamefont
  {Aksoy}},\ and\ \bibinfo {author} {\bibfnamefont {X.-G.}\ \bibnamefont
  {Wen}},\ }\bibfield  {title} {\bibinfo {title} {Quantum phases and
  transitions in spin chains with non-invertible symmetries},\ }\href@noop {}
  {\bibfield  {journal} {\bibinfo  {journal} {arXiv preprint arXiv:2405.05331}\
  } (\bibinfo {year} {2024})}\BibitemShut {NoStop}%
\bibitem [{\citenamefont {Pace}\ \emph {et~al.}(2023)\citenamefont {Pace},
  \citenamefont {Zhu}, \citenamefont {Beaudry},\ and\ \citenamefont
  {Wen}}]{pace2023generalized}%
  \BibitemOpen
  \bibfield  {author} {\bibinfo {author} {\bibfnamefont {S.~D.}\ \bibnamefont
  {Pace}}, \bibinfo {author} {\bibfnamefont {C.}~\bibnamefont {Zhu}}, \bibinfo
  {author} {\bibfnamefont {A.}~\bibnamefont {Beaudry}},\ and\ \bibinfo {author}
  {\bibfnamefont {X.-G.}\ \bibnamefont {Wen}},\ }\href
  {https://arxiv.org/abs/2310.08554} {\bibinfo {title} {Generalized symmetries
  in singularity-free nonlinear $\sigma$-models and their disordered phases}}
  (\bibinfo {year} {2023}),\ \Eprint {https://arxiv.org/abs/2310.08554}
  {arXiv:2310.08554 [cond-mat.str-el]} \BibitemShut {NoStop}%
\bibitem [{\citenamefont {Cao}\ \emph {et~al.}(2024)\citenamefont {Cao},
  \citenamefont {Li},\ and\ \citenamefont {Yamazaki}}]{cao2024generating}%
  \BibitemOpen
  \bibfield  {author} {\bibinfo {author} {\bibfnamefont {W.}~\bibnamefont
  {Cao}}, \bibinfo {author} {\bibfnamefont {L.}~\bibnamefont {Li}},\ and\
  \bibinfo {author} {\bibfnamefont {M.}~\bibnamefont {Yamazaki}},\ }\href
  {https://arxiv.org/abs/2406.05454} {\bibinfo {title} {Generating lattice
  non-invertible symmetries}} (\bibinfo {year} {2024}),\ \Eprint
  {https://arxiv.org/abs/2406.05454} {arXiv:2406.05454 [cond-mat.str-el]}
  \BibitemShut {NoStop}%
\bibitem [{\citenamefont {Inamura}\ and\ \citenamefont
  {Ohmori}(2024)}]{Inamura_2024}%
  \BibitemOpen
  \bibfield  {author} {\bibinfo {author} {\bibfnamefont {K.}~\bibnamefont
  {Inamura}}\ and\ \bibinfo {author} {\bibfnamefont {K.}~\bibnamefont
  {Ohmori}},\ }\bibfield  {title} {\bibinfo {title} {Fusion surface models:
  2+1d lattice models from fusion 2-categories},\ }\bibfield  {journal}
  {\bibinfo  {journal} {SciPost Physics}\ }\textbf {\bibinfo {volume} {16}},\
  \href {https://doi.org/10.21468/scipostphys.16.6.143}
  {10.21468/scipostphys.16.6.143} (\bibinfo {year} {2024})\BibitemShut
  {NoStop}%
\bibitem [{\citenamefont {Bhardwaj}\ \emph
  {et~al.}(2024{\natexlab{b}})\citenamefont {Bhardwaj}, \citenamefont
  {Bottini}, \citenamefont {Schafer-Nameki},\ and\ \citenamefont
  {Tiwari}}]{bhardwaj2024illustrating}%
  \BibitemOpen
  \bibfield  {author} {\bibinfo {author} {\bibfnamefont {L.}~\bibnamefont
  {Bhardwaj}}, \bibinfo {author} {\bibfnamefont {L.~E.}\ \bibnamefont
  {Bottini}}, \bibinfo {author} {\bibfnamefont {S.}~\bibnamefont
  {Schafer-Nameki}},\ and\ \bibinfo {author} {\bibfnamefont {A.}~\bibnamefont
  {Tiwari}},\ }\bibfield  {title} {\bibinfo {title} {Illustrating the
  categorical landau paradigm in lattice models},\ }\href@noop {} {\bibfield
  {journal} {\bibinfo  {journal} {arXiv preprint arXiv:2405.05302}\ } (\bibinfo
  {year} {2024}{\natexlab{b}})}\BibitemShut {NoStop}%
\bibitem [{\citenamefont {Gorantla}\ \emph {et~al.}(2024)\citenamefont
  {Gorantla}, \citenamefont {Shao},\ and\ \citenamefont
  {Tantivasadakarn}}]{gorantla2024tensor}%
  \BibitemOpen
  \bibfield  {author} {\bibinfo {author} {\bibfnamefont {P.}~\bibnamefont
  {Gorantla}}, \bibinfo {author} {\bibfnamefont {S.-H.}\ \bibnamefont {Shao}},\
  and\ \bibinfo {author} {\bibfnamefont {N.}~\bibnamefont {Tantivasadakarn}},\
  }\bibfield  {title} {\bibinfo {title} {Tensor networks for non-invertible
  symmetries in 3+ 1d and beyond},\ }\href@noop {} {\bibfield  {journal}
  {\bibinfo  {journal} {arXiv preprint arXiv:2406.12978}\ } (\bibinfo {year}
  {2024})}\BibitemShut {NoStop}%
\bibitem [{\citenamefont {Delcamp}\ and\ \citenamefont
  {Tiwari}(2024)}]{Delcamp_2024}%
  \BibitemOpen
  \bibfield  {author} {\bibinfo {author} {\bibfnamefont {C.}~\bibnamefont
  {Delcamp}}\ and\ \bibinfo {author} {\bibfnamefont {A.}~\bibnamefont
  {Tiwari}},\ }\bibfield  {title} {\bibinfo {title} {Higher categorical
  symmetries and gauging in two-dimensional spin systems},\ }\bibfield
  {journal} {\bibinfo  {journal} {SciPost Physics}\ }\textbf {\bibinfo {volume}
  {16}},\ \href {https://doi.org/10.21468/scipostphys.16.4.110}
  {10.21468/scipostphys.16.4.110} (\bibinfo {year} {2024})\BibitemShut
  {NoStop}%
\bibitem [{\citenamefont {Fechisin}\ \emph {et~al.}(2023)\citenamefont
  {Fechisin}, \citenamefont {Tantivasadakarn},\ and\ \citenamefont
  {Albert}}]{fechisin2023non}%
  \BibitemOpen
  \bibfield  {author} {\bibinfo {author} {\bibfnamefont {C.}~\bibnamefont
  {Fechisin}}, \bibinfo {author} {\bibfnamefont {N.}~\bibnamefont
  {Tantivasadakarn}},\ and\ \bibinfo {author} {\bibfnamefont {V.~V.}\
  \bibnamefont {Albert}},\ }\bibfield  {title} {\bibinfo {title}
  {Non-invertible symmetry-protected topological order in a group-based cluster
  state},\ }\href@noop {} {\bibfield  {journal} {\bibinfo  {journal} {arXiv
  preprint arXiv:2312.09272}\ } (\bibinfo {year} {2023})}\BibitemShut {NoStop}%
\bibitem [{\citenamefont {Verresen}\ \emph {et~al.}(2021)\citenamefont
  {Verresen}, \citenamefont {Thorngren}, \citenamefont {Jones},\ and\
  \citenamefont {Pollmann}}]{Verresen_2021}%
  \BibitemOpen
  \bibfield  {author} {\bibinfo {author} {\bibfnamefont {R.}~\bibnamefont
  {Verresen}}, \bibinfo {author} {\bibfnamefont {R.}~\bibnamefont {Thorngren}},
  \bibinfo {author} {\bibfnamefont {N.~G.}\ \bibnamefont {Jones}},\ and\
  \bibinfo {author} {\bibfnamefont {F.}~\bibnamefont {Pollmann}},\ }\bibfield
  {title} {\bibinfo {title} {Gapless topological phases and symmetry-enriched
  quantum criticality},\ }\bibfield  {journal} {\bibinfo  {journal} {Physical
  Review X}\ }\textbf {\bibinfo {volume} {11}},\ \href
  {https://doi.org/10.1103/physrevx.11.041059} {10.1103/physrevx.11.041059}
  (\bibinfo {year} {2021})\BibitemShut {NoStop}%
\bibitem [{\citenamefont {Ye}\ \emph {et~al.}(2022)\citenamefont {Ye},
  \citenamefont {Guo}, \citenamefont {He}, \citenamefont {Wang},\ and\
  \citenamefont {Zou}}]{Ye_2022}%
  \BibitemOpen
  \bibfield  {author} {\bibinfo {author} {\bibfnamefont {W.}~\bibnamefont
  {Ye}}, \bibinfo {author} {\bibfnamefont {M.}~\bibnamefont {Guo}}, \bibinfo
  {author} {\bibfnamefont {Y.-C.}\ \bibnamefont {He}}, \bibinfo {author}
  {\bibfnamefont {C.}~\bibnamefont {Wang}},\ and\ \bibinfo {author}
  {\bibfnamefont {L.}~\bibnamefont {Zou}},\ }\bibfield  {title} {\bibinfo
  {title} {Topological characterization of lieb-schultz-mattis constraints and
  applications to symmetry-enriched quantum criticality},\ }\bibfield
  {journal} {\bibinfo  {journal} {SciPost Physics}\ }\textbf {\bibinfo {volume}
  {13}},\ \href {https://doi.org/10.21468/scipostphys.13.3.066}
  {10.21468/scipostphys.13.3.066} (\bibinfo {year} {2022})\BibitemShut
  {NoStop}%
\bibitem [{\citenamefont {Mondal}\ \emph {et~al.}(2023)\citenamefont {Mondal},
  \citenamefont {Agarwala}, \citenamefont {Mishra},\ and\ \citenamefont
  {Prakash}}]{Mondal_2023}%
  \BibitemOpen
  \bibfield  {author} {\bibinfo {author} {\bibfnamefont {S.}~\bibnamefont
  {Mondal}}, \bibinfo {author} {\bibfnamefont {A.}~\bibnamefont {Agarwala}},
  \bibinfo {author} {\bibfnamefont {T.}~\bibnamefont {Mishra}},\ and\ \bibinfo
  {author} {\bibfnamefont {A.}~\bibnamefont {Prakash}},\ }\bibfield  {title}
  {\bibinfo {title} {Symmetry-enriched criticality in a coupled spin ladder},\
  }\bibfield  {journal} {\bibinfo  {journal} {Physical Review B}\ }\textbf
  {\bibinfo {volume} {108}},\ \href
  {https://doi.org/10.1103/physrevb.108.245135} {10.1103/physrevb.108.245135}
  (\bibinfo {year} {2023})\BibitemShut {NoStop}%
\bibitem [{\citenamefont {Yu}\ \emph {et~al.}(2022)\citenamefont {Yu},
  \citenamefont {Huang}, \citenamefont {Song}, \citenamefont {Xu},
  \citenamefont {Ding},\ and\ \citenamefont {Zhang}}]{Yu_2022}%
  \BibitemOpen
  \bibfield  {author} {\bibinfo {author} {\bibfnamefont {X.-J.}\ \bibnamefont
  {Yu}}, \bibinfo {author} {\bibfnamefont {R.-Z.}\ \bibnamefont {Huang}},
  \bibinfo {author} {\bibfnamefont {H.-H.}\ \bibnamefont {Song}}, \bibinfo
  {author} {\bibfnamefont {L.}~\bibnamefont {Xu}}, \bibinfo {author}
  {\bibfnamefont {C.}~\bibnamefont {Ding}},\ and\ \bibinfo {author}
  {\bibfnamefont {L.}~\bibnamefont {Zhang}},\ }\bibfield  {title} {\bibinfo
  {title} {Conformal boundary conditions of symmetry-enriched quantum critical
  spin chains},\ }\bibfield  {journal} {\bibinfo  {journal} {Physical Review
  Letters}\ }\textbf {\bibinfo {volume} {129}},\ \href
  {https://doi.org/10.1103/physrevlett.129.210601}
  {10.1103/physrevlett.129.210601} (\bibinfo {year} {2022})\BibitemShut
  {NoStop}%
\bibitem [{\citenamefont {Hidaka}\ \emph {et~al.}(2022)\citenamefont {Hidaka},
  \citenamefont {Furuya}, \citenamefont {Ueda},\ and\ \citenamefont
  {Tada}}]{Hidaka_2022}%
  \BibitemOpen
  \bibfield  {author} {\bibinfo {author} {\bibfnamefont {Y.}~\bibnamefont
  {Hidaka}}, \bibinfo {author} {\bibfnamefont {S.~C.}\ \bibnamefont {Furuya}},
  \bibinfo {author} {\bibfnamefont {A.}~\bibnamefont {Ueda}},\ and\ \bibinfo
  {author} {\bibfnamefont {Y.}~\bibnamefont {Tada}},\ }\bibfield  {title}
  {\bibinfo {title} {Gapless symmetry-protected topological phase of quantum
  antiferromagnets on anisotropic triangular strip},\ }\bibfield  {journal}
  {\bibinfo  {journal} {Physical Review B}\ }\textbf {\bibinfo {volume}
  {106}},\ \href {https://doi.org/10.1103/physrevb.106.144436}
  {10.1103/physrevb.106.144436} (\bibinfo {year} {2022})\BibitemShut {NoStop}%
\bibitem [{\citenamefont {Tantivasadakarn}\ \emph
  {et~al.}(2023{\natexlab{a}})\citenamefont {Tantivasadakarn}, \citenamefont
  {Thorngren}, \citenamefont {Vishwanath},\ and\ \citenamefont
  {Verresen}}]{Tantivasadakarn_2023}%
  \BibitemOpen
  \bibfield  {author} {\bibinfo {author} {\bibfnamefont {N.}~\bibnamefont
  {Tantivasadakarn}}, \bibinfo {author} {\bibfnamefont {R.}~\bibnamefont
  {Thorngren}}, \bibinfo {author} {\bibfnamefont {A.}~\bibnamefont
  {Vishwanath}},\ and\ \bibinfo {author} {\bibfnamefont {R.}~\bibnamefont
  {Verresen}},\ }\bibfield  {title} {\bibinfo {title} {Building models of
  topological quantum criticality from pivot hamiltonians},\ }\bibfield
  {journal} {\bibinfo  {journal} {SciPost Physics}\ }\textbf {\bibinfo {volume}
  {14}},\ \href {https://doi.org/10.21468/scipostphys.14.2.013}
  {10.21468/scipostphys.14.2.013} (\bibinfo {year}
  {2023}{\natexlab{a}})\BibitemShut {NoStop}%
\bibitem [{\citenamefont {Wang}\ \emph {et~al.}(2023)\citenamefont {Wang},
  \citenamefont {Li},\ and\ \citenamefont {Wu}}]{wang2023stability}%
  \BibitemOpen
  \bibfield  {author} {\bibinfo {author} {\bibfnamefont {X.}~\bibnamefont
  {Wang}}, \bibinfo {author} {\bibfnamefont {L.}~\bibnamefont {Li}},\ and\
  \bibinfo {author} {\bibfnamefont {J.}~\bibnamefont {Wu}},\ }\href
  {https://arxiv.org/abs/2306.11446} {\bibinfo {title} {Stability and fine
  structure of symmetry-enriched quantum criticality in a spin ladder
  triangular model}} (\bibinfo {year} {2023}),\ \Eprint
  {https://arxiv.org/abs/2306.11446} {arXiv:2306.11446 [cond-mat.str-el]}
  \BibitemShut {NoStop}%
\bibitem [{\citenamefont {Ye}\ and\ \citenamefont {Zou}(2024)}]{Ye_2024}%
  \BibitemOpen
  \bibfield  {author} {\bibinfo {author} {\bibfnamefont {W.}~\bibnamefont
  {Ye}}\ and\ \bibinfo {author} {\bibfnamefont {L.}~\bibnamefont {Zou}},\
  }\bibfield  {title} {\bibinfo {title} {Classification of symmetry-enriched
  topological quantum spin liquids},\ }\bibfield  {journal} {\bibinfo
  {journal} {Physical Review X}\ }\textbf {\bibinfo {volume} {14}},\ \href
  {https://doi.org/10.1103/physrevx.14.021053} {10.1103/physrevx.14.021053}
  (\bibinfo {year} {2024})\BibitemShut {NoStop}%
\bibitem [{\citenamefont {Prembabu}\ \emph {et~al.}(2024)\citenamefont
  {Prembabu}, \citenamefont {Thorngren},\ and\ \citenamefont
  {Verresen}}]{Prembabu_2024}%
  \BibitemOpen
  \bibfield  {author} {\bibinfo {author} {\bibfnamefont {S.}~\bibnamefont
  {Prembabu}}, \bibinfo {author} {\bibfnamefont {R.}~\bibnamefont
  {Thorngren}},\ and\ \bibinfo {author} {\bibfnamefont {R.}~\bibnamefont
  {Verresen}},\ }\bibfield  {title} {\bibinfo {title} {Boundary-deconfined
  quantum criticality at transitions between symmetry-protected topological
  chains},\ }\bibfield  {journal} {\bibinfo  {journal} {Physical Review B}\
  }\textbf {\bibinfo {volume} {109}},\ \href
  {https://doi.org/10.1103/physrevb.109.l201112} {10.1103/physrevb.109.l201112}
  (\bibinfo {year} {2024})\BibitemShut {NoStop}%
\bibitem [{\citenamefont {Tantivasadakarn}\ \emph
  {et~al.}(2023{\natexlab{b}})\citenamefont {Tantivasadakarn}, \citenamefont
  {Thorngren}, \citenamefont {Vishwanath},\ and\ \citenamefont
  {Verresen}}]{tantivasadakarn2023building}%
  \BibitemOpen
  \bibfield  {author} {\bibinfo {author} {\bibfnamefont {N.}~\bibnamefont
  {Tantivasadakarn}}, \bibinfo {author} {\bibfnamefont {R.}~\bibnamefont
  {Thorngren}}, \bibinfo {author} {\bibfnamefont {A.}~\bibnamefont
  {Vishwanath}},\ and\ \bibinfo {author} {\bibfnamefont {R.}~\bibnamefont
  {Verresen}},\ }\bibfield  {title} {\bibinfo {title} {Building models of
  topological quantum criticality from pivot hamiltonians},\ }\href@noop {}
  {\bibfield  {journal} {\bibinfo  {journal} {SciPost Physics}\ }\textbf
  {\bibinfo {volume} {14}},\ \bibinfo {pages} {013} (\bibinfo {year}
  {2023}{\natexlab{b}})}\BibitemShut {NoStop}%
\bibitem [{\citenamefont {Scaffidi}\ \emph {et~al.}(2017)\citenamefont
  {Scaffidi}, \citenamefont {Parker},\ and\ \citenamefont
  {Vasseur}}]{Scaffidi_2017}%
  \BibitemOpen
  \bibfield  {author} {\bibinfo {author} {\bibfnamefont {T.}~\bibnamefont
  {Scaffidi}}, \bibinfo {author} {\bibfnamefont {D.~E.}\ \bibnamefont
  {Parker}},\ and\ \bibinfo {author} {\bibfnamefont {R.}~\bibnamefont
  {Vasseur}},\ }\bibfield  {title} {\bibinfo {title} {Gapless
  symmetry-protected topological order},\ }\bibfield  {journal} {\bibinfo
  {journal} {Physical Review X}\ }\textbf {\bibinfo {volume} {7}},\ \href
  {https://doi.org/10.1103/physrevx.7.041048} {10.1103/physrevx.7.041048}
  (\bibinfo {year} {2017})\BibitemShut {NoStop}%
\bibitem [{\citenamefont {Thorngren}\ \emph {et~al.}(2021)\citenamefont
  {Thorngren}, \citenamefont {Vishwanath},\ and\ \citenamefont
  {Verresen}}]{Thorngren_2021}%
  \BibitemOpen
  \bibfield  {author} {\bibinfo {author} {\bibfnamefont {R.}~\bibnamefont
  {Thorngren}}, \bibinfo {author} {\bibfnamefont {A.}~\bibnamefont
  {Vishwanath}},\ and\ \bibinfo {author} {\bibfnamefont {R.}~\bibnamefont
  {Verresen}},\ }\bibfield  {title} {\bibinfo {title} {Intrinsically gapless
  topological phases},\ }\bibfield  {journal} {\bibinfo  {journal} {Physical
  Review B}\ }\textbf {\bibinfo {volume} {104}},\ \href
  {https://doi.org/10.1103/physrevb.104.075132} {10.1103/physrevb.104.075132}
  (\bibinfo {year} {2021})\BibitemShut {NoStop}%
\bibitem [{\citenamefont {Li}\ \emph {et~al.}(2023)\citenamefont {Li},
  \citenamefont {Oshikawa},\ and\ \citenamefont {Zheng}}]{li2023intrinsicall}%
  \BibitemOpen
  \bibfield  {author} {\bibinfo {author} {\bibfnamefont {L.}~\bibnamefont
  {Li}}, \bibinfo {author} {\bibfnamefont {M.}~\bibnamefont {Oshikawa}},\ and\
  \bibinfo {author} {\bibfnamefont {Y.}~\bibnamefont {Zheng}},\ }\href
  {https://arxiv.org/abs/2307.04788} {\bibinfo {title} {Intrinsically/purely
  gapless-spt from non-invertible duality transformations}} (\bibinfo {year}
  {2023}),\ \Eprint {https://arxiv.org/abs/2307.04788} {arXiv:2307.04788
  [cond-mat.str-el]} \BibitemShut {NoStop}%
\bibitem [{\citenamefont {Wen}\ and\ \citenamefont
  {Potter}(2023{\natexlab{a}})}]{Wen_2023}%
  \BibitemOpen
  \bibfield  {author} {\bibinfo {author} {\bibfnamefont {R.}~\bibnamefont
  {Wen}}\ and\ \bibinfo {author} {\bibfnamefont {A.~C.}\ \bibnamefont
  {Potter}},\ }\bibfield  {title} {\bibinfo {title} {Bulk-boundary
  correspondence for intrinsically gapless symmetry-protected topological
  phases from group cohomology},\ }\bibfield  {journal} {\bibinfo  {journal}
  {Physical Review B}\ }\textbf {\bibinfo {volume} {107}},\ \href
  {https://doi.org/10.1103/physrevb.107.245127} {10.1103/physrevb.107.245127}
  (\bibinfo {year} {2023}{\natexlab{a}})\BibitemShut {NoStop}%
\bibitem [{\citenamefont {Wen}\ and\ \citenamefont
  {Potter}(2023{\natexlab{b}})}]{wen2023classification}%
  \BibitemOpen
  \bibfield  {author} {\bibinfo {author} {\bibfnamefont {R.}~\bibnamefont
  {Wen}}\ and\ \bibinfo {author} {\bibfnamefont {A.~C.}\ \bibnamefont
  {Potter}},\ }\href {https://arxiv.org/abs/2311.00050} {\bibinfo {title}
  {Classification of 1+1d gapless symmetry protected phases via topological
  holography}} (\bibinfo {year} {2023}{\natexlab{b}}),\ \Eprint
  {https://arxiv.org/abs/2311.00050} {arXiv:2311.00050 [cond-mat.str-el]}
  \BibitemShut {NoStop}%
\bibitem [{\citenamefont {Li}\ \emph {et~al.}(2024)\citenamefont {Li},
  \citenamefont {Oshikawa},\ and\ \citenamefont {Zheng}}]{Li_2024}%
  \BibitemOpen
  \bibfield  {author} {\bibinfo {author} {\bibfnamefont {L.}~\bibnamefont
  {Li}}, \bibinfo {author} {\bibfnamefont {M.}~\bibnamefont {Oshikawa}},\ and\
  \bibinfo {author} {\bibfnamefont {Y.}~\bibnamefont {Zheng}},\ }\bibfield
  {title} {\bibinfo {title} {Decorated defect construction of gapless-spt
  states},\ }\bibfield  {journal} {\bibinfo  {journal} {SciPost Physics}\
  }\textbf {\bibinfo {volume} {17}},\ \href
  {https://doi.org/10.21468/scipostphys.17.1.013}
  {10.21468/scipostphys.17.1.013} (\bibinfo {year} {2024})\BibitemShut
  {NoStop}%
\bibitem [{\citenamefont {Ando}(2024)}]{ando2024gauging}%
  \BibitemOpen
  \bibfield  {author} {\bibinfo {author} {\bibfnamefont {T.}~\bibnamefont
  {Ando}},\ }\href {https://arxiv.org/abs/2402.03566} {\bibinfo {title}
  {Gauging on the lattice and gapped/gapless topological phases}} (\bibinfo
  {year} {2024}),\ \Eprint {https://arxiv.org/abs/2402.03566} {arXiv:2402.03566
  [cond-mat.str-el]} \BibitemShut {NoStop}%
\bibitem [{\citenamefont {Su}\ and\ \citenamefont {Zeng}(2024)}]{Su_2024}%
  \BibitemOpen
  \bibfield  {author} {\bibinfo {author} {\bibfnamefont {L.}~\bibnamefont
  {Su}}\ and\ \bibinfo {author} {\bibfnamefont {M.}~\bibnamefont {Zeng}},\
  }\bibfield  {title} {\bibinfo {title} {Gapless symmetry-protected topological
  phases and generalized deconfined critical points from gauging a finite
  subgroup},\ }\bibfield  {journal} {\bibinfo  {journal} {Physical Review B}\
  }\textbf {\bibinfo {volume} {109}},\ \href
  {https://doi.org/10.1103/physrevb.109.245108} {10.1103/physrevb.109.245108}
  (\bibinfo {year} {2024})\BibitemShut {NoStop}%
\bibitem [{\citenamefont {Yu}\ \emph {et~al.}(2024)\citenamefont {Yu},
  \citenamefont {Yang}, \citenamefont {Lin},\ and\ \citenamefont
  {Jian}}]{Yu_2024}%
  \BibitemOpen
  \bibfield  {author} {\bibinfo {author} {\bibfnamefont {X.-J.}\ \bibnamefont
  {Yu}}, \bibinfo {author} {\bibfnamefont {S.}~\bibnamefont {Yang}}, \bibinfo
  {author} {\bibfnamefont {H.-Q.}\ \bibnamefont {Lin}},\ and\ \bibinfo {author}
  {\bibfnamefont {S.-K.}\ \bibnamefont {Jian}},\ }\bibfield  {title} {\bibinfo
  {title} {Universal entanglement spectrum in one-dimensional gapless symmetry
  protected topological states},\ }\bibfield  {journal} {\bibinfo  {journal}
  {Physical Review Letters}\ }\textbf {\bibinfo {volume} {133}},\ \href
  {https://doi.org/10.1103/physrevlett.133.026601}
  {10.1103/physrevlett.133.026601} (\bibinfo {year} {2024})\BibitemShut
  {NoStop}%
\bibitem [{\citenamefont {Zhang}\ \emph {et~al.}(2024)\citenamefont {Zhang},
  \citenamefont {Li}, \citenamefont {Yang},\ and\ \citenamefont
  {Yu}}]{Zhang_2024}%
  \BibitemOpen
  \bibfield  {author} {\bibinfo {author} {\bibfnamefont {H.-L.}\ \bibnamefont
  {Zhang}}, \bibinfo {author} {\bibfnamefont {H.-Z.}\ \bibnamefont {Li}},
  \bibinfo {author} {\bibfnamefont {S.}~\bibnamefont {Yang}},\ and\ \bibinfo
  {author} {\bibfnamefont {X.-J.}\ \bibnamefont {Yu}},\ }\bibfield  {title}
  {\bibinfo {title} {Quantum phase transition and critical behavior between the
  gapless topological phases},\ }\bibfield  {journal} {\bibinfo  {journal}
  {Physical Review A}\ }\textbf {\bibinfo {volume} {109}},\ \href
  {https://doi.org/10.1103/physreva.109.062226} {10.1103/physreva.109.062226}
  (\bibinfo {year} {2024})\BibitemShut {NoStop}%
\bibitem [{\citenamefont {Myerson-Jain}\ \emph {et~al.}(2024)\citenamefont
  {Myerson-Jain}, \citenamefont {Wu},\ and\ \citenamefont
  {Xu}}]{myersonjain2024pristine}%
  \BibitemOpen
  \bibfield  {author} {\bibinfo {author} {\bibfnamefont {N.}~\bibnamefont
  {Myerson-Jain}}, \bibinfo {author} {\bibfnamefont {X.-C.}\ \bibnamefont
  {Wu}},\ and\ \bibinfo {author} {\bibfnamefont {C.}~\bibnamefont {Xu}},\
  }\href {https://arxiv.org/abs/2405.18481} {\bibinfo {title} {Pristine and
  pseudo-gapped boundaries of the deconfined quantum critical points}}
  (\bibinfo {year} {2024}),\ \Eprint {https://arxiv.org/abs/2405.18481}
  {arXiv:2405.18481 [cond-mat.str-el]} \BibitemShut {NoStop}%
\bibitem [{\citenamefont {Bhardwaj}\ \emph
  {et~al.}(2023{\natexlab{b}})\citenamefont {Bhardwaj}, \citenamefont
  {Bottini}, \citenamefont {Pajer},\ and\ \citenamefont
  {Schafer-Nameki}}]{bhardwaj2023clubsandwich}%
  \BibitemOpen
  \bibfield  {author} {\bibinfo {author} {\bibfnamefont {L.}~\bibnamefont
  {Bhardwaj}}, \bibinfo {author} {\bibfnamefont {L.~E.}\ \bibnamefont
  {Bottini}}, \bibinfo {author} {\bibfnamefont {D.}~\bibnamefont {Pajer}},\
  and\ \bibinfo {author} {\bibfnamefont {S.}~\bibnamefont {Schafer-Nameki}},\
  }\href {https://arxiv.org/abs/2312.17322} {\bibinfo {title} {The club
  sandwich: Gapless phases and phase transitions with non-invertible
  symmetries}} (\bibinfo {year} {2023}{\natexlab{b}}),\ \Eprint
  {https://arxiv.org/abs/2312.17322} {arXiv:2312.17322 [hep-th]} \BibitemShut
  {NoStop}%
\bibitem [{\citenamefont {Bhardwaj}\ \emph
  {et~al.}(2024{\natexlab{c}})\citenamefont {Bhardwaj}, \citenamefont {Pajer},
  \citenamefont {Schafer-Nameki},\ and\ \citenamefont
  {Warman}}]{bhardwaj2024hassediagrams}%
  \BibitemOpen
  \bibfield  {author} {\bibinfo {author} {\bibfnamefont {L.}~\bibnamefont
  {Bhardwaj}}, \bibinfo {author} {\bibfnamefont {D.}~\bibnamefont {Pajer}},
  \bibinfo {author} {\bibfnamefont {S.}~\bibnamefont {Schafer-Nameki}},\ and\
  \bibinfo {author} {\bibfnamefont {A.}~\bibnamefont {Warman}},\ }\href
  {https://arxiv.org/abs/2403.00905} {\bibinfo {title} {Hasse diagrams for
  gapless spt and ssb phases with non-invertible symmetries}} (\bibinfo {year}
  {2024}{\natexlab{c}}),\ \Eprint {https://arxiv.org/abs/2403.00905}
  {arXiv:2403.00905 [cond-mat.str-el]} \BibitemShut {NoStop}%
\bibitem [{\citenamefont {Huang}\ and\ \citenamefont
  {Cheng}(2023)}]{huang2023topological}%
  \BibitemOpen
  \bibfield  {author} {\bibinfo {author} {\bibfnamefont {S.-J.}\ \bibnamefont
  {Huang}}\ and\ \bibinfo {author} {\bibfnamefont {M.}~\bibnamefont {Cheng}},\
  }\href {https://arxiv.org/abs/2310.16878} {\bibinfo {title} {Topological
  holography, quantum criticality, and boundary states}} (\bibinfo {year}
  {2023}),\ \Eprint {https://arxiv.org/abs/2310.16878} {arXiv:2310.16878
  [cond-mat.str-el]} \BibitemShut {NoStop}%
\bibitem [{\citenamefont {Wen}\ \emph {et~al.}(2024)\citenamefont {Wen},
  \citenamefont {Ye},\ and\ \citenamefont {Potter}}]{wen2024fermions}%
  \BibitemOpen
  \bibfield  {author} {\bibinfo {author} {\bibfnamefont {R.}~\bibnamefont
  {Wen}}, \bibinfo {author} {\bibfnamefont {W.}~\bibnamefont {Ye}},\ and\
  \bibinfo {author} {\bibfnamefont {A.~C.}\ \bibnamefont {Potter}},\ }\href
  {https://arxiv.org/abs/2404.19004} {\bibinfo {title} {Topological holography
  for fermions}} (\bibinfo {year} {2024}),\ \Eprint
  {https://arxiv.org/abs/2404.19004} {arXiv:2404.19004 [cond-mat.str-el]}
  \BibitemShut {NoStop}%
\bibitem [{\citenamefont {Bhardwaj}\ \emph
  {et~al.}(2024{\natexlab{d}})\citenamefont {Bhardwaj}, \citenamefont
  {Inamura},\ and\ \citenamefont {Tiwari}}]{bhardwaj2024fermionic}%
  \BibitemOpen
  \bibfield  {author} {\bibinfo {author} {\bibfnamefont {L.}~\bibnamefont
  {Bhardwaj}}, \bibinfo {author} {\bibfnamefont {K.}~\bibnamefont {Inamura}},\
  and\ \bibinfo {author} {\bibfnamefont {A.}~\bibnamefont {Tiwari}},\ }\href
  {https://arxiv.org/abs/2405.09754} {\bibinfo {title} {Fermionic
  non-invertible symmetries in (1+1)d: Gapped and gapless phases, transitions,
  and symmetry tfts}} (\bibinfo {year} {2024}{\natexlab{d}}),\ \Eprint
  {https://arxiv.org/abs/2405.09754} {arXiv:2405.09754 [hep-th]} \BibitemShut
  {NoStop}%
\bibitem [{\citenamefont {Huang}(2024)}]{huang2024fermionic}%
  \BibitemOpen
  \bibfield  {author} {\bibinfo {author} {\bibfnamefont {S.-J.}\ \bibnamefont
  {Huang}},\ }\href {https://arxiv.org/abs/2405.09611} {\bibinfo {title}
  {Fermionic quantum criticality through the lens of topological holography}}
  (\bibinfo {year} {2024}),\ \Eprint {https://arxiv.org/abs/2405.09611}
  {arXiv:2405.09611 [cond-mat.str-el]} \BibitemShut {NoStop}%
\bibitem [{\citenamefont {Zhao}\ \emph {et~al.}(2023)\citenamefont {Zhao},
  \citenamefont {Lou}, \citenamefont {Zhang}, \citenamefont {Hung},
  \citenamefont {Kong},\ and\ \citenamefont {Tian}}]{Zhao_2023}%
  \BibitemOpen
  \bibfield  {author} {\bibinfo {author} {\bibfnamefont {J.}~\bibnamefont
  {Zhao}}, \bibinfo {author} {\bibfnamefont {J.-Q.}\ \bibnamefont {Lou}},
  \bibinfo {author} {\bibfnamefont {Z.-H.}\ \bibnamefont {Zhang}}, \bibinfo
  {author} {\bibfnamefont {L.-Y.}\ \bibnamefont {Hung}}, \bibinfo {author}
  {\bibfnamefont {L.}~\bibnamefont {Kong}},\ and\ \bibinfo {author}
  {\bibfnamefont {Y.}~\bibnamefont {Tian}},\ }\bibfield  {title} {\bibinfo
  {title} {String condensations in $3+1d$ and lagrangian algebras},\ }\href
  {https://doi.org/10.4310/atmp.2023.v27.n2.a5} {\bibfield  {journal} {\bibinfo
   {journal} {Advances in Theoretical and Mathematical Physics}\ }\textbf
  {\bibinfo {volume} {27}},\ \bibinfo {pages} {583–622} (\bibinfo {year}
  {2023})}\BibitemShut {NoStop}%
\bibitem [{\citenamefont {Kong}\ \emph {et~al.}(2024)\citenamefont {Kong},
  \citenamefont {Zhang}, \citenamefont {Zhao},\ and\ \citenamefont
  {Zheng}}]{kong2024highercondensationtheory}%
  \BibitemOpen
  \bibfield  {author} {\bibinfo {author} {\bibfnamefont {L.}~\bibnamefont
  {Kong}}, \bibinfo {author} {\bibfnamefont {Z.-H.}\ \bibnamefont {Zhang}},
  \bibinfo {author} {\bibfnamefont {J.}~\bibnamefont {Zhao}},\ and\ \bibinfo
  {author} {\bibfnamefont {H.}~\bibnamefont {Zheng}},\ }\href
  {https://arxiv.org/abs/2403.07813} {\bibinfo {title} {Higher condensation
  theory}} (\bibinfo {year} {2024}),\ \Eprint
  {https://arxiv.org/abs/2403.07813} {arXiv:2403.07813 [cond-mat.str-el]}
  \BibitemShut {NoStop}%
\bibitem [{\citenamefont {Décoppet}\ and\ \citenamefont
  {Xu}(2024)}]{D_coppet_2024}%
  \BibitemOpen
  \bibfield  {author} {\bibinfo {author} {\bibfnamefont {T.~D.}\ \bibnamefont
  {Décoppet}}\ and\ \bibinfo {author} {\bibfnamefont {H.}~\bibnamefont {Xu}},\
  }\bibfield  {title} {\bibinfo {title} {Local modules in braided monoidal
  2-categories},\ }\bibfield  {journal} {\bibinfo  {journal} {Journal of
  Mathematical Physics}\ }\textbf {\bibinfo {volume} {65}},\ \href
  {https://doi.org/10.1063/5.0172042} {10.1063/5.0172042} (\bibinfo {year}
  {2024})\BibitemShut {NoStop}%
\bibitem [{\citenamefont {Naidu}(2011)}]{naidu2011crossed}%
  \BibitemOpen
  \bibfield  {author} {\bibinfo {author} {\bibfnamefont {D.}~\bibnamefont
  {Naidu}},\ }\href {https://arxiv.org/abs/1111.5246} {\bibinfo {title}
  {Crossed pointed categories and their equivariantizations}} (\bibinfo {year}
  {2011}),\ \Eprint {https://arxiv.org/abs/1111.5246} {arXiv:1111.5246
  [math.QA]} \BibitemShut {NoStop}%
\bibitem [{\citenamefont {Zhu}\ \emph {et~al.}(2022)\citenamefont {Zhu},
  \citenamefont {Jochym-O’Connor},\ and\ \citenamefont {Dua}}]{Zhu_2022}%
  \BibitemOpen
  \bibfield  {author} {\bibinfo {author} {\bibfnamefont {G.}~\bibnamefont
  {Zhu}}, \bibinfo {author} {\bibfnamefont {T.}~\bibnamefont
  {Jochym-O’Connor}},\ and\ \bibinfo {author} {\bibfnamefont
  {A.}~\bibnamefont {Dua}},\ }\bibfield  {title} {\bibinfo {title} {Topological
  order, quantum codes, and quantum computation on fractal geometries},\
  }\bibfield  {journal} {\bibinfo  {journal} {PRX Quantum}\ }\textbf {\bibinfo
  {volume} {3}},\ \href {https://doi.org/10.1103/prxquantum.3.030338}
  {10.1103/prxquantum.3.030338} (\bibinfo {year} {2022})\BibitemShut {NoStop}%
\bibitem [{\citenamefont {Song}\ and\ \citenamefont
  {Zhu}(2024)}]{song2024magic}%
  \BibitemOpen
  \bibfield  {author} {\bibinfo {author} {\bibfnamefont {Z.}~\bibnamefont
  {Song}}\ and\ \bibinfo {author} {\bibfnamefont {G.}~\bibnamefont {Zhu}},\
  }\href {https://arxiv.org/abs/2404.05033} {\bibinfo {title} {Magic boundaries
  of 3d color codes}} (\bibinfo {year} {2024}),\ \Eprint
  {https://arxiv.org/abs/2404.05033} {arXiv:2404.05033 [quant-ph]} \BibitemShut
  {NoStop}%
\bibitem [{\citenamefont {Barkeshli}\ \emph {et~al.}(2019)\citenamefont
  {Barkeshli}, \citenamefont {Bonderson}, \citenamefont {Cheng},\ and\
  \citenamefont {Wang}}]{barkeshli2019symmetry}%
  \BibitemOpen
  \bibfield  {author} {\bibinfo {author} {\bibfnamefont {M.}~\bibnamefont
  {Barkeshli}}, \bibinfo {author} {\bibfnamefont {P.}~\bibnamefont
  {Bonderson}}, \bibinfo {author} {\bibfnamefont {M.}~\bibnamefont {Cheng}},\
  and\ \bibinfo {author} {\bibfnamefont {Z.}~\bibnamefont {Wang}},\ }\bibfield
  {title} {\bibinfo {title} {Symmetry fractionalization, defects, and gauging
  of topological phases},\ }\href@noop {} {\bibfield  {journal} {\bibinfo
  {journal} {Physical Review B}\ }\textbf {\bibinfo {volume} {100}},\ \bibinfo
  {pages} {115147} (\bibinfo {year} {2019})}\BibitemShut {NoStop}%
\bibitem [{\citenamefont {Lan}\ \emph {et~al.}(2016{\natexlab{b}})\citenamefont
  {Lan}, \citenamefont {Kong},\ and\ \citenamefont
  {Wen}}]{lan2016classification}%
  \BibitemOpen
  \bibfield  {author} {\bibinfo {author} {\bibfnamefont {T.}~\bibnamefont
  {Lan}}, \bibinfo {author} {\bibfnamefont {L.}~\bibnamefont {Kong}},\ and\
  \bibinfo {author} {\bibfnamefont {X.-G.}\ \bibnamefont {Wen}},\ }\bibfield
  {title} {\bibinfo {title} {Classification of 2+ 1d topological orders and spt
  orders for bosonic and fermionic systems with on-site symmetries},\
  }\href@noop {} {\bibfield  {journal} {\bibinfo  {journal} {arXiv preprint
  arXiv:1602.05946}\ } (\bibinfo {year} {2016}{\natexlab{b}})}\BibitemShut
  {NoStop}%
\bibitem [{\citenamefont {Kong}\ \emph
  {et~al.}(2020{\natexlab{c}})\citenamefont {Kong}, \citenamefont {Lan},
  \citenamefont {Wen}, \citenamefont {Zhang},\ and\ \citenamefont
  {Zheng}}]{kong2020classification}%
  \BibitemOpen
  \bibfield  {author} {\bibinfo {author} {\bibfnamefont {L.}~\bibnamefont
  {Kong}}, \bibinfo {author} {\bibfnamefont {T.}~\bibnamefont {Lan}}, \bibinfo
  {author} {\bibfnamefont {X.-G.}\ \bibnamefont {Wen}}, \bibinfo {author}
  {\bibfnamefont {Z.-H.}\ \bibnamefont {Zhang}},\ and\ \bibinfo {author}
  {\bibfnamefont {H.}~\bibnamefont {Zheng}},\ }\bibfield  {title} {\bibinfo
  {title} {Classification of topological phases with finite internal symmetries
  in all dimensions},\ }\href@noop {} {\bibfield  {journal} {\bibinfo
  {journal} {Journal of High Energy Physics}\ }\textbf {\bibinfo {volume}
  {2020}},\ \bibinfo {pages} {1} (\bibinfo {year}
  {2020}{\natexlab{c}})}\BibitemShut {NoStop}%
\bibitem [{\citenamefont {Wen}()}]{Wen_algebra}%
  \BibitemOpen
  \bibfield  {author} {\bibinfo {author} {\bibfnamefont {R.}~\bibnamefont
  {Wen}},\ }\bibfield  {title} {\bibinfo {title} {Algebras and modules in 3+1d
  dijkgraaf-witten theories. in preparation.},\ }\href@noop {} {\ }\BibitemShut
  {NoStop}%
\bibitem [{\citenamefont {Bhardwaj}\ and\ \citenamefont
  {Schafer-Nameki}(2023)}]{bhardwaj2023generalizedcharges}%
  \BibitemOpen
  \bibfield  {author} {\bibinfo {author} {\bibfnamefont {L.}~\bibnamefont
  {Bhardwaj}}\ and\ \bibinfo {author} {\bibfnamefont {S.}~\bibnamefont
  {Schafer-Nameki}},\ }\href {https://arxiv.org/abs/2305.17159} {\bibinfo
  {title} {Generalized charges, part ii: Non-invertible symmetries and the
  symmetry tft}} (\bibinfo {year} {2023}),\ \Eprint
  {https://arxiv.org/abs/2305.17159} {arXiv:2305.17159 [hep-th]} \BibitemShut
  {NoStop}%
\bibitem [{\citenamefont {Bhardwaj}\ and\ \citenamefont
  {Schäfer-Nameki}(2024)}]{Bhardwaj_2024}%
  \BibitemOpen
  \bibfield  {author} {\bibinfo {author} {\bibfnamefont {L.}~\bibnamefont
  {Bhardwaj}}\ and\ \bibinfo {author} {\bibfnamefont {S.}~\bibnamefont
  {Schäfer-Nameki}},\ }\bibfield  {title} {\bibinfo {title} {Generalized
  charges, part i: Invertible symmetries and higher representations},\
  }\bibfield  {journal} {\bibinfo  {journal} {SciPost Physics}\ }\textbf
  {\bibinfo {volume} {16}},\ \href
  {https://doi.org/10.21468/scipostphys.16.4.093}
  {10.21468/scipostphys.16.4.093} (\bibinfo {year} {2024})\BibitemShut
  {NoStop}%
\bibitem [{\citenamefont {Bartsch}\ \emph {et~al.}(2024)\citenamefont
  {Bartsch}, \citenamefont {Bullimore}, \citenamefont {Ferrari},\ and\
  \citenamefont {Pearson}}]{Bartsch_2024}%
  \BibitemOpen
  \bibfield  {author} {\bibinfo {author} {\bibfnamefont {T.}~\bibnamefont
  {Bartsch}}, \bibinfo {author} {\bibfnamefont {M.}~\bibnamefont {Bullimore}},
  \bibinfo {author} {\bibfnamefont {A.~E.~V.}\ \bibnamefont {Ferrari}},\ and\
  \bibinfo {author} {\bibfnamefont {J.}~\bibnamefont {Pearson}},\ }\bibfield
  {title} {\bibinfo {title} {Non-invertible symmetries and higher
  representation theory i},\ }\bibfield  {journal} {\bibinfo  {journal}
  {SciPost Physics}\ }\textbf {\bibinfo {volume} {17}},\ \href
  {https://doi.org/10.21468/scipostphys.17.1.015}
  {10.21468/scipostphys.17.1.015} (\bibinfo {year} {2024})\BibitemShut
  {NoStop}%
\bibitem [{\citenamefont {Bartsch}\ \emph
  {et~al.}(2023{\natexlab{a}})\citenamefont {Bartsch}, \citenamefont
  {Bullimore}, \citenamefont {Ferrari},\ and\ \citenamefont
  {Pearson}}]{bartsch2023noninvertible}%
  \BibitemOpen
  \bibfield  {author} {\bibinfo {author} {\bibfnamefont {T.}~\bibnamefont
  {Bartsch}}, \bibinfo {author} {\bibfnamefont {M.}~\bibnamefont {Bullimore}},
  \bibinfo {author} {\bibfnamefont {A.~E.~V.}\ \bibnamefont {Ferrari}},\ and\
  \bibinfo {author} {\bibfnamefont {J.}~\bibnamefont {Pearson}},\ }\href
  {https://arxiv.org/abs/2212.07393} {\bibinfo {title} {Non-invertible
  symmetries and higher representation theory ii}} (\bibinfo {year}
  {2023}{\natexlab{a}}),\ \Eprint {https://arxiv.org/abs/2212.07393}
  {arXiv:2212.07393 [hep-th]} \BibitemShut {NoStop}%
\bibitem [{\citenamefont {Bartsch}\ \emph
  {et~al.}(2023{\natexlab{b}})\citenamefont {Bartsch}, \citenamefont
  {Bullimore},\ and\ \citenamefont {Grigoletto}}]{bartsch2023representation}%
  \BibitemOpen
  \bibfield  {author} {\bibinfo {author} {\bibfnamefont {T.}~\bibnamefont
  {Bartsch}}, \bibinfo {author} {\bibfnamefont {M.}~\bibnamefont {Bullimore}},\
  and\ \bibinfo {author} {\bibfnamefont {A.}~\bibnamefont {Grigoletto}},\
  }\href {https://arxiv.org/abs/2305.17165} {\bibinfo {title} {Representation
  theory for categorical symmetries}} (\bibinfo {year} {2023}{\natexlab{b}}),\
  \Eprint {https://arxiv.org/abs/2305.17165} {arXiv:2305.17165 [hep-th]}
  \BibitemShut {NoStop}%
\bibitem [{\citenamefont {Kong}\ and\ \citenamefont
  {Wen}(2014)}]{kong2014braided}%
  \BibitemOpen
  \bibfield  {author} {\bibinfo {author} {\bibfnamefont {L.}~\bibnamefont
  {Kong}}\ and\ \bibinfo {author} {\bibfnamefont {X.-G.}\ \bibnamefont {Wen}},\
  }\href {https://arxiv.org/abs/1405.5858} {\bibinfo {title} {Braided fusion
  categories, gravitational anomalies, and the mathematical framework for
  topological orders in any dimensions}} (\bibinfo {year} {2014}),\ \Eprint
  {https://arxiv.org/abs/1405.5858} {arXiv:1405.5858 [cond-mat.str-el]}
  \BibitemShut {NoStop}%
\bibitem [{\citenamefont {Kong}\ \emph {et~al.}(2022)\citenamefont {Kong},
  \citenamefont {Wen},\ and\ \citenamefont {Zheng}}]{Kong_2022}%
  \BibitemOpen
  \bibfield  {author} {\bibinfo {author} {\bibfnamefont {L.}~\bibnamefont
  {Kong}}, \bibinfo {author} {\bibfnamefont {X.-G.}\ \bibnamefont {Wen}},\ and\
  \bibinfo {author} {\bibfnamefont {H.}~\bibnamefont {Zheng}},\ }\bibfield
  {title} {\bibinfo {title} {One dimensional gapped quantum phases and enriched
  fusion categories},\ }\bibfield  {journal} {\bibinfo  {journal} {Journal of
  High Energy Physics}\ }\textbf {\bibinfo {volume} {2022}},\ \href
  {https://doi.org/10.1007/jhep03(2022)022} {10.1007/jhep03(2022)022} (\bibinfo
  {year} {2022})\BibitemShut {NoStop}%
\bibitem [{\citenamefont {Luo}(2023)}]{Luo_2023}%
  \BibitemOpen
  \bibfield  {author} {\bibinfo {author} {\bibfnamefont {Z.-X.}\ \bibnamefont
  {Luo}},\ }\bibfield  {title} {\bibinfo {title} {Gapped boundaries of
  3+1-dimensional topological order},\ }\bibfield  {journal} {\bibinfo
  {journal} {Physical Review B}\ }\textbf {\bibinfo {volume} {107}},\ \href
  {https://doi.org/10.1103/physrevb.107.125425} {10.1103/physrevb.107.125425}
  (\bibinfo {year} {2023})\BibitemShut {NoStop}%
\bibitem [{\citenamefont {Ji}\ \emph {et~al.}(2023)\citenamefont {Ji},
  \citenamefont {Tantivasadakarn},\ and\ \citenamefont {Xu}}]{Ji_2023}%
  \BibitemOpen
  \bibfield  {author} {\bibinfo {author} {\bibfnamefont {W.}~\bibnamefont
  {Ji}}, \bibinfo {author} {\bibfnamefont {N.}~\bibnamefont
  {Tantivasadakarn}},\ and\ \bibinfo {author} {\bibfnamefont {C.}~\bibnamefont
  {Xu}},\ }\bibfield  {title} {\bibinfo {title} {Boundary states of three
  dimensional topological order and the deconfined quantum critical point},\
  }\bibfield  {journal} {\bibinfo  {journal} {SciPost Physics}\ }\textbf
  {\bibinfo {volume} {15}},\ \href
  {https://doi.org/10.21468/scipostphys.15.6.231}
  {10.21468/scipostphys.15.6.231} (\bibinfo {year} {2023})\BibitemShut
  {NoStop}%
\bibitem [{\citenamefont {Barkeshli}\ \emph {et~al.}(2024)\citenamefont
  {Barkeshli}, \citenamefont {Chen}, \citenamefont {Hsin},\ and\ \citenamefont
  {Kobayashi}}]{barkeshli2024higher}%
  \BibitemOpen
  \bibfield  {author} {\bibinfo {author} {\bibfnamefont {M.}~\bibnamefont
  {Barkeshli}}, \bibinfo {author} {\bibfnamefont {Y.-A.}\ \bibnamefont {Chen}},
  \bibinfo {author} {\bibfnamefont {P.-S.}\ \bibnamefont {Hsin}},\ and\
  \bibinfo {author} {\bibfnamefont {R.}~\bibnamefont {Kobayashi}},\ }\bibfield
  {title} {\bibinfo {title} {Higher-group symmetry in finite gauge theory and
  stabilizer codes},\ }\href@noop {} {\bibfield  {journal} {\bibinfo  {journal}
  {SciPost Physics}\ }\textbf {\bibinfo {volume} {16}},\ \bibinfo {pages} {089}
  (\bibinfo {year} {2024})}\BibitemShut {NoStop}%
\bibitem [{\citenamefont {Barkeshli}\ \emph {et~al.}(2023)\citenamefont
  {Barkeshli}, \citenamefont {Chen}, \citenamefont {Huang}, \citenamefont
  {Kobayashi}, \citenamefont {Tantivasadakarn},\ and\ \citenamefont
  {Zhu}}]{barkeshli2023codimension}%
  \BibitemOpen
  \bibfield  {author} {\bibinfo {author} {\bibfnamefont {M.}~\bibnamefont
  {Barkeshli}}, \bibinfo {author} {\bibfnamefont {Y.-A.}\ \bibnamefont {Chen}},
  \bibinfo {author} {\bibfnamefont {S.-J.}\ \bibnamefont {Huang}}, \bibinfo
  {author} {\bibfnamefont {R.}~\bibnamefont {Kobayashi}}, \bibinfo {author}
  {\bibfnamefont {N.}~\bibnamefont {Tantivasadakarn}},\ and\ \bibinfo {author}
  {\bibfnamefont {G.}~\bibnamefont {Zhu}},\ }\bibfield  {title} {\bibinfo
  {title} {Codimension-2 defects and higher symmetries in (3+ 1) d topological
  phases},\ }\href@noop {} {\bibfield  {journal} {\bibinfo  {journal} {SciPost
  Physics}\ }\textbf {\bibinfo {volume} {14}},\ \bibinfo {pages} {065}
  (\bibinfo {year} {2023})}\BibitemShut {NoStop}%
\bibitem [{\citenamefont {Tantivasadakarn}\ and\ \citenamefont
  {Chen}(2024)}]{tantivasadakarn2024string}%
  \BibitemOpen
  \bibfield  {author} {\bibinfo {author} {\bibfnamefont {N.}~\bibnamefont
  {Tantivasadakarn}}\ and\ \bibinfo {author} {\bibfnamefont {X.}~\bibnamefont
  {Chen}},\ }\bibfield  {title} {\bibinfo {title} {String operators for
  cheshire strings in topological phases},\ }\href@noop {} {\bibfield
  {journal} {\bibinfo  {journal} {Physical Review B}\ }\textbf {\bibinfo
  {volume} {109}},\ \bibinfo {pages} {165149} (\bibinfo {year}
  {2024})}\BibitemShut {NoStop}%
\bibitem [{\citenamefont {Douglas}\ and\ \citenamefont
  {Reutter}(2018)}]{douglas2018fusion2}%
  \BibitemOpen
  \bibfield  {author} {\bibinfo {author} {\bibfnamefont {C.~L.}\ \bibnamefont
  {Douglas}}\ and\ \bibinfo {author} {\bibfnamefont {D.~J.}\ \bibnamefont
  {Reutter}},\ }\href {https://arxiv.org/abs/1812.11933} {\bibinfo {title}
  {Fusion 2-categories and a state-sum invariant for 4-manifolds}} (\bibinfo
  {year} {2018}),\ \Eprint {https://arxiv.org/abs/1812.11933} {arXiv:1812.11933
  [math.QA]} \BibitemShut {NoStop}%
\bibitem [{\citenamefont {Kong}\ \emph
  {et~al.}(2020{\natexlab{d}})\citenamefont {Kong}, \citenamefont {Tian},\ and\
  \citenamefont {Zhang}}]{Kong_2020defect}%
  \BibitemOpen
  \bibfield  {author} {\bibinfo {author} {\bibfnamefont {L.}~\bibnamefont
  {Kong}}, \bibinfo {author} {\bibfnamefont {Y.}~\bibnamefont {Tian}},\ and\
  \bibinfo {author} {\bibfnamefont {Z.-H.}\ \bibnamefont {Zhang}},\ }\bibfield
  {title} {\bibinfo {title} {Defects in the 3-dimensional toric code model form
  a braided fusion 2-category},\ }\bibfield  {journal} {\bibinfo  {journal}
  {Journal of High Energy Physics}\ }\textbf {\bibinfo {volume} {2020}},\ \href
  {https://doi.org/10.1007/jhep12(2020)078} {10.1007/jhep12(2020)078} (\bibinfo
  {year} {2020}{\natexlab{d}})\BibitemShut {NoStop}%
\bibitem [{\citenamefont {Wang}\ and\ \citenamefont {Levin}(2014)}]{Wang_2014}%
  \BibitemOpen
  \bibfield  {author} {\bibinfo {author} {\bibfnamefont {C.}~\bibnamefont
  {Wang}}\ and\ \bibinfo {author} {\bibfnamefont {M.}~\bibnamefont {Levin}},\
  }\bibfield  {title} {\bibinfo {title} {Braiding statistics of loop
  excitations in three dimensions},\ }\bibfield  {journal} {\bibinfo  {journal}
  {Physical Review Letters}\ }\textbf {\bibinfo {volume} {113}},\ \href
  {https://doi.org/10.1103/physrevlett.113.080403}
  {10.1103/physrevlett.113.080403} (\bibinfo {year} {2014})\BibitemShut
  {NoStop}%
\bibitem [{\citenamefont {Wang}\ and\ \citenamefont {Wen}(2015)}]{Wang_2015}%
  \BibitemOpen
  \bibfield  {author} {\bibinfo {author} {\bibfnamefont {J.~C.}\ \bibnamefont
  {Wang}}\ and\ \bibinfo {author} {\bibfnamefont {X.-G.}\ \bibnamefont {Wen}},\
  }\bibfield  {title} {\bibinfo {title} {Non-abelian string and particle
  braiding in topological order: Modular sl (3, z) representation and (3+
  1)-dimensional twisted gauge theory},\ }\href@noop {} {\bibfield  {journal}
  {\bibinfo  {journal} {Physical Review B}\ }\textbf {\bibinfo {volume} {91}},\
  \bibinfo {pages} {035134} (\bibinfo {year} {2015})}\BibitemShut {NoStop}%
\bibitem [{\citenamefont {Wan}\ \emph {et~al.}(2015)\citenamefont {Wan},
  \citenamefont {Wang},\ and\ \citenamefont {He}}]{Wan_2015}%
  \BibitemOpen
  \bibfield  {author} {\bibinfo {author} {\bibfnamefont {Y.}~\bibnamefont
  {Wan}}, \bibinfo {author} {\bibfnamefont {J.~C.}\ \bibnamefont {Wang}},\ and\
  \bibinfo {author} {\bibfnamefont {H.}~\bibnamefont {He}},\ }\bibfield
  {title} {\bibinfo {title} {Twisted gauge theory model of topological phases
  in three dimensions},\ }\bibfield  {journal} {\bibinfo  {journal} {Physical
  Review B}\ }\textbf {\bibinfo {volume} {92}},\ \href
  {https://doi.org/10.1103/physrevb.92.045101} {10.1103/physrevb.92.045101}
  (\bibinfo {year} {2015})\BibitemShut {NoStop}%
\bibitem [{\citenamefont {Lin}\ and\ \citenamefont {Levin}(2015)}]{Lin_2015}%
  \BibitemOpen
  \bibfield  {author} {\bibinfo {author} {\bibfnamefont {C.-H.}\ \bibnamefont
  {Lin}}\ and\ \bibinfo {author} {\bibfnamefont {M.}~\bibnamefont {Levin}},\
  }\bibfield  {title} {\bibinfo {title} {Loop braiding statistics in exactly
  soluble three-dimensional lattice models},\ }\bibfield  {journal} {\bibinfo
  {journal} {Physical Review B}\ }\textbf {\bibinfo {volume} {92}},\ \href
  {https://doi.org/10.1103/physrevb.92.035115} {10.1103/physrevb.92.035115}
  (\bibinfo {year} {2015})\BibitemShut {NoStop}%
\bibitem [{\citenamefont {Putrov}\ \emph {et~al.}(2017)\citenamefont {Putrov},
  \citenamefont {Wang},\ and\ \citenamefont {Yau}}]{Putrov_2017}%
  \BibitemOpen
  \bibfield  {author} {\bibinfo {author} {\bibfnamefont {P.}~\bibnamefont
  {Putrov}}, \bibinfo {author} {\bibfnamefont {J.}~\bibnamefont {Wang}},\ and\
  \bibinfo {author} {\bibfnamefont {S.-T.}\ \bibnamefont {Yau}},\ }\bibfield
  {title} {\bibinfo {title} {Braiding statistics and link invariants of
  bosonic/fermionic topological quantum matter in 2+1 and 3+1 dimensions},\
  }\href {https://doi.org/10.1016/j.aop.2017.06.019} {\bibfield  {journal}
  {\bibinfo  {journal} {Annals of Physics}\ }\textbf {\bibinfo {volume}
  {384}},\ \bibinfo {pages} {254–287} (\bibinfo {year} {2017})}\BibitemShut
  {NoStop}%
\bibitem [{\citenamefont {Wang}\ \emph {et~al.}(2019)\citenamefont {Wang},
  \citenamefont {Cheng}, \citenamefont {Wang},\ and\ \citenamefont
  {Gu}}]{Wang_2019}%
  \BibitemOpen
  \bibfield  {author} {\bibinfo {author} {\bibfnamefont {Q.-R.}\ \bibnamefont
  {Wang}}, \bibinfo {author} {\bibfnamefont {M.}~\bibnamefont {Cheng}},
  \bibinfo {author} {\bibfnamefont {C.}~\bibnamefont {Wang}},\ and\ \bibinfo
  {author} {\bibfnamefont {Z.-C.}\ \bibnamefont {Gu}},\ }\bibfield  {title}
  {\bibinfo {title} {Topological quantum field theory for abelian topological
  phases and loop braiding statistics in 3+1-dimensions},\ }\bibfield
  {journal} {\bibinfo  {journal} {Physical Review B}\ }\textbf {\bibinfo
  {volume} {99}},\ \href {https://doi.org/10.1103/physrevb.99.235137}
  {10.1103/physrevb.99.235137} (\bibinfo {year} {2019})\BibitemShut {NoStop}%
\bibitem [{\citenamefont {Cheng}\ \emph {et~al.}(2018)\citenamefont {Cheng},
  \citenamefont {Tantivasadakarn},\ and\ \citenamefont {Wang}}]{cheng2018loop}%
  \BibitemOpen
  \bibfield  {author} {\bibinfo {author} {\bibfnamefont {M.}~\bibnamefont
  {Cheng}}, \bibinfo {author} {\bibfnamefont {N.}~\bibnamefont
  {Tantivasadakarn}},\ and\ \bibinfo {author} {\bibfnamefont {C.}~\bibnamefont
  {Wang}},\ }\bibfield  {title} {\bibinfo {title} {Loop braiding statistics and
  interacting fermionic symmetry-protected topological phases in three
  dimensions},\ }\href@noop {} {\bibfield  {journal} {\bibinfo  {journal}
  {Physical Review X}\ }\textbf {\bibinfo {volume} {8}},\ \bibinfo {pages}
  {011054} (\bibinfo {year} {2018})}\BibitemShut {NoStop}%
\bibitem [{\citenamefont
  {Tantivasadakarn}(2017)}]{tantivasadakarn2017dimensional}%
  \BibitemOpen
  \bibfield  {author} {\bibinfo {author} {\bibfnamefont {N.}~\bibnamefont
  {Tantivasadakarn}},\ }\bibfield  {title} {\bibinfo {title} {Dimensional
  reduction and topological invariants of symmetry-protected topological
  phases},\ }\href@noop {} {\bibfield  {journal} {\bibinfo  {journal} {Physical
  Review B}\ }\textbf {\bibinfo {volume} {96}},\ \bibinfo {pages} {195101}
  (\bibinfo {year} {2017})}\BibitemShut {NoStop}%
\bibitem [{\citenamefont {Johnson-Freyd}(2020)}]{johnsonfreyd2020}%
  \BibitemOpen
  \bibfield  {author} {\bibinfo {author} {\bibfnamefont {T.}~\bibnamefont
  {Johnson-Freyd}},\ }\href {https://arxiv.org/abs/2011.11165} {\bibinfo
  {title} {(3+1)d topological orders with only a $\mathbb{Z}_2$-charged
  particle}} (\bibinfo {year} {2020}),\ \Eprint
  {https://arxiv.org/abs/2011.11165} {arXiv:2011.11165 [math.QA]} \BibitemShut
  {NoStop}%
\bibitem [{\citenamefont {Ferrer}(2021)}]{Giovanni_BF2}%
  \BibitemOpen
  \bibfield  {author} {\bibinfo {author} {\bibfnamefont {G.}~\bibnamefont
  {Ferrer}},\ }\bibfield  {title} {\bibinfo {title} {Braided 2 categories and
  the drinfel’d center},\ }\href
  {https://www.asc.ohio-state.edu/ferrer.40/pdfs/Braided.pdf} {\  (\bibinfo
  {year} {2021})}\BibitemShut {NoStop}%
\bibitem [{\citenamefont {Kong}\ \emph
  {et~al.}(2020{\natexlab{e}})\citenamefont {Kong}, \citenamefont {Tian},\ and\
  \citenamefont {Zhou}}]{Kong_2020center}%
  \BibitemOpen
  \bibfield  {author} {\bibinfo {author} {\bibfnamefont {L.}~\bibnamefont
  {Kong}}, \bibinfo {author} {\bibfnamefont {Y.}~\bibnamefont {Tian}},\ and\
  \bibinfo {author} {\bibfnamefont {S.}~\bibnamefont {Zhou}},\ }\bibfield
  {title} {\bibinfo {title} {The center of monoidal 2-categories in 3+1d
  dijkgraaf-witten theory},\ }\href {https://doi.org/10.1016/j.aim.2019.106928}
  {\bibfield  {journal} {\bibinfo  {journal} {Advances in Mathematics}\
  }\textbf {\bibinfo {volume} {360}},\ \bibinfo {pages} {106928} (\bibinfo
  {year} {2020}{\natexlab{e}})}\BibitemShut {NoStop}%
\bibitem [{\citenamefont {Chen}\ \emph {et~al.}(2016)\citenamefont {Chen},
  \citenamefont {Tiwari},\ and\ \citenamefont {Ryu}}]{Chen_2016}%
  \BibitemOpen
  \bibfield  {author} {\bibinfo {author} {\bibfnamefont {X.}~\bibnamefont
  {Chen}}, \bibinfo {author} {\bibfnamefont {A.}~\bibnamefont {Tiwari}},\ and\
  \bibinfo {author} {\bibfnamefont {S.}~\bibnamefont {Ryu}},\ }\bibfield
  {title} {\bibinfo {title} {Bulk-boundary correspondence in (3+1)-dimensional
  topological phases},\ }\bibfield  {journal} {\bibinfo  {journal} {Physical
  Review B}\ }\textbf {\bibinfo {volume} {94}},\ \href
  {https://doi.org/10.1103/physrevb.94.045113} {10.1103/physrevb.94.045113}
  (\bibinfo {year} {2016})\BibitemShut {NoStop}%
\bibitem [{\citenamefont {Else}\ and\ \citenamefont {Nayak}(2017)}]{Else_2017}%
  \BibitemOpen
  \bibfield  {author} {\bibinfo {author} {\bibfnamefont {D.~V.}\ \bibnamefont
  {Else}}\ and\ \bibinfo {author} {\bibfnamefont {C.}~\bibnamefont {Nayak}},\
  }\bibfield  {title} {\bibinfo {title} {Cheshire charge in (3+1)-dimensional
  topological phases},\ }\bibfield  {journal} {\bibinfo  {journal} {Physical
  Review B}\ }\textbf {\bibinfo {volume} {96}},\ \href
  {https://doi.org/10.1103/physrevb.96.045136} {10.1103/physrevb.96.045136}
  (\bibinfo {year} {2017})\BibitemShut {NoStop}%
\bibitem [{\citenamefont {Ji}\ and\ \citenamefont
  {Chen}(2024)}]{ji2024topological}%
  \BibitemOpen
  \bibfield  {author} {\bibinfo {author} {\bibfnamefont {W.}~\bibnamefont
  {Ji}}\ and\ \bibinfo {author} {\bibfnamefont {X.}~\bibnamefont {Chen}},\
  }\href {https://arxiv.org/abs/2407.02488} {\bibinfo {title} {Topological
  defects of 2+1d systems from line excitations in 3+1d bulk}} (\bibinfo {year}
  {2024}),\ \Eprint {https://arxiv.org/abs/2407.02488} {arXiv:2407.02488
  [cond-mat.str-el]} \BibitemShut {NoStop}%
\bibitem [{\citenamefont {Bhardwaj}\ \emph
  {et~al.}(2024{\natexlab{e}})\citenamefont {Bhardwaj}, \citenamefont {Pajer},
  \citenamefont {Schafer-Nameki}, \citenamefont {Tiwari}, \citenamefont
  {Warman},\ and\ \citenamefont {Wu}}]{bhardwaj2024gapped}%
  \BibitemOpen
  \bibfield  {author} {\bibinfo {author} {\bibfnamefont {L.}~\bibnamefont
  {Bhardwaj}}, \bibinfo {author} {\bibfnamefont {D.}~\bibnamefont {Pajer}},
  \bibinfo {author} {\bibfnamefont {S.}~\bibnamefont {Schafer-Nameki}},
  \bibinfo {author} {\bibfnamefont {A.}~\bibnamefont {Tiwari}}, \bibinfo
  {author} {\bibfnamefont {A.}~\bibnamefont {Warman}},\ and\ \bibinfo {author}
  {\bibfnamefont {J.}~\bibnamefont {Wu}},\ }\bibfield  {title} {\bibinfo
  {title} {Gapped phases in (2+ 1) d with non-invertible symmetries: Part i},\
  }\href@noop {} {\bibfield  {journal} {\bibinfo  {journal} {arXiv preprint
  arXiv:2408.05266}\ } (\bibinfo {year} {2024}{\natexlab{e}})}\BibitemShut
  {NoStop}%
\bibitem [{\citenamefont {Davydov}\ and\ \citenamefont
  {Simmons}(2016)}]{davydov2016algebras}%
  \BibitemOpen
  \bibfield  {author} {\bibinfo {author} {\bibfnamefont {A.}~\bibnamefont
  {Davydov}}\ and\ \bibinfo {author} {\bibfnamefont {D.}~\bibnamefont
  {Simmons}},\ }\href {https://arxiv.org/abs/1603.04650} {\bibinfo {title} {On
  lagrangian algebras in group-theoretical braided fusion categories}}
  (\bibinfo {year} {2016}),\ \Eprint {https://arxiv.org/abs/1603.04650}
  {arXiv:1603.04650 [math.QA]} \BibitemShut {NoStop}%
\end{thebibliography}%

\newpage
\appendix
\section{Condensable algebras in braided monoidal 2-categories\label{app:condensable algebra}}
We review algebras in braided monoidal 2-categories here. We follow~\cite{D_coppet_2024}. Let $\cc{C}$ be a braided fusion 2-category, with monoidal product denoted by $\Box$, and monoidal unit $\bb{1}$. An algebra in $\cc{C}$ is an object $\eA$ with a product: $\mu: \eA\Box \eA\to \eA$, and a unit $i: \bb{1}\to \eA$, such that there are 2-isomorphisms $\alpha,\lambda,\rho$, defined as
\begin{align}
    \begin{tikzcd}[ampersand replacement=\&]
	{\eA\Box \eA\Box \eA} \& {\eA\Box \eA} \& \eA \&\& \eA \&\& {\eA\Box \eA} \\
	{\eA\Box \eA} \& \eA \&\& {\eA\Box \eA} \&\& \eA \&\& \eA
	\arrow["{1_{\eA}\Box \mu}", from=1-1, to=1-2]
	\arrow["{\mu\Box 1_{\eA}}"', from=1-1, to=2-1]
	\arrow["\mu", from=1-2, to=2-2]
	\arrow[""{name=0, anchor=center, inner sep=0}, Rightarrow, no head, from=1-3, to=1-5]
	\arrow["{i\Box 1}"', from=1-3, to=2-4]
	\arrow["\mu", from=2-4, to=1-5]
	\arrow["\mu", from=1-7, to=2-8]
	\arrow["\alpha", shorten <=6pt, shorten >=6pt, Rightarrow, from=2-1, to=1-2]
	\arrow["\mu"', from=2-1, to=2-2]
	\arrow["{1\Box i}", from=2-6, to=1-7]
	\arrow[""{name=1, anchor=center, inner sep=0}, Rightarrow, no head, from=2-6, to=2-8]
	\arrow["\lambda"', shorten <=3pt, shorten >=3pt, Rightarrow, from=0, to=2-4]
	\arrow["\rho", shorten <=3pt, shorten >=3pt, Rightarrow, from=1-7, to=1]
\end{tikzcd}
\end{align}
$\alpha$ is called the 2-associator of the algebra, and $\lambda,\rho$ are called left and right 2-unitors. They satisfy the following consistency conditions.
\begin{align}
    \begin{tikzcd}[ampersand replacement=\&]
	{\eA\Box \eA\Box \eA\Box \eA} \& {\eA\Box \eA\Box \eA} \& {\eA\Box \eA} \&\& {\eA\Box \eA\Box \eA\Box \eA} \& {\eA\Box \eA\Box \eA} \& {\eA\Box \eA} \\
	\& \eA\Box\eA\Box\eA \&\& {=} \&\& \eA\Box\eA \\
	{\eA\Box \eA\Box \eA} \& \eA\Box\eA \& \eA \&\& {\eA\Box \eA\Box \eA} \& \eA\Box\eA \& \eA
	\arrow["{\mu 1 1}", from=1-1, to=1-2]
	\arrow[""{name=0, anchor=center, inner sep=0}, "1\mu1"', from=1-1, to=2-2]
	\arrow[""{name=0p, anchor=center, inner sep=0}, phantom, from=1-1, to=2-2, start anchor=center, end anchor=center]
	\arrow["11\mu"', from=1-1, to=3-1]
	\arrow["{\mu 1}", from=1-2, to=1-3]
	\arrow["\alpha", shorten <=1pt, shorten >=2pt, Rightarrow, from=1-2, to=2-2]
	\arrow["\mu", from=1-3, to=3-3]
	\arrow[""{name=1, anchor=center, inner sep=0}, "{\mu 1 1}", from=1-5, to=1-6]
	\arrow[""{name=1p, anchor=center, inner sep=0}, phantom, from=1-5, to=1-6, start anchor=center, end anchor=center]
	\arrow["11\mu"', from=1-5, to=3-5]
	\arrow[""{name=2, anchor=center, inner sep=0}, "{\mu 1}", from=1-6, to=1-7]
	\arrow[""{name=2p, anchor=center, inner sep=0}, phantom, from=1-6, to=1-7, start anchor=center, end anchor=center]
	\arrow["1\mu"', from=1-6, to=2-6]
	\arrow["\mu", from=1-7, to=3-7]
	\arrow[""{name=3, anchor=center, inner sep=0}, "{\mu 1}"', from=2-2, to=1-3]
	\arrow[""{name=3p, anchor=center, inner sep=0}, phantom, from=2-2, to=1-3, start anchor=center, end anchor=center]
	\arrow["1\mu", from=2-2, to=3-2]
	\arrow["\alpha", Rightarrow, from=2-6, to=3-6]
	\arrow[""{name=4, anchor=center, inner sep=0}, "\mu", from=2-6, to=3-7]
	\arrow[""{name=4p, anchor=center, inner sep=0}, phantom, from=2-6, to=3-7, start anchor=center, end anchor=center]
	\arrow[""{name=5, anchor=center, inner sep=0}, "1\mu"', from=3-1, to=3-2]
	\arrow[""{name=5p, anchor=center, inner sep=0}, phantom, from=3-1, to=3-2, start anchor=center, end anchor=center]
	\arrow[""{name=6, anchor=center, inner sep=0}, "\mu"', from=3-2, to=3-3]
	\arrow[""{name=6p, anchor=center, inner sep=0}, phantom, from=3-2, to=3-3, start anchor=center, end anchor=center]
	\arrow[""{name=7, anchor=center, inner sep=0}, "{\mu 1}", from=3-5, to=2-6]
	\arrow[""{name=7p, anchor=center, inner sep=0}, phantom, from=3-5, to=2-6, start anchor=center, end anchor=center]
	\arrow["1\mu"', from=3-5, to=3-6]
	\arrow["\mu"', from=3-6, to=3-7]
	\arrow["\alpha", shorten <=10pt, shorten >=10pt, Rightarrow, from=0p, to=5p]
	\arrow[shorten <=10pt, shorten >=10pt, Rightarrow, no head, from=1p, to=7p]
	\arrow["\alpha", shorten <=10pt, shorten >=10pt, Rightarrow, from=2p, to=4p]
	\arrow["\alpha", shorten <=10pt, shorten >=10pt, Rightarrow, from=3p, to=6p]
\end{tikzcd}
\end{align}
\begin{align}
    \begin{tikzcd}[ampersand replacement=\&]
	\eA\Box\eA \&\& {\eA\Box \eA\Box \eA} \&\& \eA\Box\eA \&\& \eA\Box\eA\Box\eA \\
	\&\&\& {=} \&\& \eA\Box\eA \\
	\eA \&\& \eA\Box\eA \&\& \eA \&\& \eA\Box\eA
	\arrow["1i1", from=1-1, to=1-3]
	\arrow["\mu"', from=1-1, to=3-1]
	\arrow[""{name=0, anchor=center, inner sep=0}, Rightarrow, no head, from=1-1, to=3-3]
	\arrow["1\mu", from=1-3, to=3-3]
	\arrow["1i1", from=1-5, to=1-7]
	\arrow[""{name=1, anchor=center, inner sep=0}, shorten <=2pt, shorten >=2pt, Rightarrow, no head, from=1-5, to=2-6]
	\arrow["\mu"', from=1-5, to=3-5]
	\arrow["{\mu 1}", from=1-7, to=2-6]
	\arrow["1\mu", from=1-7, to=3-7]
	\arrow["\mu", from=2-6, to=3-5]
	\arrow["\mu", from=3-3, to=3-1]
	\arrow[""{name=2, anchor=center, inner sep=0}, "\mu", from=3-7, to=3-5]
	\arrow["{\lambda^{-1}}", shorten <=10pt, shorten >=10pt, Rightarrow, from=1-3, to=0]
	\arrow["\rho", shorten <=16pt, shorten >=16pt, Rightarrow, from=1-7, to=1]
	\arrow["\alpha", shorten <=3pt, shorten >=3pt, Rightarrow, from=2-6, to=2]
\end{tikzcd}
\end{align}
A braided algebra is an algebra $\eA$ with a 2-isomorphism called braiding, defined as 
\begin{align}
    \begin{tikzcd}[ampersand replacement=\&]
	\eA\Box\eA \&\& \eA\Box\eA \\
	\& \eA
	\arrow["{b_{\eA,\eA}}", from=1-1, to=1-3]
	\arrow[""{name=0, anchor=center, inner sep=0}, "\mu"', from=1-1, to=2-2]
	\arrow["\mu", from=1-3, to=2-2]
	\arrow["\beta", shorten <=12pt, shorten >=12pt, Rightarrow, from=1-3, to=0]
\end{tikzcd},
\end{align}
such that the following conditions hold.
\begin{align}
    \begin{tikzcd}[ampersand replacement=\&,column sep=small]
	{\eA\Box \eA\Box \eA} \& {} \& {} \& {\eA\Box \eA\Box \eA} \&\& {\eA\Box \eA\Box \eA} \&\&\& {\eA\Box \eA\Box \eA} \\
	\& {\eA\Box \eA\Box \eA} \&\& {\eA\Box \eA} \& {=} \&\& {\eA\Box \eA} \& {\eA\Box \eA} \& {\eA\Box \eA} \\
	{\eA\Box \eA} \&\& {} \& \eA \&\& {\eA\Box \eA} \& {} \&\& \eA
	\arrow["{b_{1,23}}", from=1-1, to=1-4]
	\arrow["{b_{1,2}}", from=1-1, to=2-2]
	\arrow[""{name=0, anchor=center, inner sep=0}, "\mu1"', from=1-1, to=3-1]
	\arrow["R", shorten <=1pt, shorten >=2pt, Rightarrow, from=1-2, to=2-2]
	\arrow["1\mu", from=1-4, to=2-4]
	\arrow[""{name=1, anchor=center, inner sep=0}, "{b_{1,23}}", from=1-6, to=1-9]
	\arrow["1\mu", from=1-6, to=2-7]
	\arrow["{\mu 1}"', from=1-6, to=3-6]
	\arrow["{\mu 1}"', from=1-9, to=2-8]
	\arrow["1\mu", from=1-9, to=2-9]
	\arrow["{b_{2,3}}", from=2-2, to=1-4]
	\arrow[""{name=2, anchor=center, inner sep=0}, "1\mu", from=2-2, to=2-4]
	\arrow[""{name=2p, anchor=center, inner sep=0}, phantom, from=2-2, to=2-4, start anchor=center, end anchor=center]
	\arrow[""{name=2p, anchor=center, inner sep=0}, phantom, from=2-2, to=2-4, start anchor=center, end anchor=center]
	\arrow["{\mu 1}", from=2-2, to=3-1]
	\arrow["\mu", from=2-4, to=3-4]
	\arrow[""{name=3, anchor=center, inner sep=0}, "b", from=2-7, to=2-8]
	\arrow["{\alpha^{-1}}"', shift right=5, shorten <=2pt, shorten >=2pt, Rightarrow, from=2-7, to=3-7]
	\arrow[""{name=4, anchor=center, inner sep=0}, "\mu"'{pos=0.3}, curve={height=6pt}, from=2-7, to=3-9]
	\arrow[""{name=4p, anchor=center, inner sep=0}, phantom, from=2-7, to=3-9, start anchor=center, end anchor=center, curve={height=6pt}]
	\arrow["\alpha", shorten <=4pt, shorten >=4pt, Rightarrow, from=2-8, to=2-9]
	\arrow["\mu", from=2-8, to=3-9]
	\arrow["\mu", from=2-9, to=3-9]
	\arrow["\mu"', from=3-1, to=3-4]
	\arrow["\mu"', from=3-6, to=3-9]
	\arrow["\beta", shift left=5, shorten <=5pt, shorten >=6pt, Rightarrow, from=1-3, to=2p]
	\arrow["{b_{\mu,\eA}}"{pos=0.4}, shorten <=4pt, shorten >=9pt, Rightarrow, from=1, to=3]
	\arrow["\beta", shorten <=6pt, shorten >=6pt, Rightarrow, from=2-2, to=0]
	\arrow["{\alpha^{-1}}", shorten <=3pt, shorten >=2pt, Rightarrow, from=2p, to=3-3]
	\arrow["\beta"'{pos=0.2}, shorten >=4pt, Rightarrow, from=2-8, to=4p]
\end{tikzcd}
\end{align}
\begin{align}
\begin{tikzcd}[ampersand replacement=\&,column sep=small]
	{A\Box A\Box A} \& {A\Box A\Box A} \&\& {A\Box A\Box A} \&\& {A\Box A\Box A} \& {A\Box A\Box A} \&\& {A\Box A\Box A} \\
	\&\&\& {A\Box A} \& {=} \&\& {} \&\& {A\Box A} \\
	{} \&\& {A\Box A} \&\&\&\& {A\Box A} \\
	{A\Box A} \&\&\& A \&\& {A\Box A} \&\&\& A
	\arrow["{b_{2,3}}", from=1-1, to=1-2]
	\arrow[""{name=0, anchor=center, inner sep=0}, "1\mu"', from=1-1, to=4-1]
	\arrow["{b_{1,2}}", from=1-2, to=1-4]
	\arrow[""{name=1, anchor=center, inner sep=0}, "\mu"', from=1-2, to=3-3]
	\arrow[""{name=2, anchor=center, inner sep=0}, "1\mu", from=1-2, to=4-1]
	\arrow["1\mu", from=1-4, to=2-4]
	\arrow["{\mu 1}", shift right=2, from=1-4, to=3-3]
	\arrow["{b_{2,3}}", from=1-6, to=1-7]
	\arrow["{b_{12,3}}"', shift left, curve={height=30pt}, from=1-6, to=1-9]
	\arrow["{\mu 1}"', from=1-6, to=3-7]
	\arrow["1\mu"', from=1-6, to=4-6]
	\arrow[""{name=3, anchor=center, inner sep=0}, "{b_{1,2}}", from=1-7, to=1-9]
	\arrow["{S^{-1}}"'{pos=0.3}, shorten >=5pt, Rightarrow, from=1-7, to=2-7]
	\arrow["1\mu", from=1-9, to=2-9]
	\arrow[""{name=4, anchor=center, inner sep=0}, "\mu", from=2-4, to=4-4]
	\arrow["\mu", from=2-9, to=4-9]
	\arrow["\alpha", shorten <=20pt, shorten >=27pt, Rightarrow, from=3-3, to=3-1]
	\arrow["\mu"', from=3-3, to=4-4]
	\arrow[""{name=5, anchor=center, inner sep=0}, "b"', from=3-7, to=2-9]
	\arrow["\alpha"', shorten <=5pt, shorten >=8pt, Rightarrow, from=3-7, to=4-6]
	\arrow[""{name=6, anchor=center, inner sep=0}, "\mu"'{pos=0.3}, from=3-7, to=4-9]
	\arrow["\mu"', from=4-1, to=4-4]
	\arrow["\mu"', from=4-6, to=4-9]
	\arrow["\beta"', shorten <=7pt, shorten >=7pt, Rightarrow, from=2, to=0]
	\arrow["\beta"', shift left, shorten <=18pt, shorten >=18pt, Rightarrow, from=1-4, to=1]
	\arrow["{b_{A,\mu}}"{pos=0.8}, shift right=5, shorten <=26pt, Rightarrow, from=3, to=5]
	\arrow["{\alpha^{-1}}"', shorten <=6pt, shorten >=9pt, Rightarrow, from=4, to=3-3]
	\arrow["\beta"'{pos=0.6}, shift left=2, shorten <=15pt, shorten >=8pt, Rightarrow, from=2-9, to=6]
\end{tikzcd}
\end{align}
\begin{align}
    \begin{tikzcd}[ampersand replacement=\&]
	A \&\& {A\Box A} \&\& A \&\& {A\Box A} \\
	\&\&\& {=} \\
	A \&\& {A\Box A} \&\& A \&\& {A\Box A}
	\arrow["1i", from=1-1, to=1-3]
	\arrow[""{name=0, anchor=center, inner sep=0}, Rightarrow, no head, from=1-1, to=3-1]
	\arrow[""{name=1, anchor=center, inner sep=0}, "\mu"', from=1-3, to=3-1]
	\arrow["b", from=1-3, to=3-3]
	\arrow["1i", from=1-5, to=1-7]
	\arrow[""{name=2, anchor=center, inner sep=0}, Rightarrow, no head, from=1-5, to=3-5]
	\arrow[""{name=3, anchor=center, inner sep=0}, "i1", from=1-5, to=3-7]
	\arrow["b", from=1-7, to=3-7]
	\arrow["\mu", from=3-3, to=3-1]
	\arrow["\mu", from=3-7, to=3-5]
	\arrow["\rho"', shift right=2, shorten <=16pt, shorten >=11pt, Rightarrow, from=1-3, to=0]
	\arrow["{b_{i,A}}"', shorten <=8pt, shorten >=11pt, Rightarrow, from=1-7, to=3]
	\arrow["\beta", shorten >=5pt, Rightarrow, from=3-3, to=1]
	\arrow["{\lambda^{-1}}"', shift left=3, shorten <=22pt, shorten >=11pt, Rightarrow, from=3-7, to=2]
\end{tikzcd}
\end{align}
Here $b_{f,A}$ denotes the braiding between the 1-morphism $f$ and the object $A$, and $b_{A,f}$ denotes the braiding between the object $A$ and the 1-morphism $f$~\cite{Kong_2020center,D_coppet_2024}.  

A condensable algebra is a connected separable braided algebra. We refer readers to~\cite{D_coppet_2024} for definition of connectivity and separability of an algebra. The magnetic simple algebras $\eA[N,\phi,\sigma,\alpha,\beta]$ are by construction connected and separable.

For a magnetic simple algebra $\eA[N,\phi,\sigma,\alpha,\beta]$, $N$ and $\phi$ determine the underlying object of the algebra. $\sigma$ determines the algebra product $\mu$. $\alpha$ is the 2-associator. $\beta$ is the braiding on the algebra. The unitors $\lambda,\rho$ are identities. 
\section{Some details of $\mathcal{Z}[2\vc_G]$\label{app:Z[2vcG]}}
Here we give some details of the theory $\mathcal{Z}[2\vc_G]$. This braided fusion 2-category was first studied in details in~\cite{Kong_2020center}. However we use a slightly different(but equivalent) definition, which is the categorification of the definition of $\mathcal{Z}[\vc_G]$ in~\cite{davydov2016algebras}.
\paragraph{
(Objects)}
An object in $\mathcal{Z}[2\vc_G]$ is a triple $X=(\overline{X},\e{L},\phi)$. 

\begin{itemize}
    \item $\overline{X}\in 2\vc_G$ is a $G$-graded finite semi-simple category, with decomposition $\overline{X}=\boxplus_g \overline{X}_g,~\overline{X}_g\in 2\vc$. 
    \item $\e{L}=\{\e{L}_g|g\in G\}$ is a collection of functors: $\e{L}_g\in 2\vc[\overline{X},\overline{X}]$, such that $\e{L}_g(\overline{X}_h)\subset \overline{X}_{\cg{g}{h}}$,  where $\cg{g}{h}:=ghg^{-1}$. We denote the action of $\e{L}_g$ on objects and morphisms of $\overline{X}$ as $g*$.
    \item $\phi=\{\phi(g,h)|g,h\in G\}$ is a collection of natural equivalences: $\L_g\circ \L_h\To \L_{gh}$, such that for all $g_i\in G,~a\in \sf{Obj}(\overline{X})$ the following diagram commutes
\begin{align}
\begin{tikzcd}[ampersand replacement=\&]
	{g_1*g_2*g_3*a} \&\&\& {g_1*g_2g_3*a} \\
	{g_1g_2*g_3*a} \&\&\& {g_1g_2g_3*a}
	\arrow["{g_1*(\phi(g_2,g_3)(a))}", from=1-1, to=1-4]
	\arrow["{\phi(g_1,g_2)(g_3*a)}"', from=1-1, to=2-1]
	\arrow["{\phi(g_1,g_2g_3)(a)}", from=1-4, to=2-4]
	\arrow["{\phi(g_1g_2,g_3)(a)}"', from=2-1, to=2-4]
\end{tikzcd}\label{eq:phi eq}
\end{align}
\end{itemize}
$(\e{L}, \phi)$ equips the $G$-graded 1-category $\overline{X}$ with a $G$-crossed structure.  Therefor an object of $\mathcal{Z}[2\vc_G]$ is simply a $G$-crossed finite semi-simple category.

\paragraph{
(1-morphism.)}A 1-morphism between two objects $X=(\overline{X},\e{L}^X,\phi^X), Y=(\overline{Y},\e{L}^Y,\phi^Y)$ is a pair $f=(\overline{f},\widetilde{f})$. 

$\overline{f}\in 2\vc_G[\overline{X},\overline{Y}]$ is a 1-morphism in $2\vc_G$. $\widetilde{f}=\{\widetilde{f}_g\}$ is a collection of natural transformations: $\widetilde{f}_g: \overline{f}\circ \e{L}^X_g\To \e{L}^Y_g\circ \overline{f}$, such that for any $g,h\in G~, a\in \sf{Obj}(\overline{X})$, the following diagram commutes
\begin{align}
    \begin{tikzcd}[ampersand replacement=\&]
	{\overline{f}(g*_Xh*_Xa)} \& {g*_Y(\overline{f}(h*_Xa))} \& {g*_Yh*_Y(\overline{f}(a))} \\
	{\overline{f}(gh*_Xa)} \&\& {gh*_Y\overline{f}(a)}
	\arrow["{\widetilde{f}_g(h*_Xa)}", rightarrow, from=1-1, to=1-2]
	\arrow["{\overline{f}(\phi_X(g,h)(a))}"', rightarrow, from=1-1, to=2-1]
	\arrow["{g*_Y(\widetilde{f}_h(a))}", rightarrow, from=1-2, to=1-3]
	\arrow["{\phi_Y(g,h)(\overline{f}(a))}", rightarrow, from=1-3, to=2-3]
	\arrow["{\widetilde{f}_{gh}(a)}"', rightarrow, from=2-1, to=2-3]
\end{tikzcd}\label{eq:f2}
\end{align}
\paragraph{(2-morphism).} Let $f,g: X\to Y$ be two 1-morphisms, a 2-morphism $\eta: f\To g$ is a 2-morphism in $2\vc_G$: $\eta: \overline{f}\To \overline{g}$, such that for any $a\in \obj(\overline{X}),h\in G$, the following diagram commutes
    \begin{align}
        \begin{tikzcd}[ampersand replacement=\&]
	{\overline{f}(h*a)} \& {h*\overline{f}(a)} \\
	{\overline{g}(h*a)} \& {h*\overline{g}(a)}
	\arrow["{\widetilde{f}_h}", from=1-1, to=1-2]
	\arrow["\eta", from=1-1, to=2-1]
	\arrow["{h*\eta}", from=1-2, to=2-2]
	\arrow["{\widetilde{g}_h}"', from=2-1, to=2-2]
\end{tikzcd}
    \end{align}

\paragraph{(Braiding)}
Braiding between two objects $A,B$ is a 1-morphism $b_{A,B}=(\overline{b}_{A,B},\widetilde{b}_{A,B}): A\Box B\to B\Box A$, defined as follows. 
\begin{align}
    \overline{b}_{A,B}: &\overline{A}\boxtimes \overline{B}\to \overline{B}\boxtimes \overline{A},\\
    & a\boxtimes b\mapsto |a|*b\boxtimes a.\\
    \widetilde{b}^g_{A,B}:& \bar{b}_{A,B}(g*(a\boxtimes b))=\cg{g}{|a|}*g*b\boxtimes g*a\xrightarrow{\phi^B(g,|a|)^{-1}(b)\circ\phi^B(\cg{g}{|a|},g)(b)} g*|a|*b\boxtimes g*a=g*\widetilde{b}_{A,B}(a\boxtimes b)
\end{align}
The hexagonators are defined as follows:
\[\begin{tikzcd}[ampersand replacement=\&]
	{A\Box B\Box C} \&\& {B\Box C\Box A} \\
	\& {B\Box A\Box C}
	\arrow[""{name=0, anchor=center, inner sep=0}, "{b_{A,B\Box C}}", from=1-1, to=1-3]
	\arrow["{b_{A,B}}"', from=1-1, to=2-2]
	\arrow["{b_{A,C}}"', from=2-2, to=1-3]
	\arrow["{R_{A,B,C}}", shorten <=3pt, Rightarrow, from=0, to=2-2]
\end{tikzcd}\]
for any $a\in \sf{Obj}(\overline{A}),b\in \sf{Obj}(\overline{B}),c\in \sf{Obj}(\overline{C})$, 
\begin{align}
    R_{A,B,C}(a\boxtimes b\boxtimes c)=\sf{id}: |a|*b\boxtimes |a|*c\boxtimes a\to |a|*(b\boxtimes c)\boxtimes a
\end{align}.
\[\begin{tikzcd}[ampersand replacement=\&]
	{A\Box B\Box C} \&\& {C\Box A\Box B} \\
	\& {A\Box C\Box B}
	\arrow[""{name=0, anchor=center, inner sep=0}, "{b_{A\Box B, C}}", from=1-1, to=1-3]
	\arrow["{b_{B,C}}"', from=1-1, to=2-2]
	\arrow["{b_{A,C}}"', from=2-2, to=1-3]
	\arrow["{S_{A,B,C}}", shorten <=3pt, Rightarrow, from=0, to=2-2]
\end{tikzcd}\]
\begin{align}
    S_{A,B,C}(a\boxtimes b\boxtimes c): (|a||b|)*c\boxtimes a\boxtimes b\xrightarrow{(\phi^C(|a|,|b|))^{-1}}(c)|a|*|b|*c\boxtimes a\boxtimes b
\end{align}
The consistency conditions that should be satisfied by $R,S$ are supported by the condition Eq.\eqref{eq:phi eq} satisfied by $\phi$. 

\section{Low dimensional gSPTs\label{app:low_d_gSPT}}
Here we show that the classification of 1+1D and 0+1D gSPTs are $H^2_{qa}[(G,N),U(1)]$ and $H_{qa}^1[(G,N),U(1)]$. We may write the consistency conditions on a quasi-abelian 3-cocycle $\kappa$ as $\delta^{(3)}\kappa=1$, and the quasi-abelian coboundary condition as $\kappa=\delta^{(2)}\vartheta$, where $\vartheta=(\epsilon,\eta)$ is a pair of functions. Therefore it is natural to define a quasi-abelian 2-cocycle as a pair $\epsilon\in C^1[N,C^1[G,\bC^\times]],~\eta\in C^2[N,\bC^\times]$ such that $\delta^{(2)}\vartheta=1$, i.e. 
\begin{align}
    d_N\eta=d_G\epsilon=d_N\epsilon d_G\eta=1,~\epsilon(n_1,n_2)=\frac{\eta(n_1,n_2)}{\eta(\cg{n_1}{n_2},n_1)}.
\end{align}
This is exactly the classification of 1+1D gSPT~\cite{wen2023classification}, which is also the classification of magnetic condensable algebras in $\mathcal{Z}[\vc_G]$~\cite{davydov2016algebras}. In particular, we have $\eta\in Z^2[N,U(1)]$, which describes the SPT-class of the gapped degrees of freedom, and $\epsilon\in C^1[N,Z^1[G,U(1)]]$ describes the $G$-charges decorated on $N$-domain walls. The above conditions make sure there is no anomaly for $G$, but the IR symmetry $K$ may be anomalous. Thus 1+1D gSPTs are given by elements of $Z^2_{qa}[(G,N),U(1)]$.  Furthermore, the 1+1D gSPT is trivial if $\epsilon,\delta$ are given by
\begin{align}
    \epsilon(n,g)=\rho(\cg{g}{n})/\rho(n)=d_G\rho,~\eta(n_1,n_2)=d_N\rho, \rho\in C^1[N,\bC^\times].
\end{align}
We may write this as a coboundary condition $\vartheta=\delta^{(1)}\rho$, then equivalent classes of 1+1D gSPTs are classified by $H^2_{qa}[(G,N),U(1)]$.  

A natural definition of a quasi-abelian 1-cocyle is then a function $\rho\in C^1[N,\bC^\times]$ such that $\delta^{(1)}\rho=1$, i.e. $d_G\rho=d_N\rho=1$. This is indeed the classification of 0+1D gSPTs. A 0+1D gSPT is simply a quantum mechanical system with symmetry $G$, whose ground space has a trivial $N$-action: $(U_n)|_{GS}\propto \bb{1} $. Therefore we can write $U_n=\rho(n)\in U(1)$. The conditions $d_N\rho=d_G\rho=1$ state that the ground space has a well-defined $N$-charge, and the $N$-charge is invariant under action by $G$. The effective symmetry of the ground space is $K$ and satisfies $U_{k_1}U_{k_2}=U_{e_2(k_1,k_2)}U_{k_1k_2}=\rho(e_2(k_1,k_2))U_{k_1k_2}$, where $e_2$ is the extension class of $1\to N\to G\to K\to 1$.  In other words, the IR symmetry of the 0+1D gSPT could act projectively on the ground space, with projective phase $\rho(e_2(k_1,k_2))$. This is the emergent anomaly of the 0+1D gSPT. When the emergent anomaly is nontrivial, the ground space must be degenerated, and this degeneracy can not be lifted without closing the gap to excited states first. 

\section{Coupled layer construction for a string condensation\label{sec:coupled layer}}
\begin{figure}[h]
    \centering
    \includegraphics[width=0.8\linewidth]{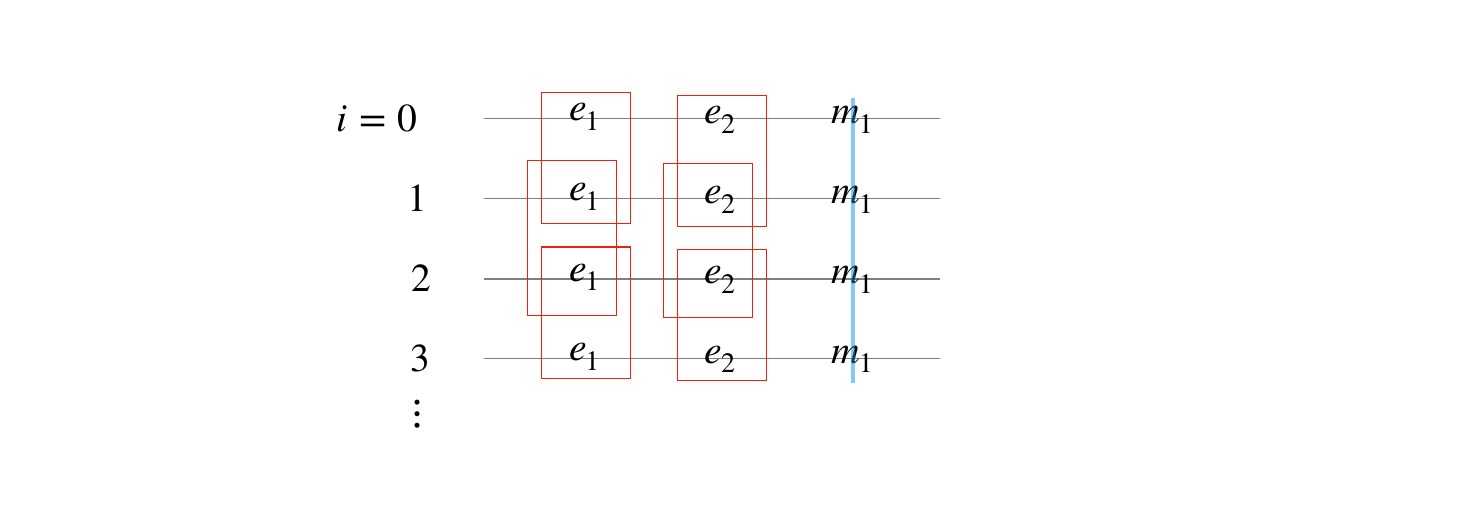}
    \caption{Coupled layer construction of a gapped boundary. The top layer is the boundary layer, hosting a twisted $\bZ_2^2$ gauge theory. Other layers have untwisted $\bZ_2^2$ gauge theory. $e_{1,2}$ pairs in adjacent layers are condensed(red boxes). The bulk is a 3+1D untwisted $\bZ_2^2$ gauge theory. An $m_1$ string(blue) ends on the top layer. Two such strings fuse to an $e_2$ in the top layer, since in the top layer we have $m_1\otimes m_1=e_2$.}
    \label{fig:coupled layer}
\end{figure}
Here we provide a coupled layer construction for the string condensation studied in Sec.~\ref{sec:type-II SPT}, dual to the 2+1D type-II SPT with symmetry $\bZ_2^2$. We consider layers of 2+1D $\bZ_2^2$ gauge theory, and denote the gauge charges and fluxes in layer $i$ as $e^{(i)}_{1,2},m^{(i)}_{1,2}$. The top layer($i=0$) is a twisted $\bZ_2^2$ gauge theory with twist given by a type-II cocycle. Other layers have untwisted $\bZ_2^2$ gauge theory. We condense the inter-layer charge pairs $e^{(i)}_{1}e^{(i+1)}_{1},e^{(i)}_{2}e^{(i+1)}_{2}$.
Then the charges $e^{(i)}_{1,2}$ in all layers are identified, and become a deconfined charge $e_{1,2}$. Strings of $m_1,m_2$ are also deconfined. The $m_{1,2}$ strings can end on the top layer. Consider an $m_1$ string that ends on the top layer with the endpoint being $m_1^{(0)}$. Since in the top layer $m_1^{(0)}\otimes m_1^{(0)} =e_2^{(0)}$, we see that fusing two such $m_1$ strings results in an $e_2$ particle in the top layer. Similarly, fusing two $m_2$ strings that end on the top boundary results in an $e_1$ particle in the top layer.  It is also clear that braiding the endpoints of an $m_1$ string and an $m_2$ strings results in a phase $i$.
\end{document}